\documentclass[preprint]{ptephy_v1}

\usepackage{braket}
\usepackage{comment}

\preprintnumber{XXXX-XXXX} 

\def\Journal#1#2#3#4{{#1} {\bf #2}, #3 (#4)}

\def\AandA{Astron. Astrophys.}

\def\CQG{Class. Quantum Grav.}
\def\CPC{Chin. Phys. C}
\def\EPJC{Eur. Phys. J. C}
\def\EPL{EPL}
\def\IJMPA{Int. J. Mod. Phys. A}

\def\JCAP{J. Cosmol. Astropart. Phys.}
\def\JHEP{J. High Energy Phys.}

\def\JPG{J. Phys. G} 
\def\JPCS{J. Phys. Conf. Ser.} 
\def\JPGNP{J. Phys. G: Nucl. Part. Phys.} 
\def\MPLA{Mod. Phys. Lett. A}

\def\NPB{Nucl. Phys. B}

\def\PLB{{Phys. Lett.} B}

\def\PREP{Phys. Rep.}

\def\PRL{Phys. Rev. Lett.}
\def\PRD{Phys. Rev. D}

\def\PTP{Prog. Theor. Phys.}
\def\PTEP{Prog. Theor. Exp. Phys.}
\def\RMP{Rev. Mod. Phys.}

\def\SCIENCE{Science}

\begin{document}

\title{New magic textures of Majorana neutrinos and baryon asymmetry of the Universe}


\author{Yuta Hyodo, Teruyuki Kitabayashi\footnote{Corresponding author}}
\affil{Department of Physics, Tokai University,
4-1-1 Kitakaname, Hiratsuka, Kanagawa 259-1292, Japan\email{teruyuki@tokai-u.jp}}


\begin{abstract}
The magic texture is one of the successful textures of the flavor neutrino mass matrix for Majorana neutrinos. In this paper, it turns out that two new types of magic textures are also consistent with the neutrino oscillation experiments, observation of cosmic microwave background radiation, and neutrinoless double beta decay experiments. The connection between these new magic textures and the leptogenesis scenario of the origin of the baryon asymmetry of the Universe is also discussed.
\end{abstract}

\subjectindex{}

\maketitle

\section{Introduction\label{section:introduction}}
Understanding the nature of the flavor structure of elementary particles is one of the outstanding problems in particle physics \cite{King2015JPG,Feruglio2019}. To solve the flavor puzzle, many texture ansatz is proposed, such as tri-bi maximal texture \cite{Harrison2002PLB,Xing2002PLB,Harrison2002PLB2,Kitabayashi2007PRD}, texture zeros \cite{Berger2001PRD,Frampton2002PLB,Xing2002PLB530,Xing2002PLB539,Kageyama2002PLB,Xing2004PRD,Grimus2004EPJC,Low2004PRD,Low2005PRD,Grimus2005JPG,Dev2007PRD,Xing2009PLB,Fritzsch2011JHEP,Kumar2011PRD,Dev2011PLB,Araki2012JHEP,Ludle2012NPB,Lashin2012PRD,Deepthi2012EPJC,Meloni2013NPB,Meloni2014PRD,Dev2014PRD,Felipe2014NPB,Ludl2014JHEP,Cebola2015PRD,Gautam2015PRD,Dev2015EPJC,Kitabayashi2016PRD1,Zhou2016CPC,Singh2016PTEP,Bora2017PRD,Barreiros2018PRD,Kitabayashi2018PRD,Barreiros2019JHEP,Capozzi2020PRD,Singh2020EPL,Barreiros2020,Kitabayashi2020PRD,Kitabayashi2017IJMPA,Kitabayashi2017IJMPA2,Kitabayashi2019IJMPA}, $\mu -\tau$ symmetric texture \cite{Fukuyama1997,Lam2001PLB,Ma2001PRL,Balaji2001PLB,Koide2002PRD,Kitabayashi2003PRD,Koide2004PRD,Aizawa2004PRD,Ghosal2004MPLA,Mohapatra2005PRD,Koide2005PLB,Kitabayashi2005PLB,Haba206PRD,Xing2006PLB,Ahn2006PRD,Joshipura2008EPJC,Gomez-Izquierdo2010PRD,He2001PRD,He2012PRD,Gomez-Izquierdo2017EPJC,Fukuyama2017PTEP,Kitabayashi2016IJMPA,Kitabayashi2016PRD,Bao2021arXiv}, and textures under discrete symmetries e.g., $A_n$ and $S_n$ \cite{Altarelli2010PMP}. 

The {\it magic texture} which is parameterized as 
\begin{eqnarray}
&& \qquad \quad
\left(
\begin{matrix}
a & b & c \\
b & d & a+c-d \\
c & a+c-d & b-c+d
\end{matrix}
\right)
\begin{array}{c}
\leftarrow a+b+c \\
\leftarrow a+b+c \\
\leftarrow a+b+c \\
\end{array}
\label{Eq:magicTextureHarrison}
\end{eqnarray}
is one of the successful textures of the flavor neutrino mass matrix for Majorana neutrinos \cite{Harrison2004PLB, Lam2006PLB}. The applications of the magic texture for Majorana neutrinos have been studied for texture zeros \cite{Gautam2016PRD,Yang2021arXiv}, with two simple extensions \cite{Channey2019JPGNP} and for baryon asymmetry of the Universe \cite{Verma2020JPGNP}. Magic textures for Dirac neutrinos are also discussed \cite{Hyodo2020IJMPA}.

Now, we would like to introduce the {\it magic square} \cite{Levitin2011}. A magic square of order $n$ is an $n \times n$ square grid filled with distinct natural numbers. The sum of the numbers in each row, column, and diagonal are equal. The sum is called a magic constant or magic sum. For example, the magic sum of the following magic square of order 3 is 15:
\begin{eqnarray}
\begin{array}{|c|c|c|}
\hline
 2 & 7 & 6  \\
\hline
 9 & 5 & 1 \ \\
\hline
 4 & 3 & 8 \\  
\hline
\end{array}
\begin{array}{c}
\leftarrow 15 \\
\leftarrow 15 \\
\leftarrow 15 \\
\end{array}
\quad
\nonumber \\
\begin{array}{ccccccc}
& \nearrow & \uparrow & \uparrow & \uparrow & \nwarrow &\\
15 & & 15 & 15 & 15 & &15 \\
\end{array}.
\end{eqnarray}

Even though the magic square has been known for a long time, discoveries about magic square are being made in the field of linear algebra \cite{Sallows1997MathIntelli,Loly2009LLA,Nordgren2012LAA,Nordgren2020,Nordgren2021}. Moreover, there is an application of the magic square in theoretical physics \cite{Borsten2017CQG}. 

The {\it magic texture} is related to the {\it magic square}. The magic texture has a part of the nature of the magic square, e.g., the sum of the elements in each row and each column is equal to $a+b+c$. This magic texture was obtained as one of the consequences of the so-called trimaximal mixing for $\nu_2$ \cite{Harrison2004PLB, Lam2006PLB}. 

In this paper, we reverse our way of thinking about the magic texture. The magic texture is required as the first principle. In this viewpoint, trimaximal mixing for $\nu_2$ is one of the consequences of the magic texture. Since the problem of the texture of the neutrino mass matrix is long-standing in particle physics, changing viewpoints may be helpful for a breakthrough in the future. 

The magic texture could be defined as a Majorana matrix in which three independent sums are the same, such as Eq. (\ref{Eq:magicTextureHarrison}).  Under this definition, not only Eq. (\ref{Eq:magicTextureHarrison}) but also new nine matrices are classified into the magic textures. 

The paper is organized as follows. In Sec.\ref{section:magicTextures}, we classify the magic textures and show some useful relations of the neutrino mixings and mass matrix. An analytical method to obtain the magic textures is also proposed. In Sec. \ref{section:Allowed Magic Textures}, first, we  show that the following two new types of magic textures
\begin{eqnarray}
&& \qquad \quad
\left(
\begin{matrix}
a & b & c \\
b & d & a-c+d \\
c & a-c+d & b-c+d
\end{matrix}
\right)
\begin{array}{c}
 \\
\leftarrow a+b-c+2d \\
\leftarrow a+b-c+2d \\
\end{array}
\nonumber \\
&&\begin{array}{ccccccc}
& &  &  & & \qquad \qquad \qquad \qquad   \qquad \  \nwarrow &\\
& & & & & &a+b-c+2d \\
\end{array}
\label{Eq:type-IV-intro}
\end{eqnarray}
and 
\begin{eqnarray}
&& \qquad \quad
\left(
\begin{matrix}
a & b & c \\
b & d & 2c-b \\
c & 2c-b & -a+2c
\end{matrix}
\right)
\begin{array}{c}
 \\
\leftarrow d+2c \\
  \\
\end{array}
\nonumber \\
&&\begin{array}{ccccccc}
\qquad \nearrow &  &   &   &  &  \qquad   \qquad  \qquad \nwarrow &\\
d+2c & & & & & &d+2c \\
\end{array}
\label{Eq:type-IX-intro}
\end{eqnarray}
are also consistent with experiments. Next, the allowed parameter region and correlations between these parameters are shown.  In Sec. \ref{section:leptogeneisis}, the connection between these new magic textures and the leptogenesis scenario of the origin of the baryon asymmetry of the Universe is discussed. Section \ref{section:summary} is devoted to a summary.

\section{Magic textures\label{section:magicTextures}}
\subsection{Classification \label{section:classification}}
Since the Majorana neutrino flavor mass matrix
\begin{eqnarray}
M =\left(
\begin{matrix}
M_{ee} & M_{e\mu} & M_{e\tau} \\
M_{\mu e} & M_{\mu\mu} & M_{\mu\tau} \\
M_{\tau e} & M_{\tau\mu} & M_{\tau\tau} \\
\end{matrix}
\right)
=
\left(
\begin{matrix}
a & b & c \\
b & d & e \\
c& e & f \\
\end{matrix}
\right),
\end{eqnarray}
is a symmetric matrix, there are five independent sums, $\{S_1, S_2, \cdots, S_5\}$, schematically:
\begin{eqnarray}
&& \qquad \quad
\left(
\begin{matrix}
a & b & c \\
b & d & e \\
c & e & f
\end{matrix}
\right)
\begin{array}{c}
\leftarrow S_1 \\
\leftarrow S_2 \\
\leftarrow S_3 \\
\end{array}
\nonumber \\
&&\begin{array}{ccccccc}
& \nearrow \ \  \uparrow \ \ \uparrow \ \ \uparrow \ \nwarrow &\\
& S_5 \quad   S_1 \  S_2 \  S_3 \ \ \ S_4 \ \\
\end{array} 
\end{eqnarray}
where
\begin{eqnarray}
S_1&=&M_{ee} + M_{e\mu} + M_{e\tau} = M_{ee} + M_{\mu e} + M_{\tau e} = a+b+c, \nonumber \\
S_2&=& M_{\mu e} + M_{\mu \mu} + M_{\mu\tau} = M_{e\mu} + M_{\mu\mu} + M_{\tau\mu} = b+d+e, \nonumber \\
S_3&=& M_{\tau e} + M_{\tau \mu} + M_{\tau \tau} = M_{e\tau} + M_{\mu\tau} + M_{\tau\tau} = c+e+f
\label{Eq:S1S2S3}
\end{eqnarray}
for $i$th raw ($i$th column) and
\begin{eqnarray}
S_4&=&M_{ee} + M_{\mu\mu} + M_{\tau\tau} = a+d+f, \nonumber \\
S_5&=&M_{e\tau} + M_{\mu\mu} + M_{\tau e} = 2c + d,
\label{Eq:S4S5}
\end{eqnarray}
for diagonal elements. 

These five sums $\{S_1, S_2, \cdots, S_5\}$ can be used to classify the type of magic texture of the Majorana neutrino mass matrix. Based on the success of the magic texture in Eq. (\ref{Eq:magicTextureHarrison}), we require that three of five sums be the same in the Majorana neutrino mass matrix \footnote{We will investigate more six types of magic textures without this requirement in the last of Sec. \ref{section:Allowed Magic Textures}.}. Under this requirement, there are $_5C_3=10$ types of magic textures. We call these ten textures type I, type II, $\cdots$, and type X magic texture with the following definitions:
\begin{description}
\item[Type I:  ] $S_1=S_2=S_3 \neq S_4 \neq S_5$
\begin{eqnarray}
M_{\rm I} =\left(
\begin{matrix}
a & b & c \\
b & d & a+c-d \\
c & a+c-d & b-c+d
\end{matrix}
\right).
\end{eqnarray}
\item[Type II: ]$S_1=S_2=S_4 \neq S_3 \neq S_5$
\begin{eqnarray}
M_{\rm II} = \left(
\begin{matrix}
a & b & c \\
b & d & a+c-d \\
c & a+c-d & b+c-d
\end{matrix}
\right).
\end{eqnarray}
\item[Type III: ]$S_1=S_3=S_4 \neq S_2 \neq S_5$
\begin{eqnarray}
M_{\rm III} = \left(
\begin{matrix}
a & b & c \\
b & d & a-c+d \\
c & a-c+d & b+c-d
\end{matrix}
\right).
\end{eqnarray}
\item[Type IV: ]$S_2=S_3=S_4 \neq S_1 \neq S_5$
\begin{eqnarray}
M_{\rm IV} = \left(
\begin{matrix}
a & b & c \\
b & d & a-c+d \\
c & a-c+d & b-c+d
\end{matrix}
\right).
\end{eqnarray}
\item[Type V: ]$S_1=S_2=S_5 \neq S_3 \neq S_4$
\begin{eqnarray}
M_{\rm V} = \left(
\begin{matrix}
a & b & c \\
b & a+b-c & -b+2c \\
c & -b+2c & f
\end{matrix}
\right).
\end{eqnarray}
\item[Type VI: ]$S_1=S_3=S_5 \neq S_2 \neq S_4$
\begin{eqnarray}
M_{\rm VI} = \left(
\begin{matrix}
a & b & c \\
b & a+b-c & e \\
c & e & a+b-e
\end{matrix}
\right).
\end{eqnarray}
\item[Type VII: ]$S_2=S_3=S_5 \neq S_1 \neq S_4$
\begin{eqnarray}
M_{\rm VII} = \left(
\begin{matrix}
a & b & c \\
b & d & -b+2c \\
c & -b+2c & b-c+d 
\end{matrix}
\right).
\end{eqnarray}
\item[Type VIII: ]$S_1=S_4=S_5 \neq S_2 \neq S_3$
\begin{eqnarray}
M_{\rm VIII} = \left(
\begin{matrix}
a & b & c \\
b & a+b-c & e \\
c & e & -a+2c
\end{matrix}
\right).
\end{eqnarray}
\item[Type IX: ]$S_2=S_4=S_5 \neq S_1 \neq S_3$
\begin{eqnarray}
M_{\rm IX} = \left(
\begin{matrix}
a & b & c \\
b & d & 2c-b \\
c & 2c-b & -a+2c
\end{matrix}
\right).
\end{eqnarray}
\item[Type X: ]$S_3=S_4=S_5 \neq S_1 \neq S_2$
\begin{eqnarray}
M_{\rm X} = \left(
\begin{matrix}
a & b & c \\
b & d & a-c+d \\
c & a-c+d & -a+2c
\end{matrix}
\right).
\end{eqnarray}
\end{description}
We note that the type I magic texture is the traditional magic texture in Eq. (\ref{Eq:magicTextureHarrison}).

\subsection{Useful relations}
\subsubsection{Mixing matrix and mass matrix \label{subsection:mixin_matrix_and_mass_matrix}}
The charged lepton mass matrix is assumed to be diagonal and real. The flavor neutrino mass matrix $M$ is related to the diagonal neutrino mass matrix
\begin{eqnarray}
M =  U^\ast {\rm diag.}(\lambda_1,\lambda_2,\lambda_3) U^\dag,
\label{Eq:M=UMU}
\end{eqnarray}
where
\begin{eqnarray}
\lambda_1=m_1e^{2i\alpha_1}, \quad
\lambda_2=m_2e^{2i\alpha_2}, \quad
\lambda_3=m_3,
\label{Eq:lambda_123}
\end{eqnarray}
$m_i$ $(i=1,2,3)$ is a neutrino mass eigenvalues and
\begin{eqnarray}
U = \left(
\begin{array}{ccc}
U_{e1}  & U_{e2} & U_{e3} \\
U_{\mu 1}  & U_{\mu 2} & U_{\mu 3} \\
U_{\tau 1}  & U_{\tau 2} & U_{\tau 3} \\
\end{array}
\right),
\label{Eq:UPMNS}
\end{eqnarray}
with
\begin{eqnarray}
U_{e1} &=& c_{12}c_{13}, \quad U_{e 2} = s_{12}c_{13}, \quad U_{e 3} = s_{13} e^{-i\delta},  \label{Eq:UPMNS_elements} \\
U_{\mu 1} &=&- s_{12}c_{23} - c_{12}s_{23}s_{13} e^{i\delta}, \quad
U_{\mu 2} =  c_{12}c_{23} - s_{12}s_{23}s_{13}e^{i\delta}, \quad U_{\mu 3} = s_{23}c_{13}, \nonumber \\
U_{\tau 1} &=& s_{12}s_{23} - c_{12}c_{23}s_{13}e^{i\delta},\quad 
U_{\tau 2} = - c_{12}s_{23} - s_{12}c_{23}s_{13}e^{i\delta}, \quad U_{\tau 3} = c_{23}c_{13},\nonumber 
\end{eqnarray}
denotes the Pontecorvo-Maki-Nakagawa-Sakata (PMNS) mixing matrix \cite{Pontecorvo1957,Pontecorvo1958,Maki1962PTP,PDG}. We use abbreviations $c_{ij}=\cos\theta_{ij}$ and $s_{ij}=\sin\theta_{ij}$  ($i,j$=1,2,3) where $\theta_{ij}$ is a neutrino mixing angle. The Dirac CP phase is denoted by $\delta$ and the Majorana CP phases are denoted by $\alpha_1$ and $\alpha_2$. 

The sine and cosine of the three mixing angles of the PMNS matrix $U$ are given by
\begin{eqnarray}
&& s_{12}^2=\frac{|U_{e2}|^2}{1-|U_{e3}|^2}, \quad s_{23}^2=\frac{|U_{\mu 3}|^2}{1-|U_{e3}|^2}, \quad s_{13}^2=|U_{e3}|^2, \nonumber \\
&& c_{12}^2= \frac{|U_{e1}|^2}{1-|U_{e3}|^2}, \quad c_{23}^2 = \frac{|U_{\tau 3}|^2}{1-|U_{e3}|^2}.
\label{Eq:mixingAngle_from_UPMNS}
\end{eqnarray}
The Jarlskog rephasing invariant which is a measure of CP violation \cite{Jarlskog1986PRL}
\begin{eqnarray}
J &=& {\rm Im} (U_{e1}U_{e2}^*U_{\mu 1}^*U_{\mu 2}) 
   = s_{12}s_{23}s_{13}c_{12}c_{23}c_{13}^2\sin\delta,
\label{Eq:Jarlskog}
\end{eqnarray}
is useful to calculate the Dirac phase $\delta$ in Eq. (\ref{Eq:UPMNS}) from any representation of the mixing matrix.  

\subsubsection{Realization of magic textures \label{subsection:realization_of_magic_textures}}
We show an analytical method to construct a magic texture. The five sums $\{S_1, S_2, \cdots, S_5\}$ of a Majorana mass matrix could be written by
\begin{eqnarray}
S_i=C_i^1 \lambda_1  + C_i^2 \lambda_2  + C_i^3\lambda_3, \quad (i=1,2,3,4,5)
\label{Eq:S_i=Clambda}
\end{eqnarray}
where
\begin{eqnarray}
C_1^1 &=&c_{12}c_{13}\left(s_{12}(-c_{23}+s_{23}) +c_{12}(c_{13}-e^{-i\delta}s_{13}(c_{23}+s_{23}) \right),
\nonumber \\
C_1^2 &=&c_{13}s_{12}\left(c_{12}(c_{23}-s_{23}) +s_{12}(c_{13}-e^{-i\delta}s_{13}(c_{23}+s_{23}))\right),
\nonumber \\
C_1^3 &=&e^{i\delta}s_{13}\left(e^{i\delta}s_{13}+c_{13}(c_{23}+s_{23})\right),
\end{eqnarray}
\begin{eqnarray}
C_2^1 &=&e^{-2i\delta}(e^{i\delta}c_{23}s_{12}+c_{12}s_{13}s_{23}) \nonumber \\
&& \times \left(c_{12}s_{13}(c_{23}+s_{23}) -e^{i\delta}(c_{13}c_{13}+s_{12}(-c_{23}+s_{23})) \right),
\nonumber \\
C_2^2 &=&e^{-2i\delta}(-e^{i\delta}c_{12}c_{23}+s_{12}s_{13}s_{23}) \nonumber \\
&& \times \left(-e^{i\delta}(c_{13}s_{12}+c_{12}(c_{23}-s_{23})) +s_{12}s_{13}(c_{23}+s_{23})\right),
\nonumber \\
C_2^3 &=&c_{13}s_{23}\left(e^{i\delta}s_{13}+c_{13}(c_{23}+s_{23})\right),
\end{eqnarray}
\begin{eqnarray}
C_3^1 &=&e^{-2i\delta}(c_{12}c_{23}s_{13}-e^{i\delta}s_{12}s_{23}) \nonumber \\
&& \times \left(c_{12}s_{13}(c_{23}+s_{23})  -e^{i\delta}(c_{12}c_{13}+s_{12}(-c_{23}+s_{23})) \right),
\nonumber \\
C_3^2 &=&e^{-2i\delta}(c_{23}s_{12}s_{12}+e^{i\delta}c_{12}s_{23}) \nonumber \\
&& \times \left(-e^{i\delta}(c_{13}s_{12}+c_{12}(c_{23}-s_{23})) +s_{12}s_{13}(c_{23}+s_{23})\right.),
\nonumber \\
C_3^3 &=&c_{13}c_{23}\left(e^{i\delta}s_{13}+c_{13}(c_{23}+s_{23})\right),
\end{eqnarray}
\begin{eqnarray}
C_4^1 &=&s_{12}^2+c_{12}^2\left(c_{13}^2+e^{-2i\delta}s_{13}^2\right),
\nonumber \\
C_4^2 &=&c_{12}^2+s_{12}^2\left(c_{13}^2+e^{-2i\delta}s_{13}^2\right),
\nonumber \\
C_4^3 &=&\frac{1}{2}\left(1+\cos 2\theta_{13} +2e^{2i\delta}s_{13}^2\right),
\end{eqnarray}
and
\begin{eqnarray}
C_5^1 &=&2c_{12}c_{13}(-e^{-i\delta}c_{12}c_{23}s_{13}+s_{12}s_{23}) +(c_{23}s_{12}+e^{-i\delta}c_{12}s_{13}s_{23})^2,
\nonumber \\
C_5^2 &=&2c_{13}s_{12}(-e^{-i\delta}c_{23}s_{12}s_{13}-c_{12}s_{23})+\left(c_{12}c_{23}-e^{-i\delta}s_{12}s_{13}s_{23}\right)^2,
\nonumber \\
C_5^3 &=&c_{13}\left(2e^{i\delta}c_{23}s_{13}+c_{13}s_{23}^2\right).
\end{eqnarray}

The required condition of type I, type II, $\cdots$, or type X magic texture is $S_i=S_j=S_k$ with the appropriate $i, j$, and $k$. This requirement $S_i=S_j=S_k$ yields
\begin{eqnarray}
\begin{cases}
(C_i^1 - C_j^1) \lambda_1 + (C_i^2 -C_j^2) \lambda_2+  (C_i^3 - C_j^3) \lambda_3 = 0, \\
(C_i^1-C_k^1) \lambda_1 + (C_i^2  -C_k^2) \lambda_2+  (C_i^3 -C_k^3) \lambda_3 =  0,
\end{cases}
\end{eqnarray}
or equivalently, 
\begin{eqnarray}
\frac{\lambda_1}{\lambda_3} &=& \frac{ (C_i^2  -C_k^2)(C_i^3 - C_j^3)-(C_i^2 -C_j^2) (C_i^3 -C_k^3)}{(C_i^1-C_k^1)(C_i^2 -C_j^2)-(C_i^1 - C_j^1)(C_i^2  -C_k^2)}, \nonumber \\
\frac{\lambda_2}{\lambda_3} &=& \frac{(C_i^1-C_k^1)(C_i^3 - C_j^3)-(C_i^1 - C_j^1)(C_i^3 -C_k^3)}{(C_i^1 - C_j^1)(C_i^2  -C_k^2)- (C_i^1-C_k^1)(C_i^2 -C_j^2)}.
\label{Eq:lambda_i/lambda_j}
\end{eqnarray}
Under this requirements, the ratio of mass eigenvalues, $(m_1/m_3, m_2/m_3)$, and the two Majorana phases, $(\alpha_1, \alpha_2)$, should be 
\begin{eqnarray}
\frac{m_1}{m_3}  = \left| \frac{ \lambda_1}{\lambda_3} \right|,
\quad
\frac{m_2}{m_3}  = \left| \frac{ \lambda_2}{\lambda_3} \right|,
\label{Eq:ratio_of_masses}
\end{eqnarray}
and
\begin{eqnarray}
\alpha_1  =\frac{1}{2}{\rm  arg}\left( \frac{ \lambda_1}{\lambda_3}  \right),
\quad
\alpha_2  =\frac{1}{2}{\rm  arg}\left(\frac{ \lambda_2}{\lambda_3} \right),
\label{Eq:alpha1_alpha2}
\end{eqnarray}
respectively.

Now we would like to comment about Eq.(\ref{Eq:lambda_i/lambda_j}). It seems better to use $\lambda_1/\lambda_2$ and $\lambda_3/\lambda_2$ instead of $\lambda_1/\lambda_3$ and $\lambda_2/\lambda_3$ in Eq. (\ref{Eq:lambda_i/lambda_j}) because $m_2=0$ has been excluded from observations. To use $\lambda_1/\lambda_2$ and $\lambda_3/\lambda_2$, we have to change our phase convention of Majorana CP phases from $(m_1e^{2i\alpha_1}, m_2e^{2i\alpha_2}, m_3)$ in Eq. (\ref{Eq:lambda_123}) to $(m_1e^{2i\alpha_1}, m_2, m_3e^{2i\alpha_3})$. Otherwise, we could not estimate two Majorana CP phases via simple relation such as Eq. (\ref{Eq:alpha1_alpha2}), e. g, we obtain $\alpha_1 - \alpha_2 = {\rm arg} (\lambda_1/\lambda_2)/2$ from $\lambda_1/\lambda_2$ with $(m_1e^{2i\alpha_1}, m_2e^{2i\alpha_2}, m_3)$. 

In addition, since we use Eqs. (\ref{Eq:lambda_i/lambda_j}) and (\ref{Eq:ratio_of_masses}), $m_3=0$ should be excluded. We take $m_3$ is the free parameter and vary within $m_3 = 0.001 - 0.12$ eV in our numerical calculation. Since we choose $m_3 = 0.001 - 0.12$ eV, the lightest neutrino mass, $m_1$ for normal mass ordering or $m_3$ for inverted mass ordering, could not to be zero by using our method of making a magic texture, Eqs.  (\ref{Eq:lambda_i/lambda_j}), (\ref{Eq:ratio_of_masses}), and (\ref{Eq:alpha1_alpha2}). If we change our phase convention from $(m_1e^{2i\alpha_1}, m_2e^{2i\alpha_2}, m_3)$ to $(m_1e^{2i\alpha_1}, m_2, m_3e^{2i\alpha_3})$, unfortunately, we encounter the same problem. In this case, we may take $m_2$ is the free parameter and vary within $m_2 = 0.001 - 0.12$ eV, then the lightest neutrino mass may not to be zero again as far as we  use Eqs.  (\ref{Eq:lambda_i/lambda_j}), (\ref{Eq:ratio_of_masses}), and (\ref{Eq:alpha1_alpha2}). 

The lightest neutrino mass is allowed to be zero in the current experiment. Thus, the range for the lightest neutrino mass should be $0 - 0.12$ eV for more precise analysis. However, as far as we use our method for constructing magic texture, we could not avoid this problem with any phase convention. This is the weakest point of our method. We have to construct other method to construct a magic texture which can allow the lightest neutrino mass to be zero. Although we attempted to construct new method, we have not good method at present. In this paper, we would like to use our current method without zero neutrino mass.  If we are allowed to use our current method without zero neutrino mass, we can use $\lambda_1/\lambda_3$ and $\lambda_2/\lambda_3$ in Eq. (\ref{Eq:lambda_i/lambda_j}) without consideration because not only $m_2=0$ but also $m_1=0$ and $m_3=0$ have been excluded.

Equations (\ref{Eq:lambda_i/lambda_j}), (\ref{Eq:ratio_of_masses}), and (\ref{Eq:alpha1_alpha2}) could be used to make a magic texture. For example, the type IV magic texture is obtained by $(i,j,k)=(2,3,4)$ as we will show in Sec. \ref{section:Allowed Magic Textures}.

Eqs.  (\ref{Eq:lambda_i/lambda_j}), (\ref{Eq:ratio_of_masses}), and (\ref{Eq:alpha1_alpha2}) can be used to make any type of magic texture except for type I. The type I magic texture is invariant under a symmetry which is related with the trimaximal mixing for $\nu_2$:
\begin{eqnarray}
U_{\rm TM} =\left(
\begin{matrix}
\sqrt{\frac{2}{3}}\cos\theta & \frac{1}{\sqrt{3}} & \sqrt{\frac{2}{3}}\sin\theta \\
-\frac{\cos\theta}{\sqrt{6}}+\frac{e^{-i\phi}\sin\theta}{\sqrt{2}} & \frac{1}{\sqrt{3}} & -\frac{\sin\theta}{\sqrt{6}}-\frac{e^{-i\phi}\cos\theta}{\sqrt{2}} \\
-\frac{\cos\theta}{\sqrt{6}}-\frac{e^{-i\phi}\sin\theta}{\sqrt{2}} & \frac{1}{\sqrt{3}} & -\frac{\sin\theta}{\sqrt{6}}+\frac{e^{-i\phi}\cos\theta}{\sqrt{2}} \\
\end{matrix}
\right), 
\label{Eq:UTM}
\end{eqnarray}
where $\theta$ and $\phi$ denote free parameters. The corresponding Majorana neutrino mass matrix for tribimaximal mixing $M_{\rm TM} $ can be obtained by Eq. (\ref{Eq:M=UMU}) and we have the following relations
\begin{eqnarray}
S_1 = S_2 = S_3 = \lambda_2.
\label{Eq:S1S2S3_TM}
\end{eqnarray}
The type I criteria $S_1+S_2+S_3$ should be dependent only on $\lambda_2$ and independent on $\lambda_1, \lambda_3$. Thus, Eq. (\ref{Eq:lambda_i/lambda_j}) does not work to obtain the type I magic texture.

\section{Allowed Magic Textures\label{section:Allowed Magic Textures}}
\subsection{Criteria \label{subsection:criteria}}
A global analysis of current data from the neutrino oscillation experiments shows the following best-fit values of the squared mass differences $\Delta m^2_{ij} = m_i^2-m_j^2$ and mixing angles for the normal mass ordering (NO), $m_1<m_2<m_3$, \cite{Esteban2020JHEP}:
\begin{eqnarray} 
\Delta m^2_{21}/(10^{-5} {\rm eV}^2) &=& 7.42^{+0.21}_{-0.20}  \quad (6.82 \sim 8.04), \nonumber \\
\Delta m^2_{31}/(10^{-3}{\rm eV}^2) &=& 2.517^{+0.026}_{-0.028}  \quad (2.435  \sim 2.598), \nonumber \\
\theta_{12}/^\circ &=& 33.44^{+0.77}_{-0.74}    \quad (31.27  \sim 35.86), \nonumber \\
\theta_{23}/^\circ &=& 49.2^{+0.9}_{-1.2}   \quad (40.1  \sim 51.7), \nonumber \\
\theta_{13}/^\circ &=& 8.57^{+0.12}_{-0.12}   \quad (8.20  \sim 8.93), \nonumber \\
\delta/^\circ &=& 197^{+27}_{-24}   \quad (120  \sim 369), 
\label{Eq:neutrino_observation_NO}
\end{eqnarray}
where the $\pm$ denote the $1 \sigma$ region and the parentheses denote the $3 \sigma$ region. For the inverted mass ordering (IO), $m_3<m_1\simeq m_2$, a global analysis shows \cite{Esteban2020JHEP}
\begin{eqnarray} 
\Delta m^2_{21}/(10^{-5} {\rm eV}^2) &=& 7.42^{+0.21}_{-0.20}   \quad (6.82 \sim 8.04), \nonumber \\
-\Delta  m^2_{32}/(10^{-3}{\rm eV}^2) &=& 2.498^{+0.028}_{-0.028}   \quad (2.581  \sim 2.414), \nonumber \\
\theta_{12}/^\circ &=& 33.45^{+0.78}_{-0.75}   \quad (31.27  \sim 35.87), \nonumber \\
\theta_{23}/^\circ &=& 49.3^{+0.9}_{-1.1}   \quad (40.3  \sim 51.8), \nonumber \\
\theta_{13}/^\circ &=& 8.60^{+0.12}_{-0.12}   \quad (8.24  \sim 8.96), \nonumber \\
\delta/^\circ &=& 282^{+26}_{-30}   \quad (193  \sim 352).
\label{Eq:neutrino_observation_IO}
\end{eqnarray}
Moreover, we have the following constraints  
\begin{eqnarray} 
\sum m_i < 0.12  \sim 0.69 ~{\rm eV},
\label{Eq:sumMObs}
\end{eqnarray}
from observation of cosmic microwave background radiation \cite{Planck2020AA, Capozzi2020PRD} and
\begin{eqnarray} 
|M_{ee}|<0.066  \sim 0.155 ~{\rm eV},
\label{Eq:MeeObs}
\end{eqnarray}
from the neutrinoless double beta decay experiments \cite{GERDA2019Science,Capozzi2020PRD}. We require that the predictions from a magic texture are consistent with the $3 \sigma$ region in Eq. (\ref{Eq:neutrino_observation_NO}) for NO and Eq. (\ref{Eq:neutrino_observation_IO}) for IO. We also require that the constraints $\sum m_i < 0.12$ eV and $|M_{ee}| < 0.066$ eV are satisfied. 

Since there are only two measured squared mass differences, one of the neutrino mass eigenstates is a free parameter in the neutrino oscillation experiments. As we said, we chose a neutrino mass within $0.001 - 0.12$ eV. 


\begin{figure}[t]
\begin{center}
\includegraphics{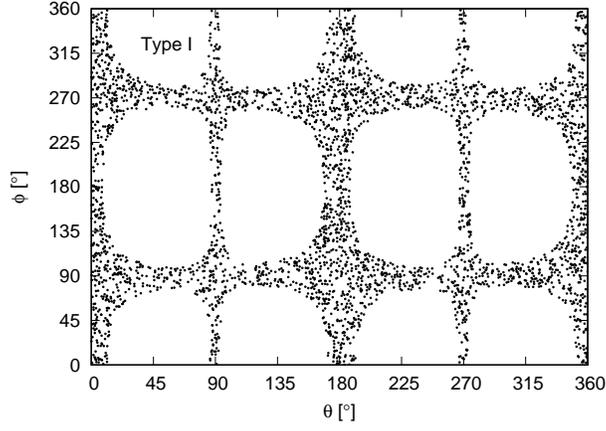}
\caption{Allowed regions of $\phi$ and $\theta$ for $40.1^\circ \le \theta_{23} \le 51.7^\circ$ in type I magic texture.}
\label{fig:I_allowed_phi_delta}
\end{center}
\end{figure}

\begin{figure}[t]
\begin{center}
\includegraphics[scale=0.45]{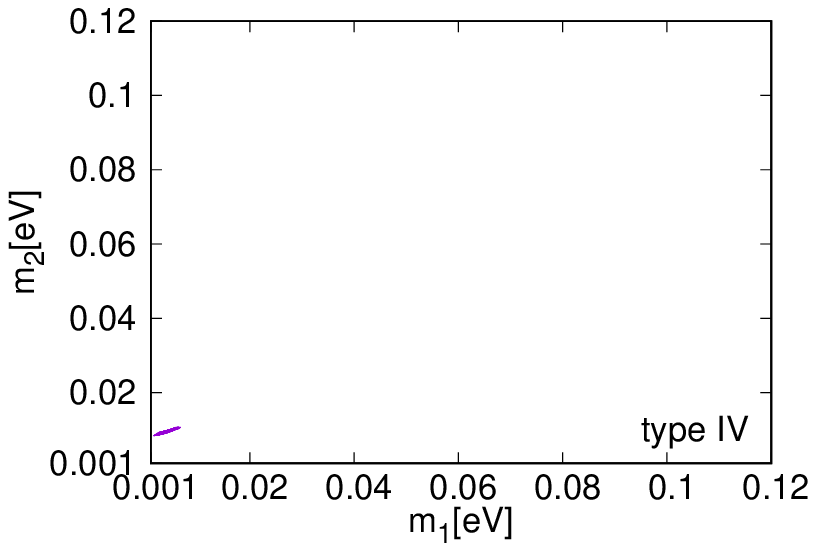}
\includegraphics[scale=0.45]{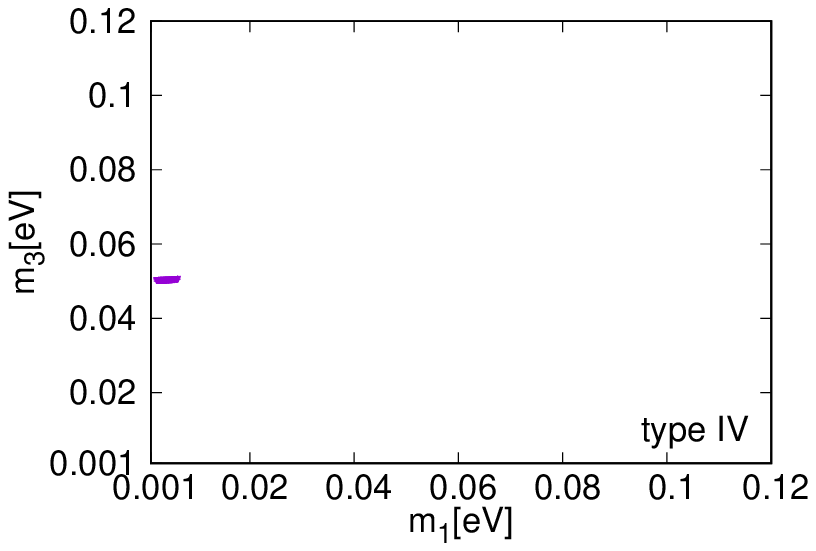}
\includegraphics[scale=0.45]{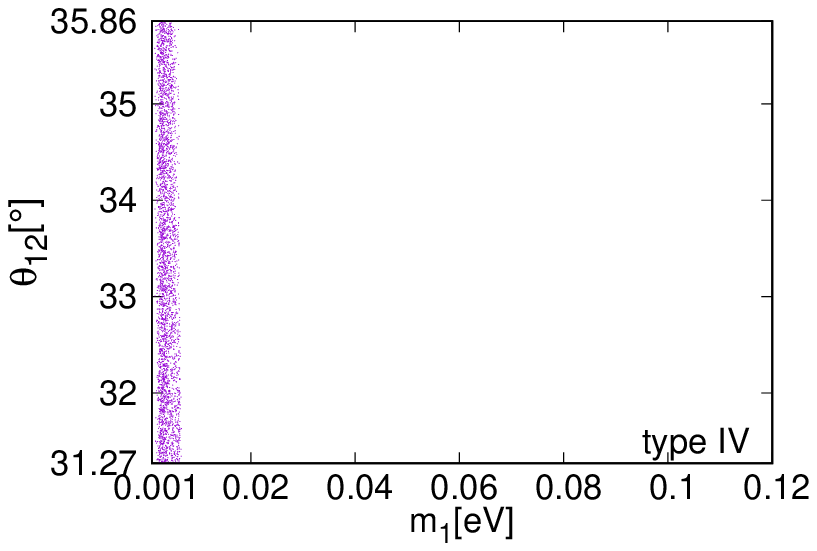}
\includegraphics[scale=0.45]{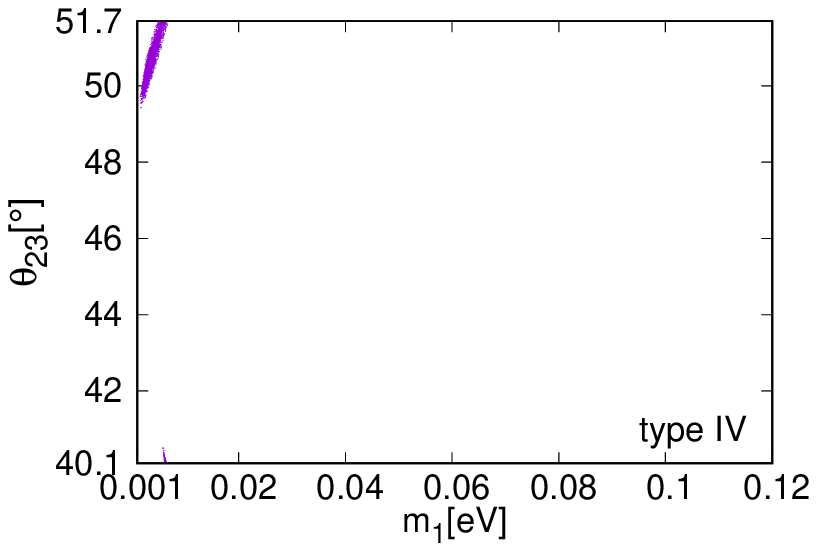}
\includegraphics[scale=0.45]{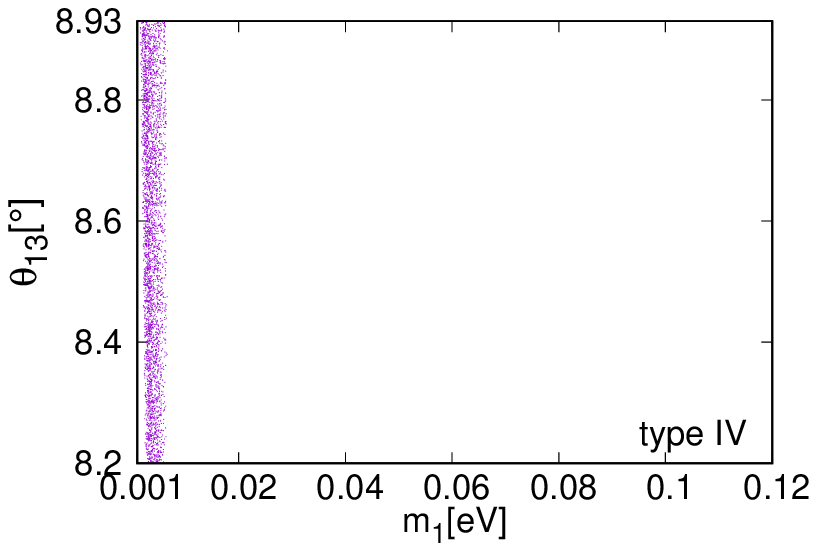}
\includegraphics[scale=0.45]{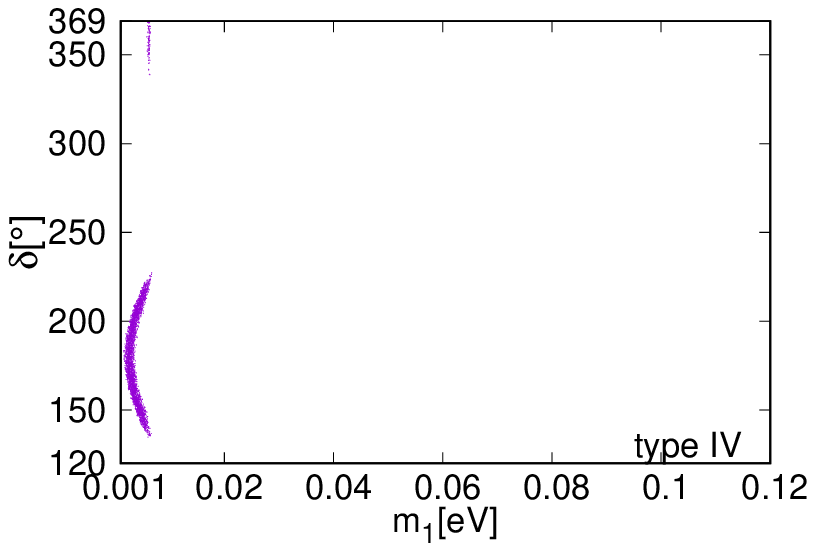}
\includegraphics[scale=0.45]{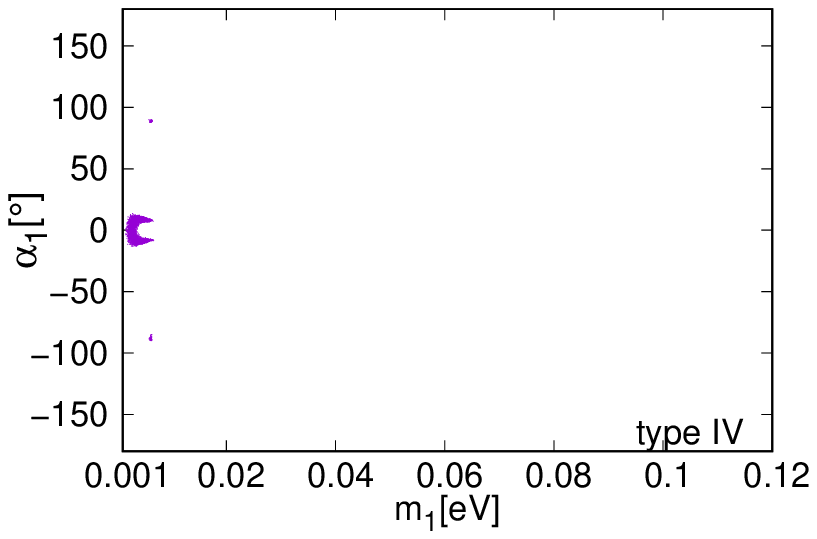}
\includegraphics[scale=0.45]{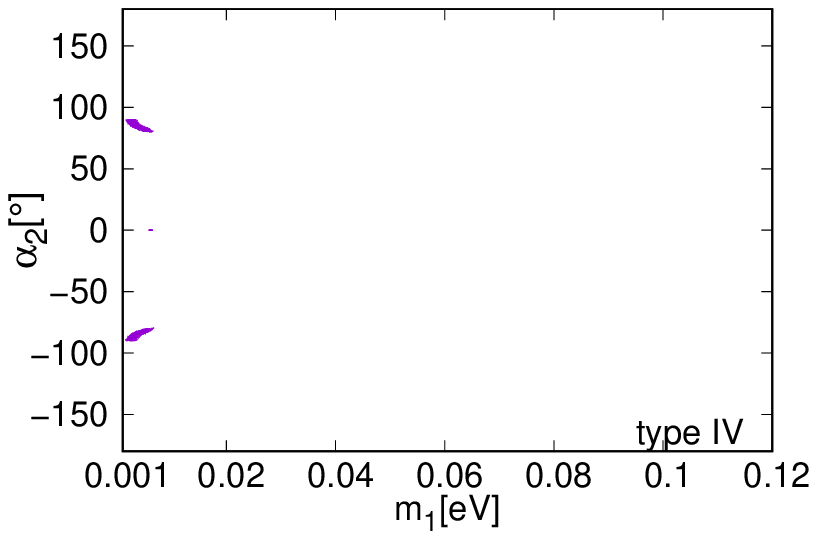}
\includegraphics[scale=0.45]{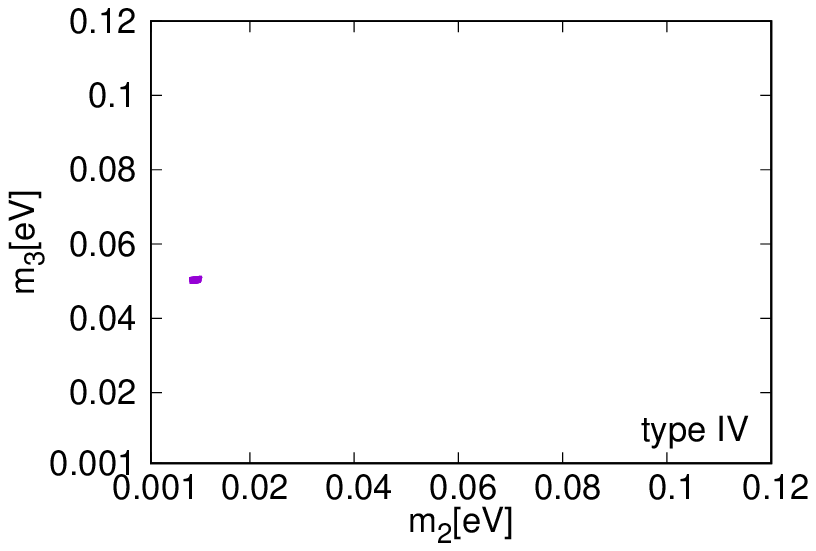}
\includegraphics[scale=0.45]{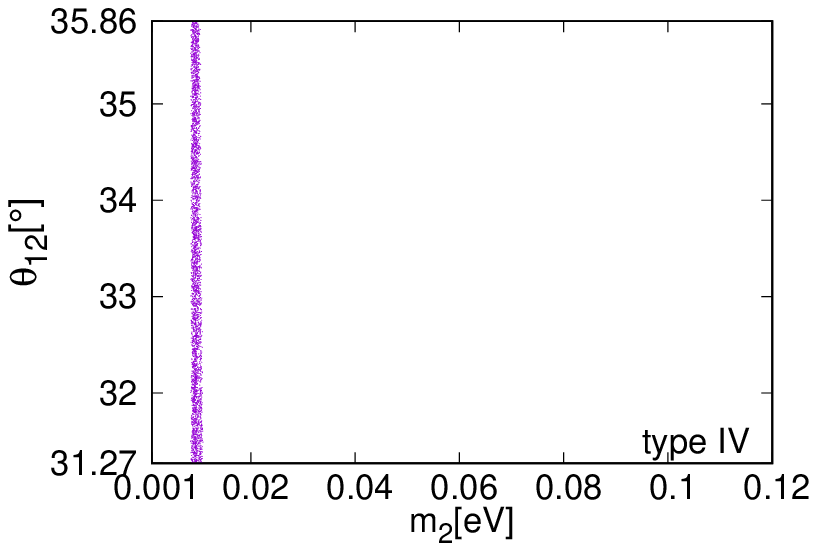}
\includegraphics[scale=0.45]{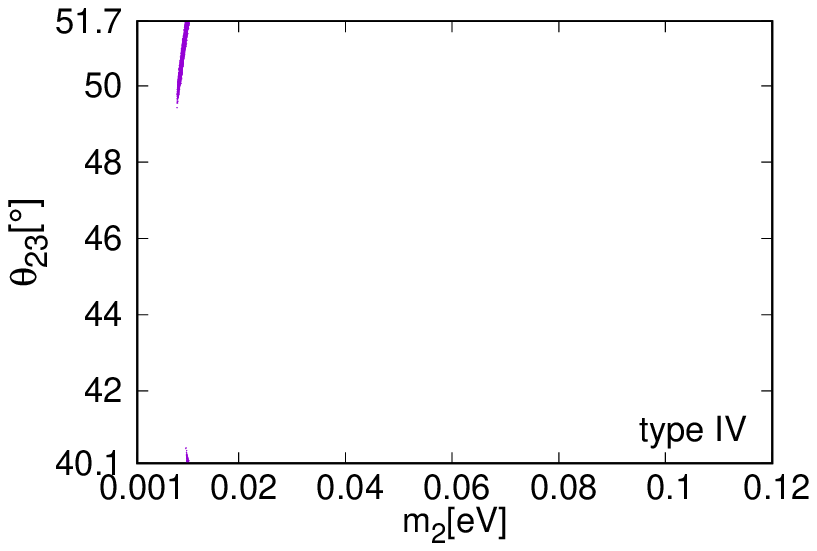}
\includegraphics[scale=0.45]{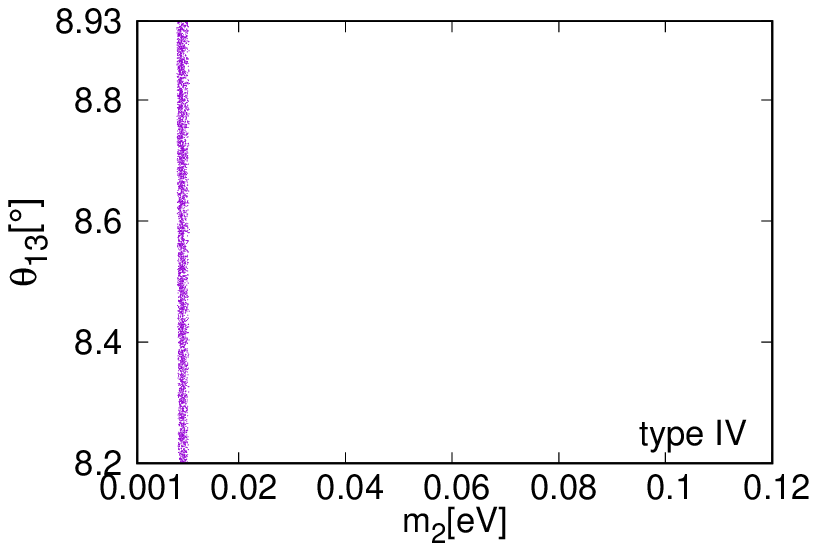}
\includegraphics[scale=0.45]{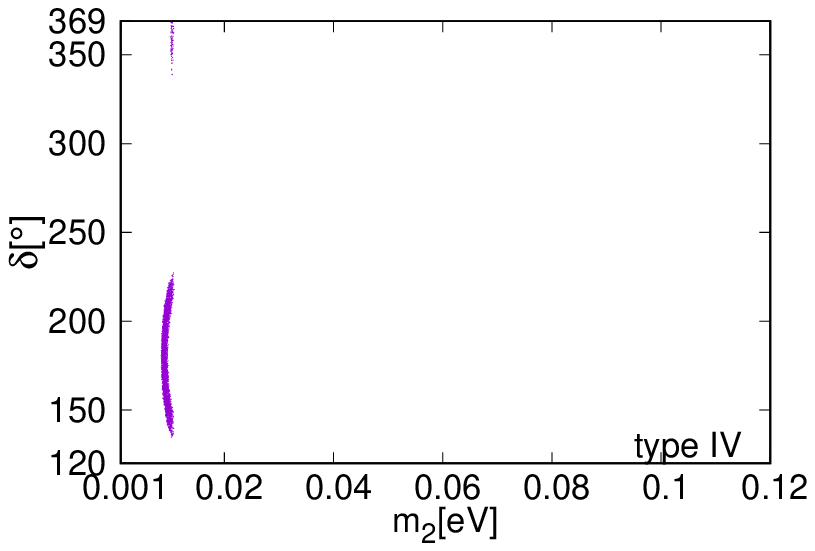}
\includegraphics[scale=0.45]{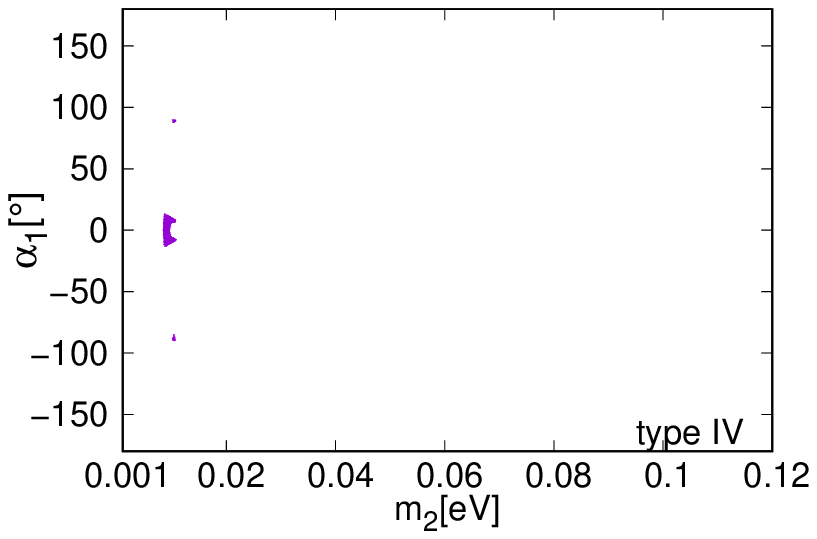}
\includegraphics[scale=0.45]{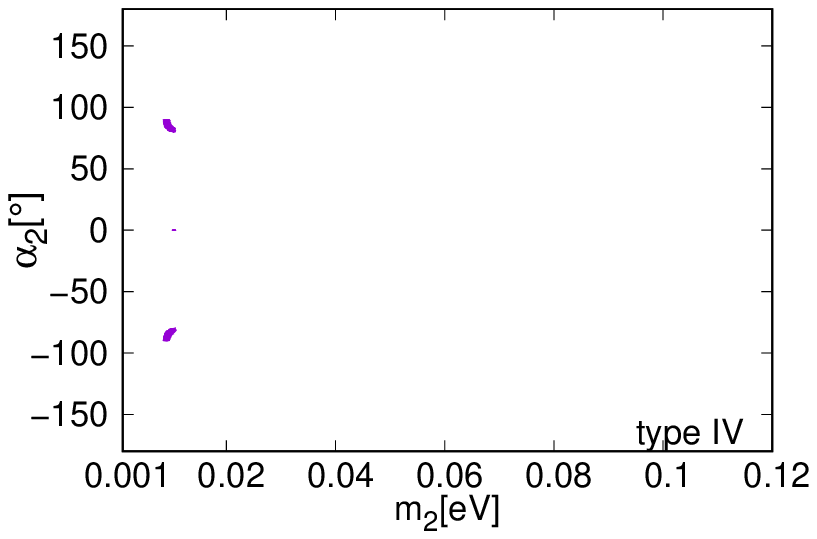}
\includegraphics[scale=0.45]{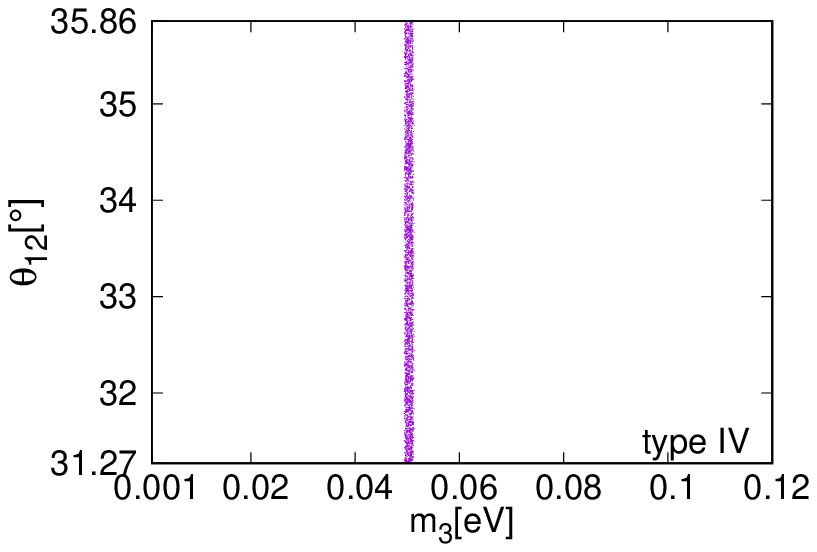}
\includegraphics[scale=0.45]{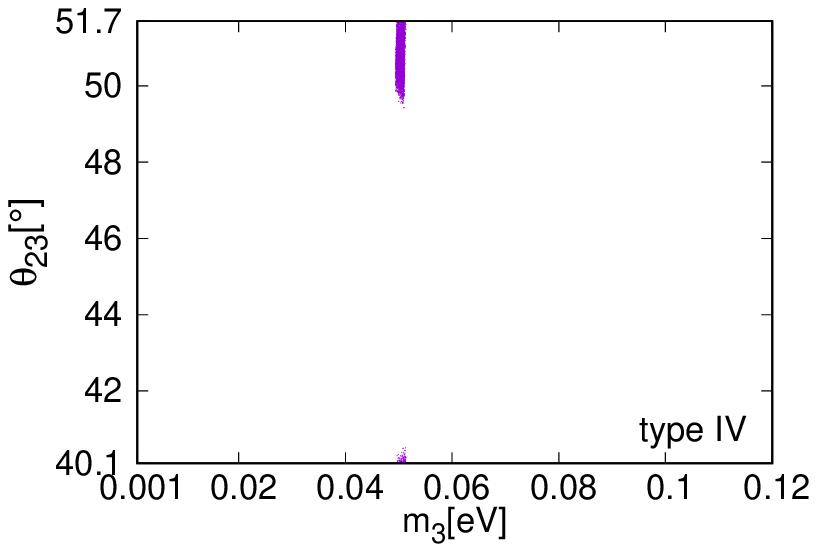}
\includegraphics[scale=0.45]{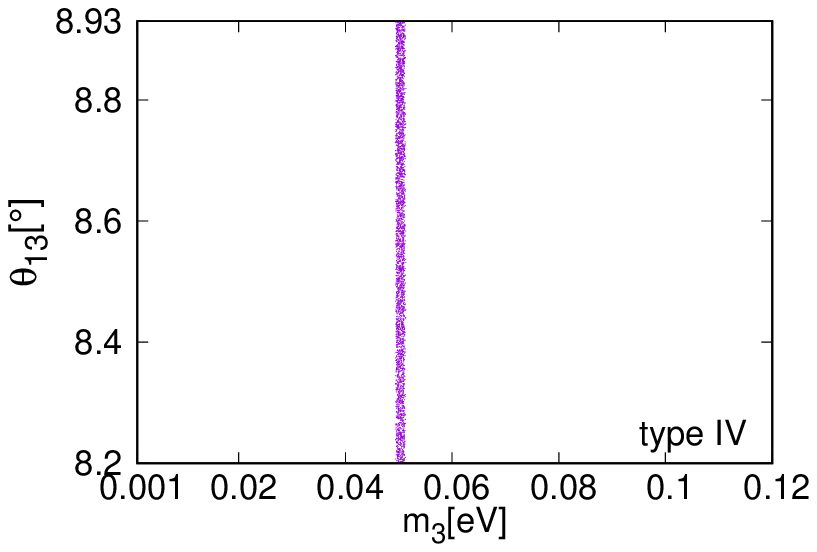}
\includegraphics[scale=0.45]{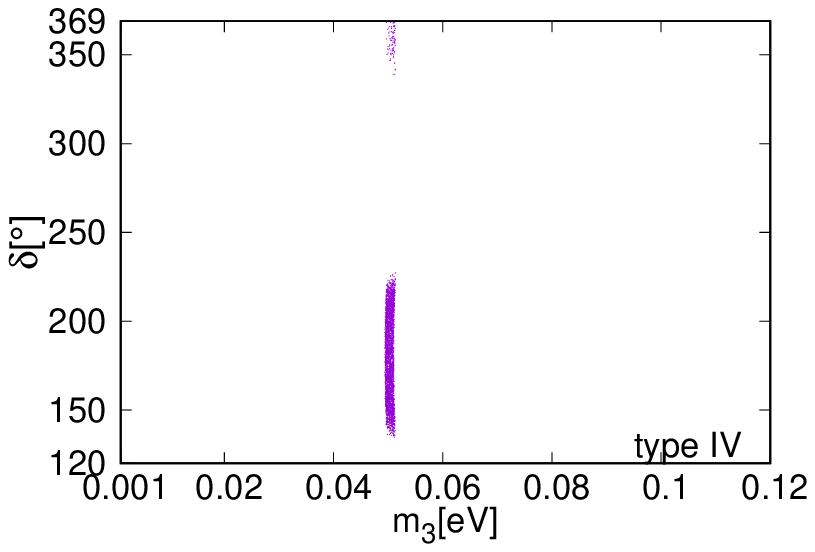}
\includegraphics[scale=0.45]{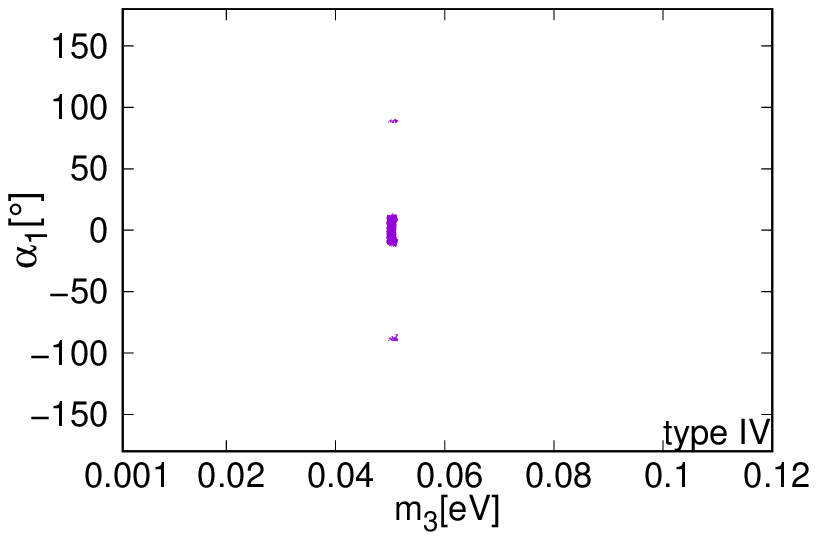}
\includegraphics[scale=0.45]{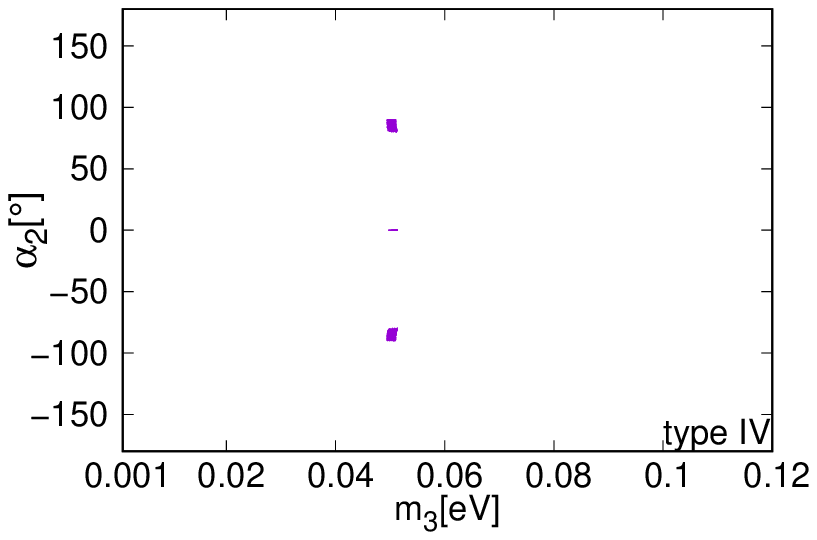}
\caption{Variation in the neutrino oscillation parameters with the masses for type IV magic texture.}
\label{fig:IV_param_mass}
\end{center}
\end{figure}
\begin{figure}[t]
\begin{center}
\includegraphics[scale=0.45]{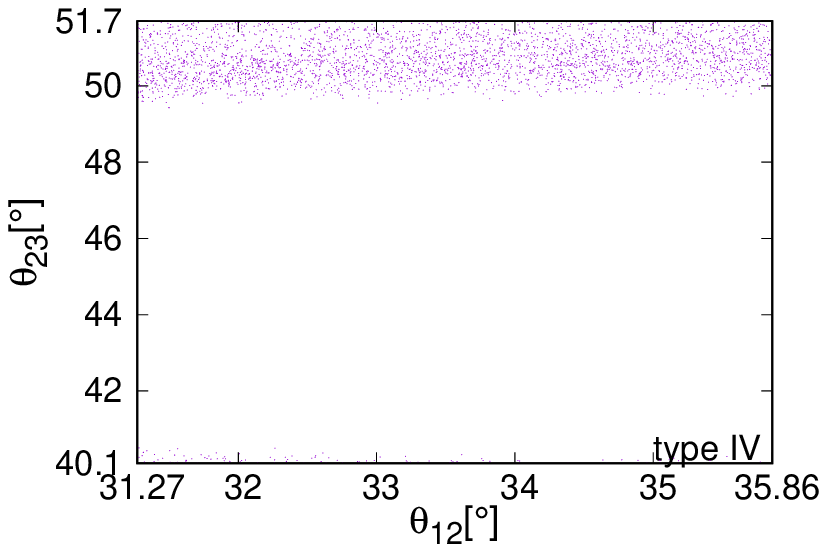}
\includegraphics[scale=0.45]{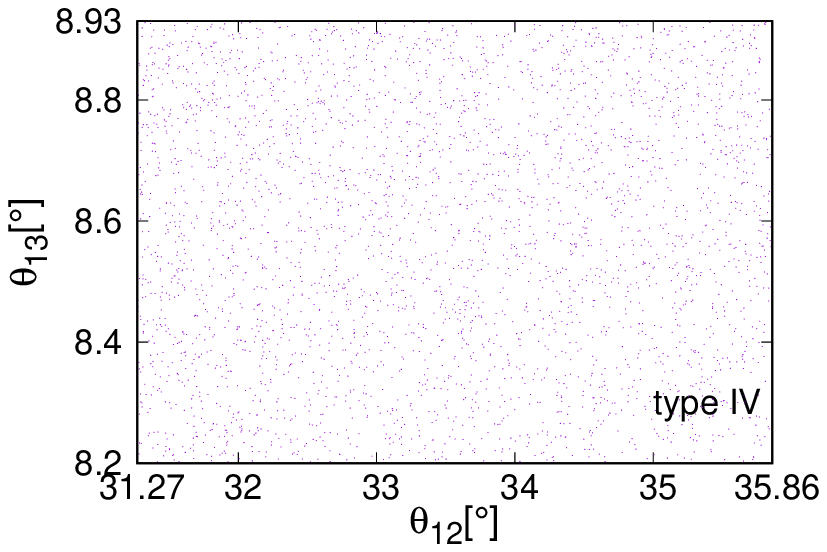}
\includegraphics[scale=0.45]{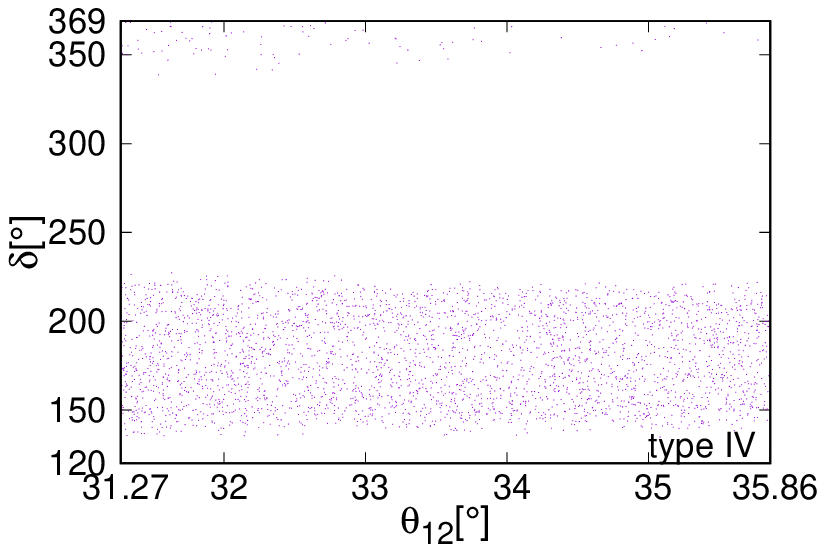}
\includegraphics[scale=0.45]{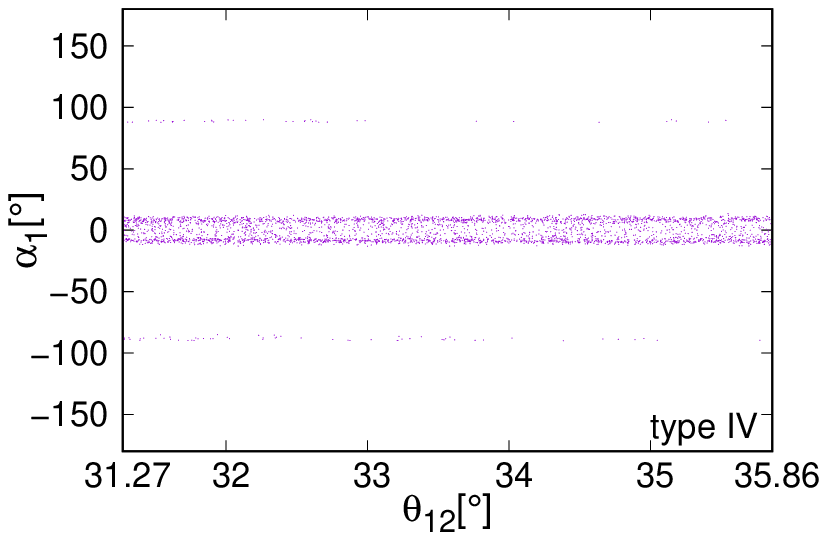}
\includegraphics[scale=0.45]{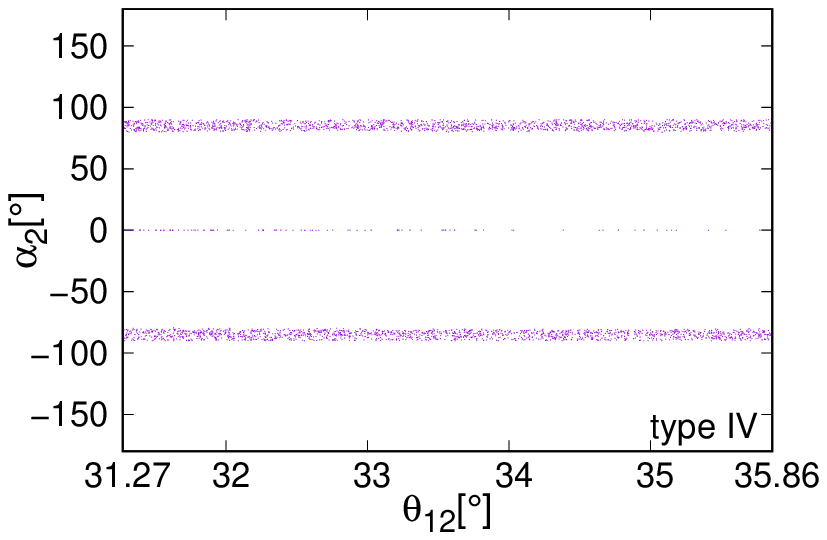}
\includegraphics[scale=0.45]{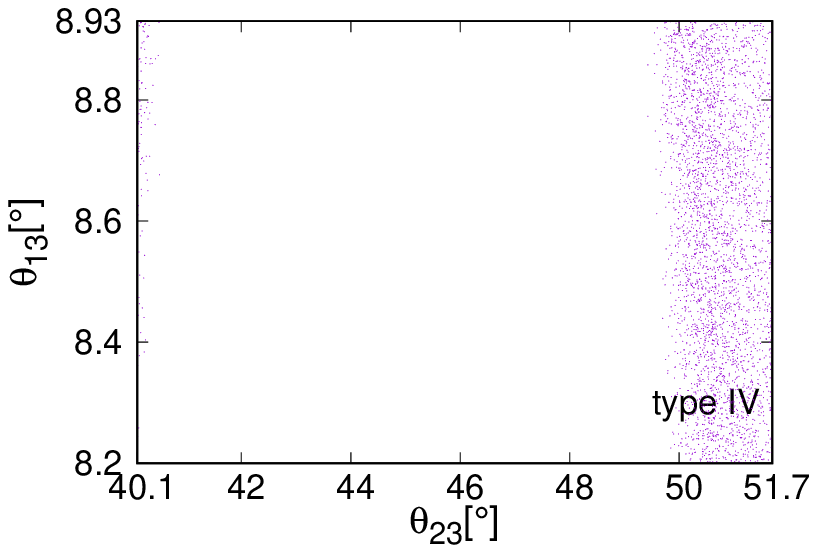}
\includegraphics[scale=0.45]{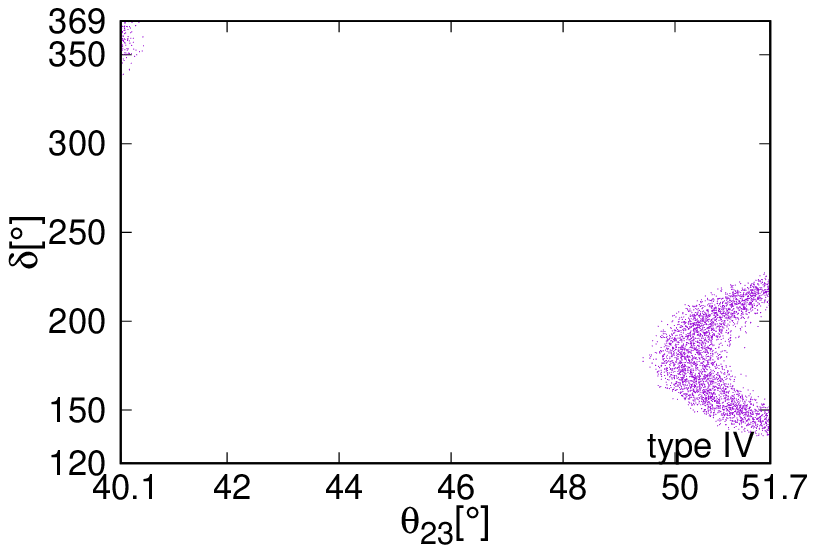}
\includegraphics[scale=0.45]{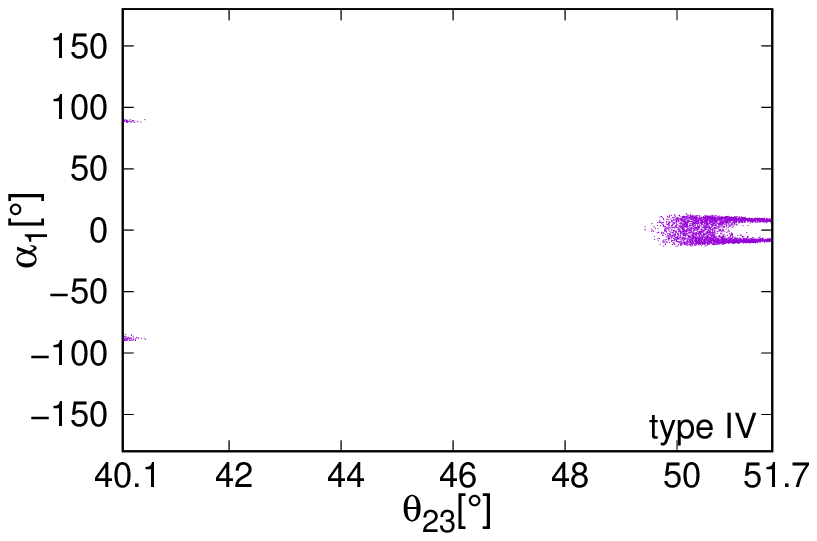}
\includegraphics[scale=0.45]{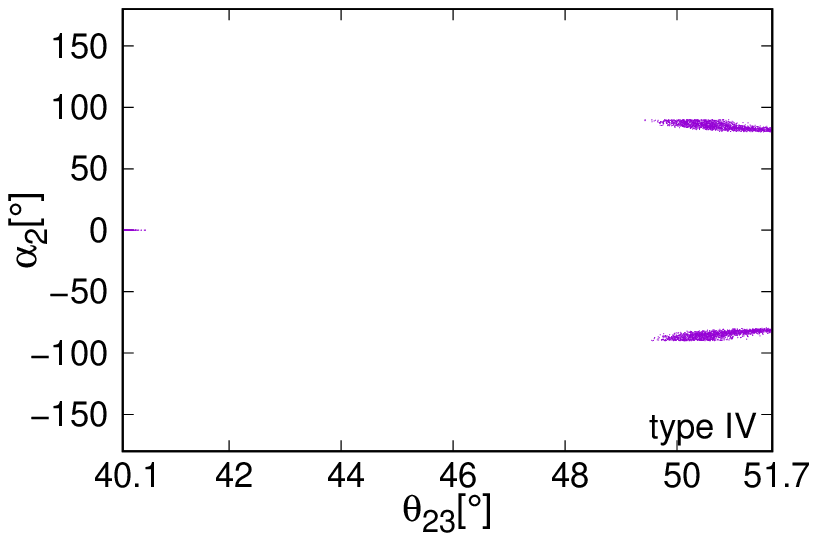}
\includegraphics[scale=0.45]{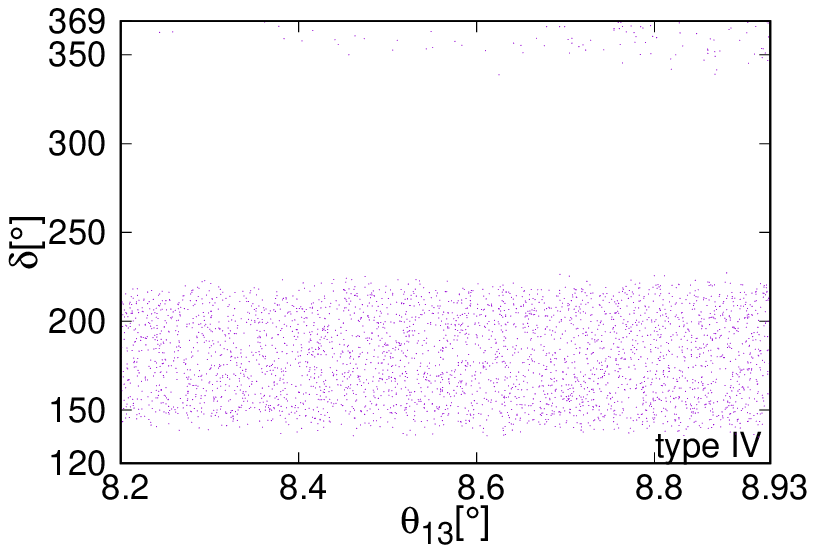}
\includegraphics[scale=0.45]{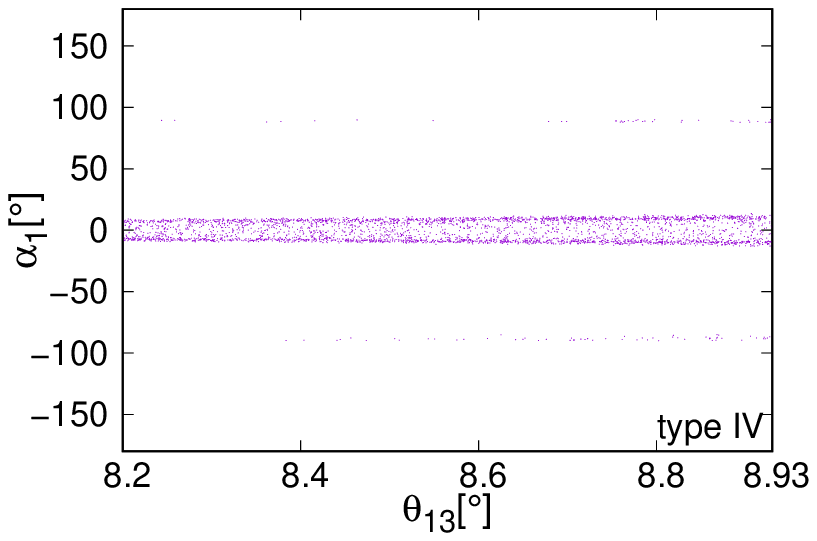}
\includegraphics[scale=0.45]{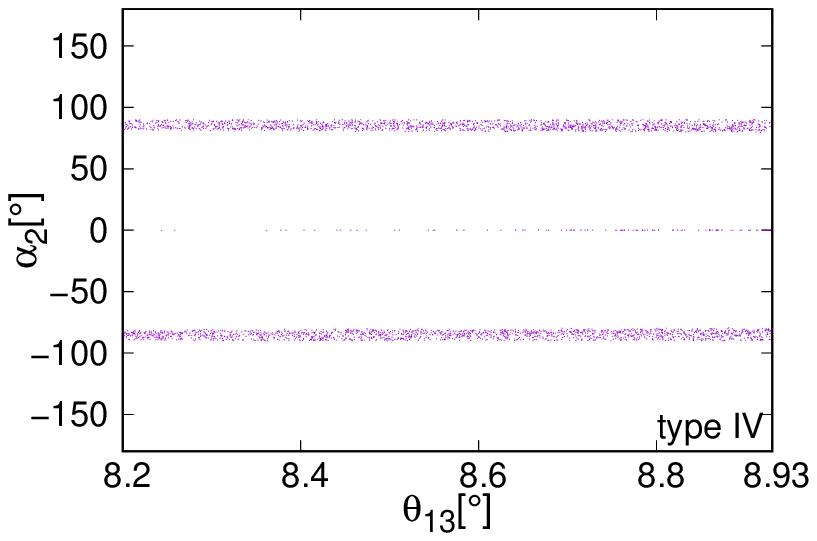}
\includegraphics[scale=0.45]{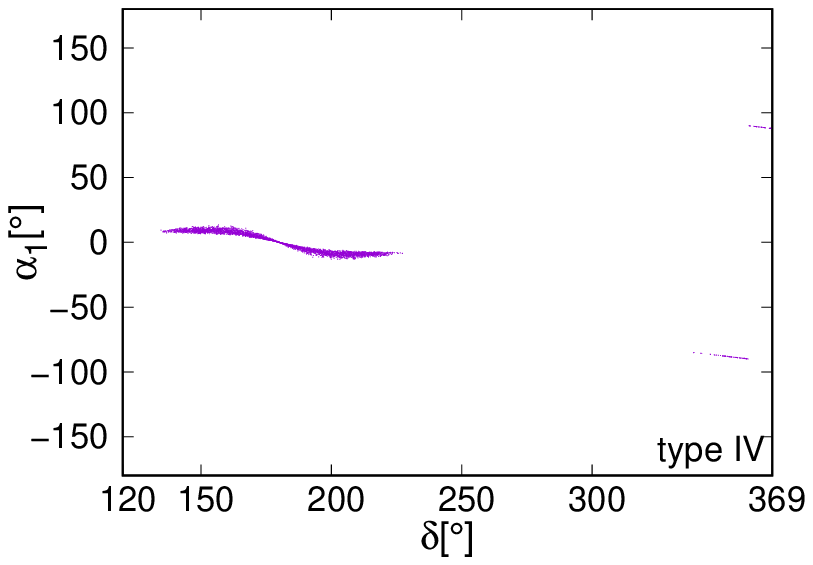}
\includegraphics[scale=0.45]{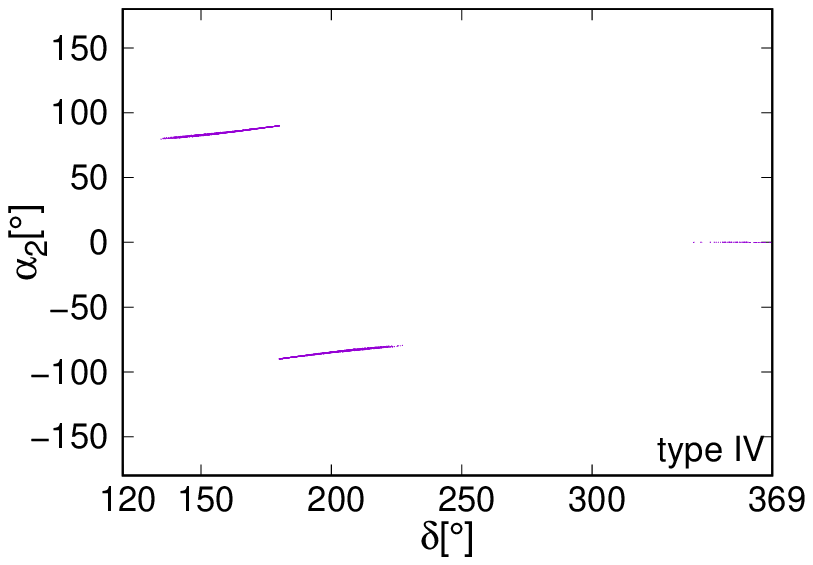}
\includegraphics[scale=0.45]{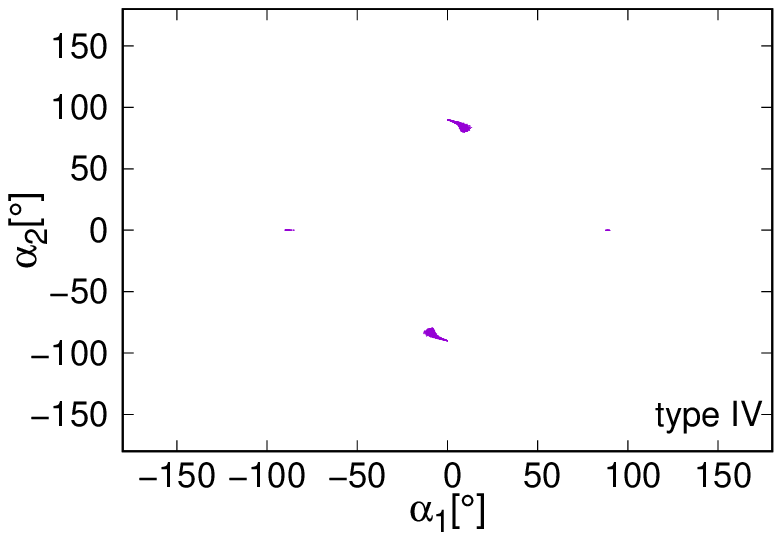}
\caption{Variation in the neutrino oscillation parameters with the mixing angles and CP phases for type IV magic texture.}
\label{fig:IV_param_phases}
\end{center}
\end{figure}

\begin{figure}[t]
\begin{center}
\includegraphics[scale=0.6]{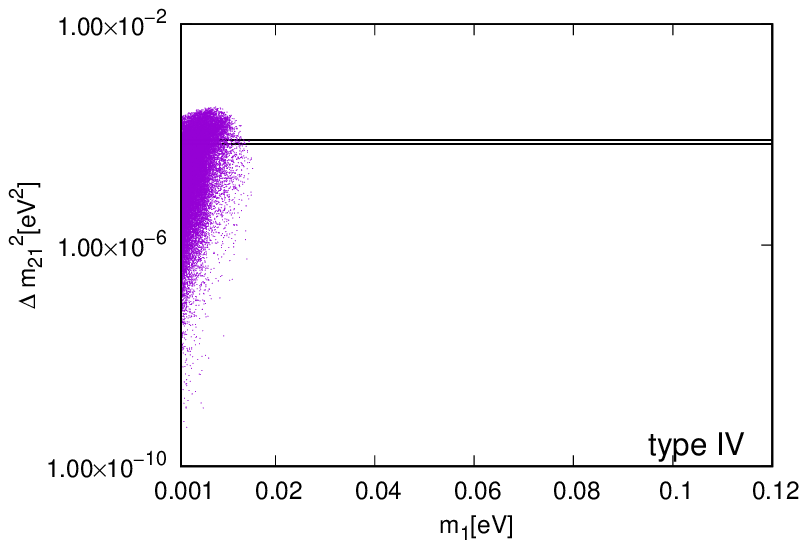}
\includegraphics[scale=0.6]{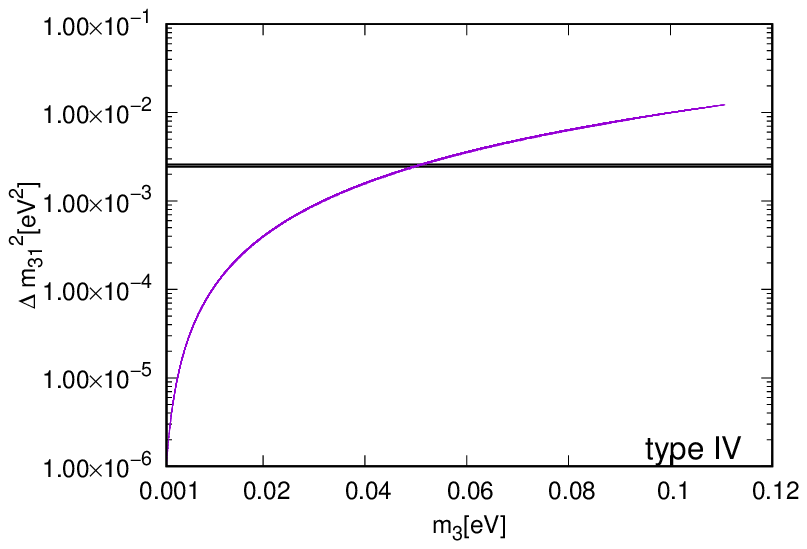}
\caption{Predicted square mass deferences $\Delta m_{ij}^2$ in type IV magic texture. The upper and lower horizontal lines show the observed $\Delta m_{ij}^2$ in 3 $\sigma$ region.}
\label{fig:IV_Dm2}
\end{center}
\end{figure}

\begin{figure}[t]
\begin{center}
\includegraphics[scale=0.6]{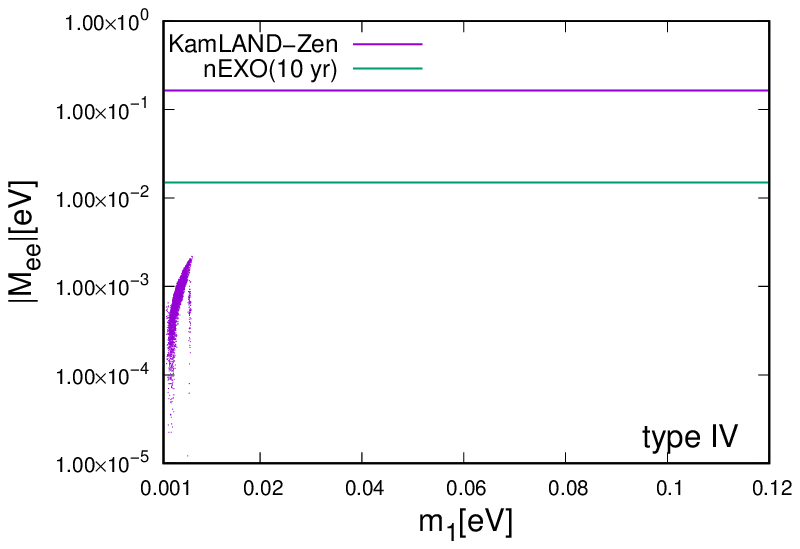}
\includegraphics[scale=0.6]{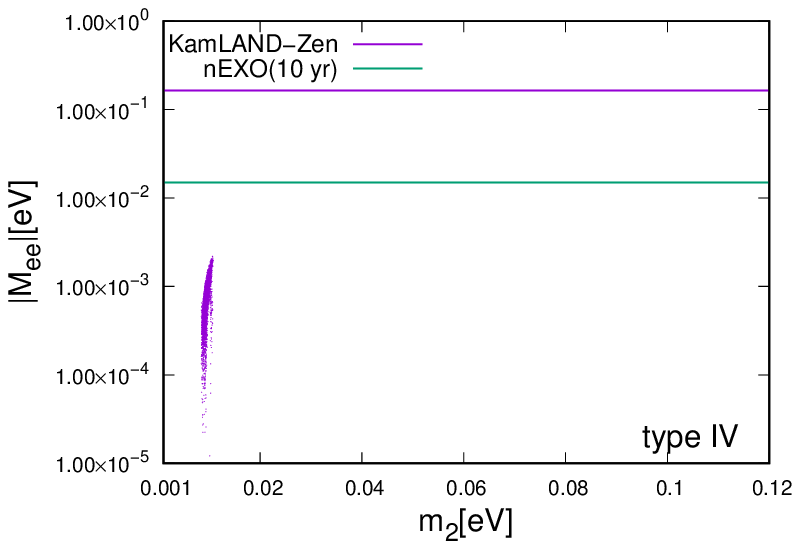}
\includegraphics[scale=0.6]{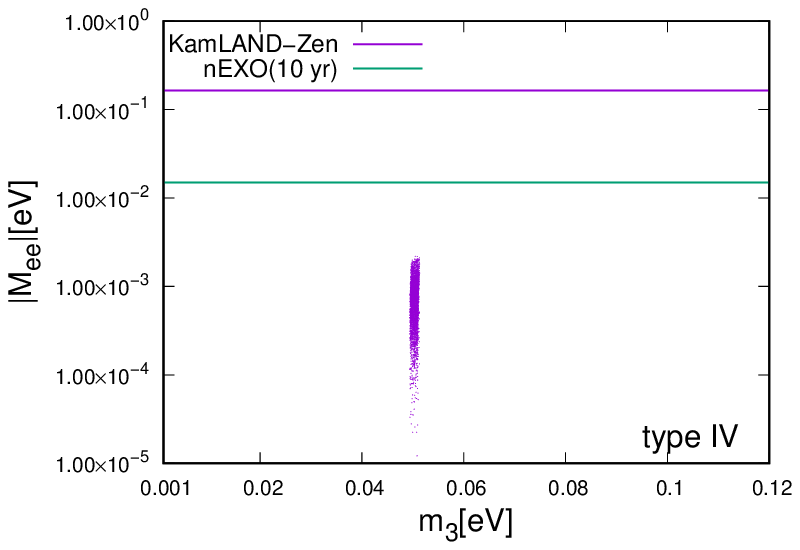}
\caption{Predicted effective neutrino mass for neutrinoless double $\beta$ decay $|M_{ee}|$ in type IV magic texture. The upper  (lower) horizontal line shows the upper bound from KamLAND-Zen (nEXO) experiment.}
\label{fig:IV_Mee}
\end{center}
\end{figure}

\subsection{Type I \label{subsection:type-I}}
It is already known that the type I magic texture is consistent with observations \cite{Harrison2004PLB,Lam2006PLB,Gautam2016PRD,Channey2019JPGNP,Verma2020JPGNP}. For example, if we set 
\begin{eqnarray}
&& (\theta, \phi) = (10.5^\circ, 20.0^\circ),
\label{Eq:theta_ph_type1ex}
\end{eqnarray}
then we obtain
\begin{eqnarray}
(\theta_{12}, \theta_{23}, \theta_{13}, \delta) = (35.7^\circ, 50.7^\circ,8.56^\circ, 159.6^\circ),
\label{Eq:theta_type1ex}
\end{eqnarray}
by Eqs. (\ref{Eq:mixingAngle_from_UPMNS}), (\ref{Eq:Jarlskog}), and (\ref{Eq:UTM}). These predicted magnitudes of the mixing angles and the Dirac CP phase are consistent with Eq. (\ref{Eq:neutrino_observation_NO}). Figure \ref{fig:I_allowed_phi_delta} shows allowed regions of $\phi$ and $\theta$ for $40.1^\circ \le \theta_{23} \le 51.76^\circ$ in the type I magic texture.

Since the nature of the type I magic square is controlled by only a trimaximal mixing for $\nu_2$ scheme, there is no prediction of the neutrino masses. In addition, the type I magic texture could not predict the Majorana CP-violating phases. Thus, we can take any values of neutrino masses and the Majorana CP phases in the type I magic texture. If we take 
\begin{eqnarray}
(m_1, m_2, m_3)= (0.001, 0.0088, 0.05) ~{\rm eV}, \quad (\alpha_1, \alpha_2)= (5^\circ, 5^\circ),
\label{Eq:m1_m2_m3_alpha_type1ex_NO}
\end{eqnarray}
then we obtain
\begin{eqnarray}
&& \Delta m^2_{21} = 7.64 \times 10^{-5}~ {\rm eV^2}, \quad
\Delta m^2_{31} = 2.50\times 10^{-3} ~{\rm eV^2}, \nonumber \\
&& \sum m_i = 0.06~{\rm eV}, \quad |M_{ee}| = 0.0047~{\rm eV}.
\label{Eq:DeltaMetc_type1ex}
\end{eqnarray}
These squared mass differences etc. are also consistent with Eq. (\ref{Eq:neutrino_observation_NO}). Thus, the type I magic texture is consistent with experiments. In addition, if we take $ (m_1, m_2, m_3)= (0.0492, 0.05, 0.005)$ {\rm eV}, we can show that the type I magic texture is also consistent with observations for IO.

\subsection{Type IV  \label{subsection:type-IV}}
The type IV magic texture, $S_2=S_3=S_4 \neq S_1 \neq S_5$, could be constructed using Eq. (\ref{Eq:lambda_i/lambda_j}) with $(i,j,k)=(2,3,4)$:
\begin{eqnarray}
\frac{\lambda_1}{\lambda_3} &=& \frac{ (C_2^2  -C_4^2)(C_2^3 - C_3^3)-(C_2^2 -C_3^2) (C_2^3 -C_4^3)}{(C_2^1-C_4^1)(C_2^2 -C_3^2)-(C_2^1 - C_3^1)(C_2^2  -C_4^2)}, \nonumber \\
\frac{\lambda_2}{\lambda_3} &=& \frac{(C_2^1-C_4^1)(C_2^3 - C_3^3)-(C_2^1 - C_3^1)(C_2^3 -C_4^3)}{(C_2^1 - C_3^1)(C_2^2  -C_4^2)- (C_2^1-C_4^1)(C_2^2 -C_3^2)}, \nonumber \\
\label{Eq:lambda_i/lambda_j_type4}
\end{eqnarray}
with Eqs.  (\ref{Eq:ratio_of_masses}), (\ref{Eq:alpha1_alpha2}) and (\ref{Eq:lambda_i/lambda_j_type4}).

For example, if we set
\begin{eqnarray}
 (\theta_{12},\theta_{23},\theta_{13},\delta) = (32.8^\circ, 50.4^\circ, 8.47^\circ, 169^\circ), \quad  m_3= 0.05~{\rm eV},
\label{Eq:theta_delta_m3_type4ex}
\end{eqnarray}
then, the masses and two Majorana phases should be
\begin{eqnarray}
 (m_1,m_2)=(0.00305, 0.00914)~{\rm eV},  \quad (\alpha_1, \alpha_2) =(5.49^\circ, 87.1^\circ),
\label{Eq:m1_m2_alpha_type4ex}
\end{eqnarray}
for the type IV magic texture by Eqs.  (\ref{Eq:ratio_of_masses}), (\ref{Eq:alpha1_alpha2}), and (\ref{Eq:lambda_i/lambda_j_type4}). These specific values of the neutrino parameters in Eqs. (\ref{Eq:theta_delta_m3_type4ex}) and (\ref{Eq:m1_m2_alpha_type4ex}) occur in the type IV magic texture. Indeed, by inserting numerical values in Eqs. (\ref{Eq:theta_delta_m3_type4ex}) and (\ref{Eq:m1_m2_alpha_type4ex}) into Eqs. (\ref{Eq:M=UMU}), (\ref{Eq:lambda_123}), (\ref{Eq:UPMNS}), and (\ref{Eq:UPMNS_elements}), we obtain the following elements of the Majorana flavor neutrino mass matrix
\begin{eqnarray}
(M_{\rm IV})_{ee} &=& 0.000465+0.000253i, \nonumber \\
(M_{\rm IV})_{e\mu} &=& -0.00907+0.00125i,  \nonumber \\
(M_{\rm IV})_{e\tau} &=& -0.000445+0.000835i,  \nonumber \\
(M_{\rm IV})_{\mu\mu} &=& 0.0261+0.000208i,  \nonumber \\ 
(M_{\rm IV})_{\mu\tau} &=& 0.0270-0.000375i,  \nonumber \\
(M_{\rm IV})_{\tau\tau} &=& 0.0175+0.000624i.
\label{Eq:M_IV_ex}
\end{eqnarray}
The matrix $M_{\rm IV}$ in Eq. (\ref{Eq:M_IV_ex}) is satisfied with the criteria of the type IV magic texture ($S_2=S_3=S_4 \neq S_1 \neq S_5$) as follows: 
\begin{eqnarray}
&& S_2=S_3=S_4=0.0440+0.00108i, \nonumber \\
&& S_1=-0.00905+0.00234i, \quad
 S_5=0.0252+0.00188i
\end{eqnarray}

The mixing angles and Dirac CP phase in Eq. (\ref{Eq:theta_delta_m3_type4ex}) are consistent with Eq. (\ref{Eq:neutrino_observation_NO}). Moreover, we obtain 
\begin{eqnarray}
&& \Delta m^2_{21} = 7.44 \times 10^{-5}~ {\rm eV^2}, \quad
\Delta m^2_{31} = 2.51 \times 10^{-3} ~{\rm eV^2}, \nonumber \\
&& \sum m_i = 0.0624~{\rm eV}, \quad
|M_{ee}| = 0.0005~{\rm eV},
\label{Eq:DeltaMetc_type4ex}
\end{eqnarray}
from Eqs. (\ref{Eq:m1_m2_alpha_type4ex}) and (\ref{Eq:M_IV_ex}). These predicted values in Eq. (\ref{Eq:DeltaMetc_type4ex}) are consistent with Eq. (\ref{Eq:neutrino_observation_NO}). Thus, the type IV magic texture for NO is consistent with experiments. 

Figures \ref{fig:IV_param_mass} and \ref{fig:IV_param_phases} show the variation in the neutrino oscillation parameters with the masses (Fig. \ref{fig:IV_param_mass})  and phases (Fig. \ref{fig:IV_param_phases}) for type IV magic texture. The following allowed regions and neutrino parameters correlations for the type IV magic texture for NO are obtained:
\begin{itemize}
\item Neutrino masses should be constrained in the following narrow regions:
\begin{eqnarray}
&&m_1/{\rm eV} = 0.00157 \sim 0.00666, \quad
m_2/{\rm eV} = 0.00842 \sim 0.0108, \nonumber \\
&&m_3/{\rm eV} = 0.0494 \sim 0.0513.
\end{eqnarray}
\item Full $3 \sigma$ regions of mixing angles $\theta_{12}$ and of $\theta_{13}$ are allowed:
\begin{eqnarray}
\theta_{12}/^\circ = 31.27 \sim 35.86, \quad
\theta_{13}/^\circ = 8.200 \sim 8.930.
\end{eqnarray}
\item Mixing angle $\theta_{23}$ is mainly in the upper octant region  ($\theta_{23} > 45^\circ$):
\begin{eqnarray}
\theta_{23}/^\circ = 40.10 \sim 40.50, \quad
  49.43 \sim 51.70.
\end{eqnarray}
\item The allowed region of the Dirac CP phase is separated into two regions:
\begin{eqnarray}
\delta/^\circ = 134.7 \sim 227.3,\quad
 338.3 \sim 368.9.
\end{eqnarray}
\item Majorana CP phases should be constrained in the following narrow and separated regions:
\begin{eqnarray}
\alpha_1/^{\circ} = -89.94 \sim -85.00,\quad
 -13.09 \sim 13.43,\quad
87.80 \sim89.96, 
\end{eqnarray}
and
\begin{eqnarray}
\alpha_2/^{\circ} = -90.00 \sim -79.46,\quad
-0.05248 \sim  0.07798, \quad
79.64 \sim 90.00.
\end{eqnarray}
\item The correlation between $\theta_{23}$ and $\delta$ is remarkable. The detail of this topic will be shown in section \ref{section:typeIV_typeIX}.
\end{itemize}

Figure \ref{fig:IV_Dm2} shows the predicted square mass deferences $\Delta m_{ij}^2$ in type IV magic texture. The upper and lower horizontal lines show the observed $\Delta m_{ij}^2$ in 3 $\sigma$ region. Figure \ref{fig:IV_Mee} shows the predicted effective neutrino mass for neutrinoless double $\beta$ decay $|M_{ee}|$ in type IV magic texture. The upper  (lower) horizontal line shows the upper bound from KamLAND-Zen (nEXO) experiment \cite{KamLAND-Zen2016PRL,Licciardi2017JPCS}. From these figures we observed following results.
\begin{itemize}
\item Full 3 $\sigma$ region of the squared mass differences $\Delta m_{21}^{2}$ ($\Delta m_{31}^{2}$) is allowed for $m_1/{\rm eV} = 0.00939 \sim 0.0144$ ($m_3/{\rm eV} = 0.0494 \sim 0.0513$). 
\item The magnitude of the effective neutrino mass of the neutrino less double beta decay $|M_{ee}|$ is predicted as
\begin{eqnarray}
|M_{ee}|/{\rm eV} = 1.216 \times 10^{-5} \sim 2.177\times 10^{-3}.
\end{eqnarray}
We can expect that this prediction may be tested in the next-to-next future experiments.
\end{itemize}
%

\begin{figure}[t]
\begin{center}
\includegraphics[scale=0.45]{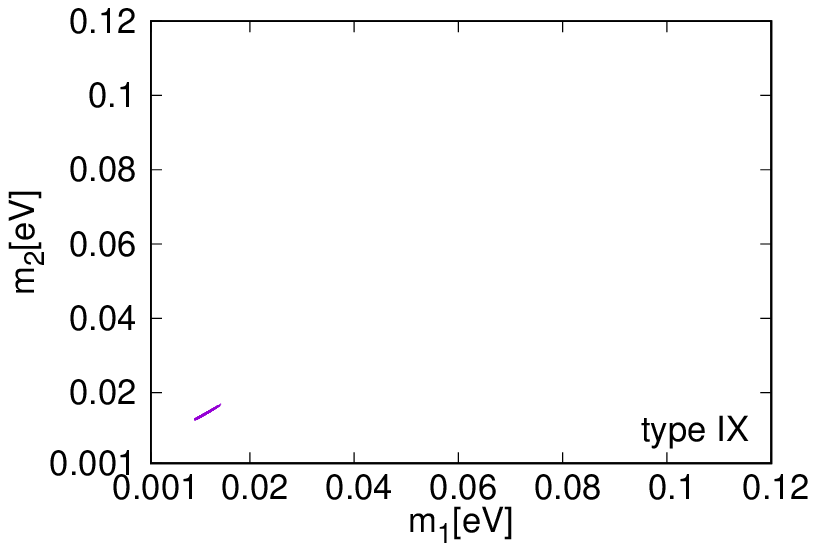}
\includegraphics[scale=0.45]{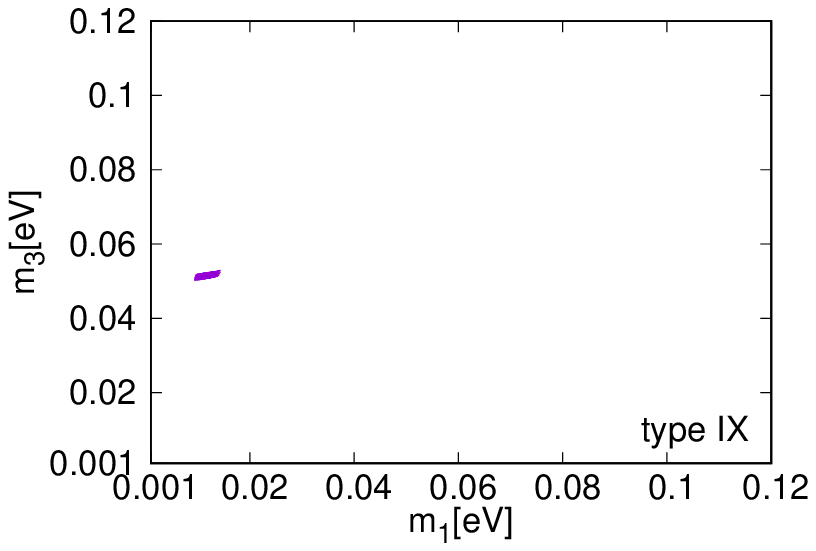}
\includegraphics[scale=0.45]{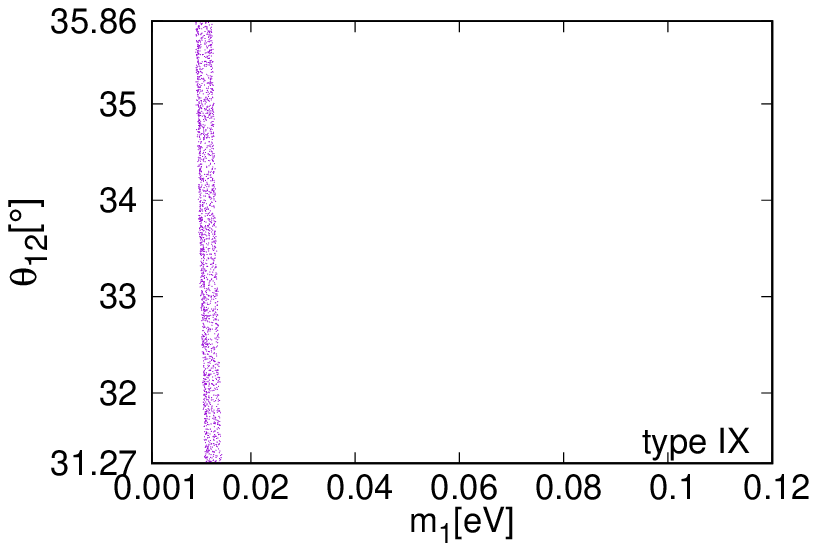}
\includegraphics[scale=0.45]{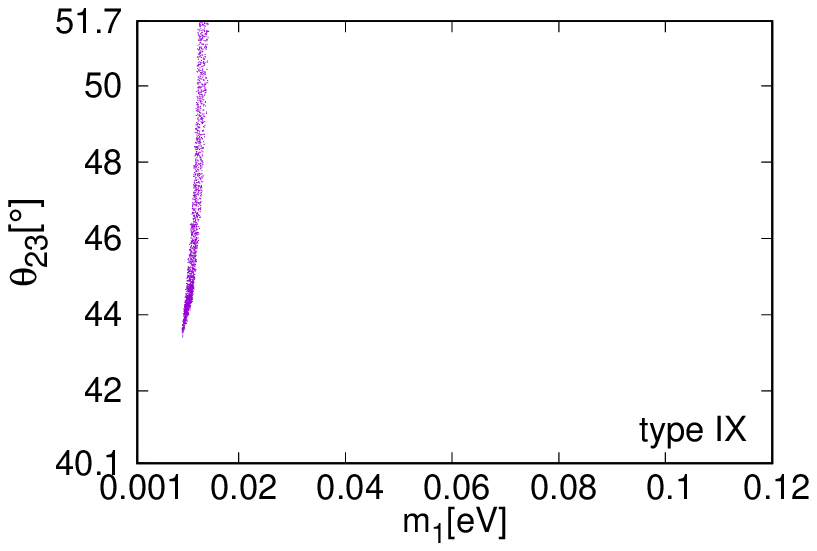}
\includegraphics[scale=0.45]{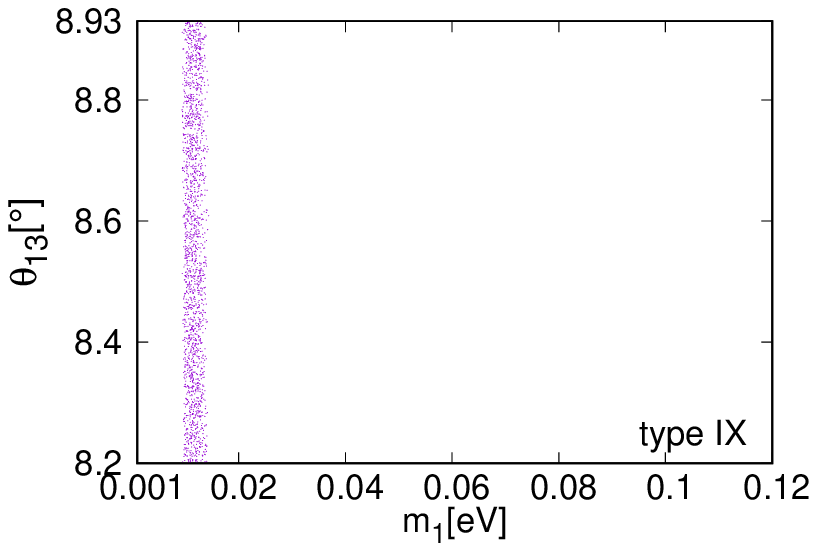}
\includegraphics[scale=0.45]{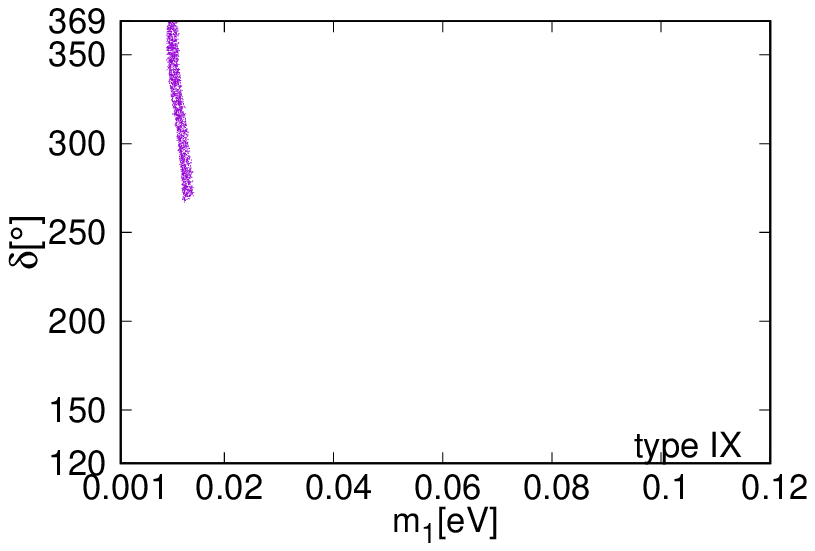}
\includegraphics[scale=0.45]{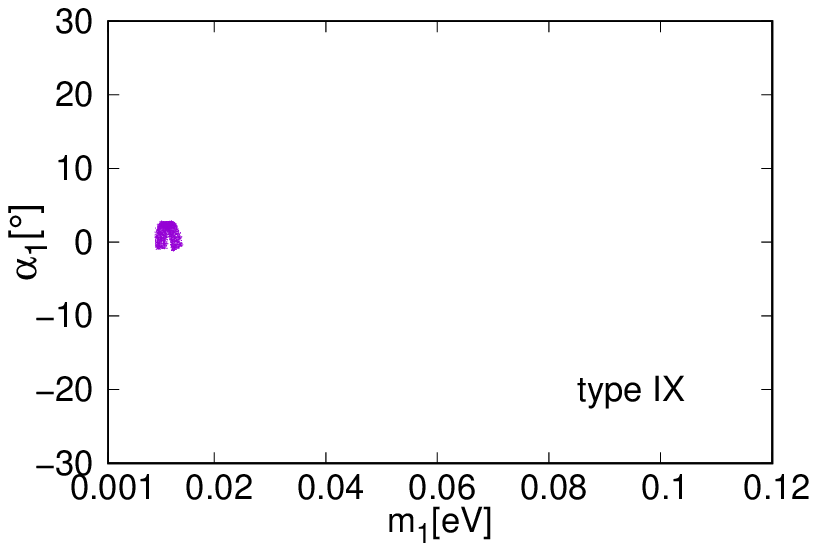}
\includegraphics[scale=0.45]{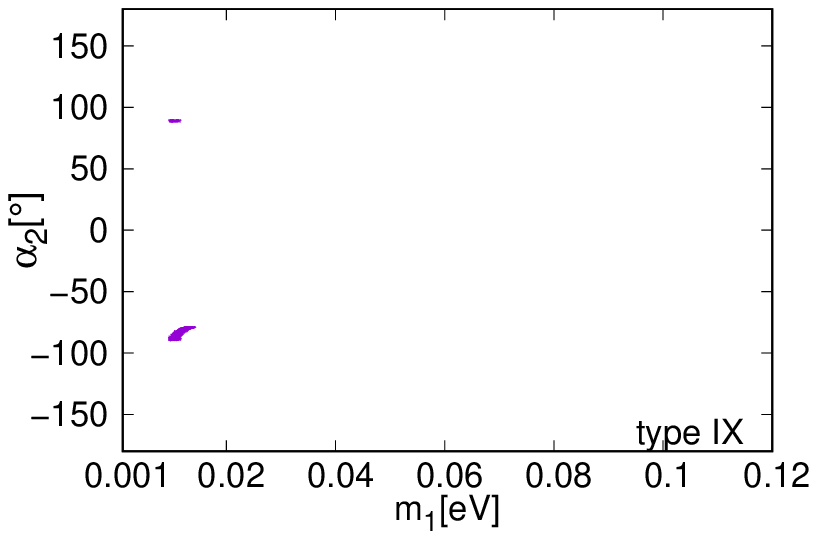}
\includegraphics[scale=0.45]{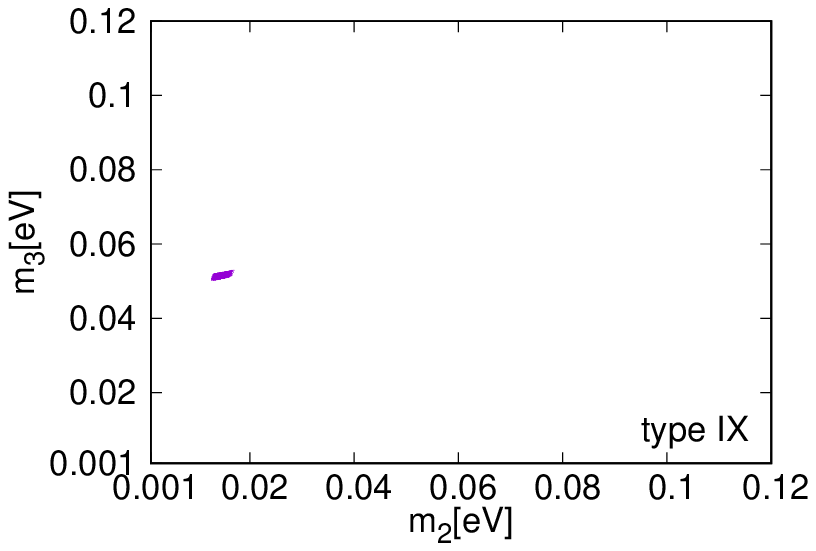}
\includegraphics[scale=0.45]{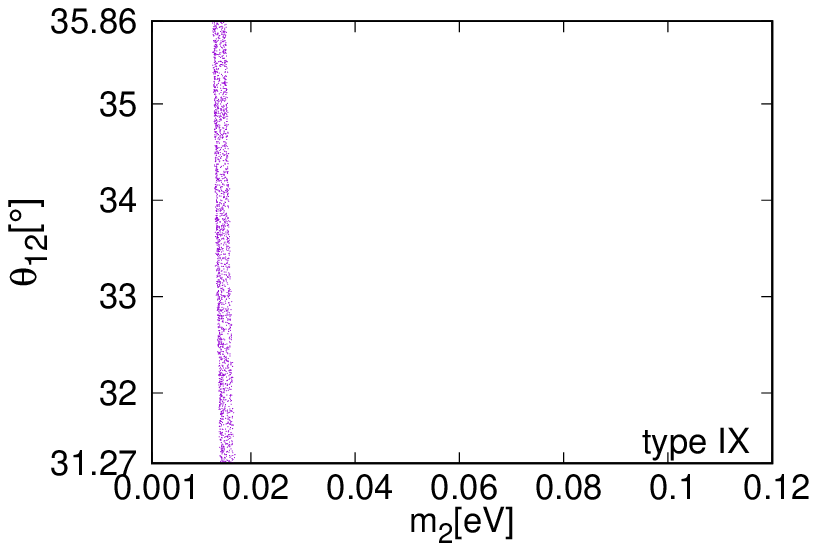}
\includegraphics[scale=0.45]{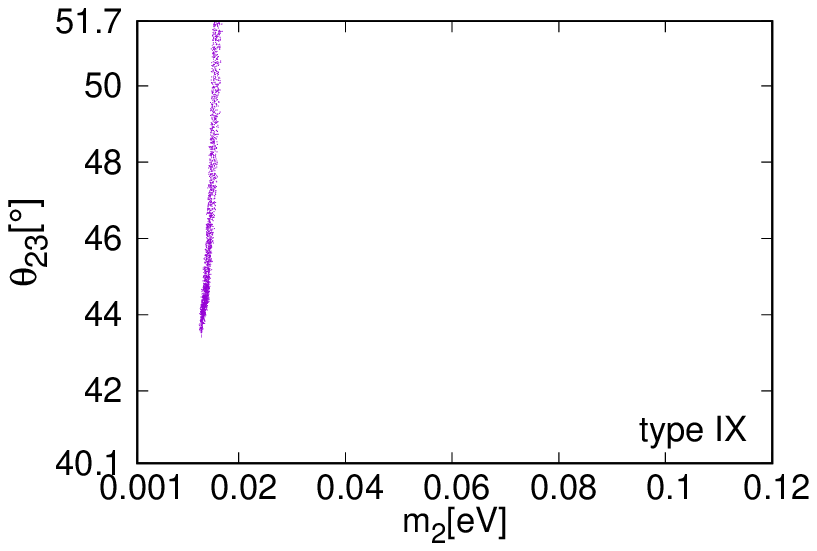}
\includegraphics[scale=0.45]{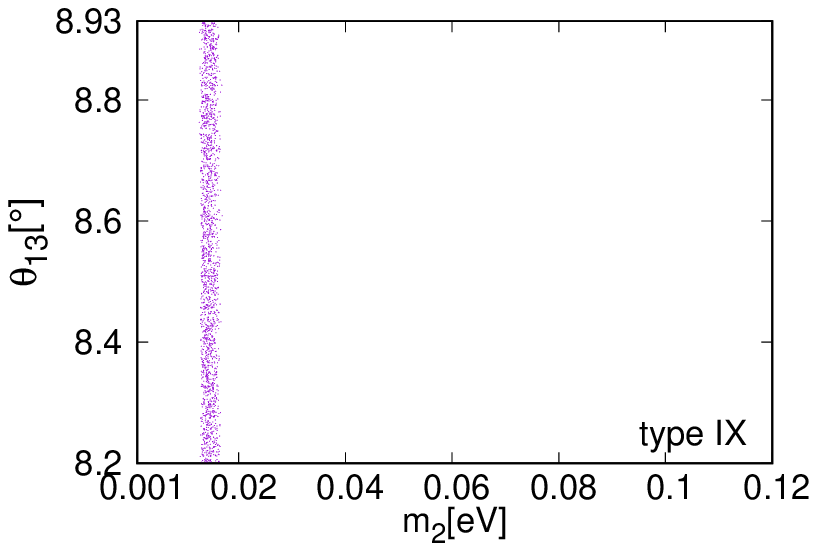}
\includegraphics[scale=0.45]{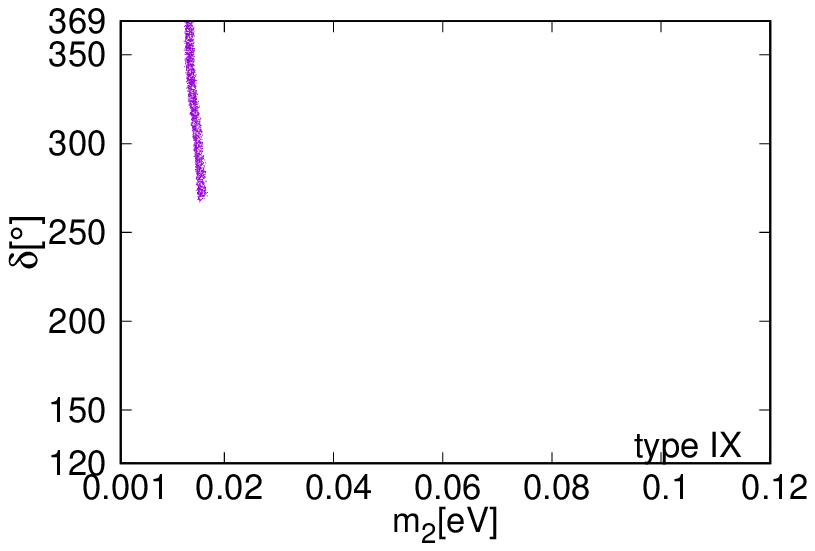}
\includegraphics[scale=0.45]{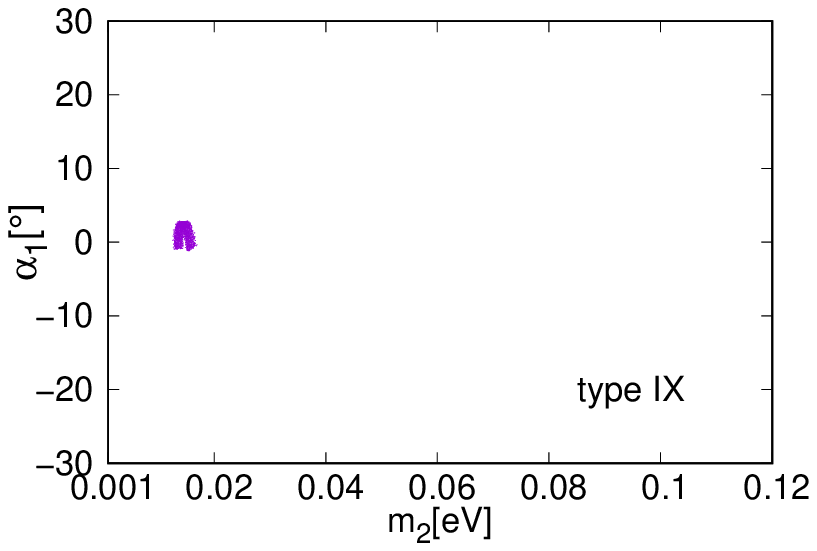}
\includegraphics[scale=0.45]{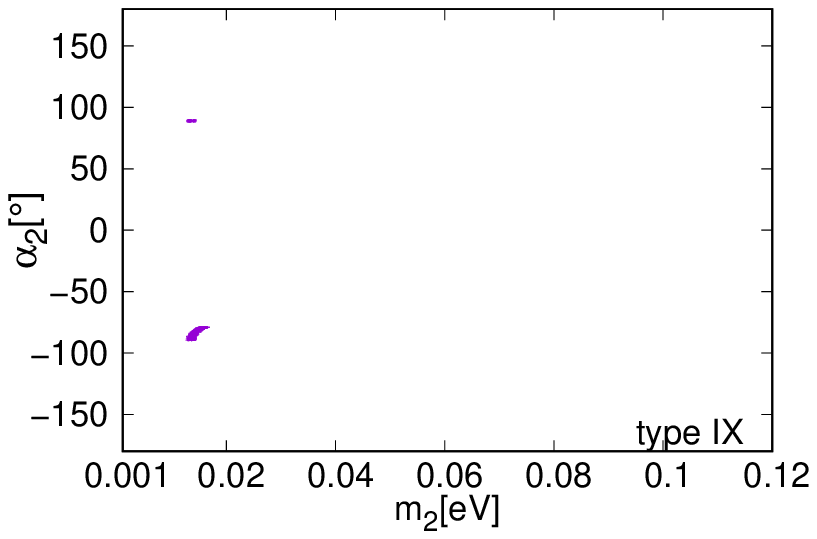}
\includegraphics[scale=0.45]{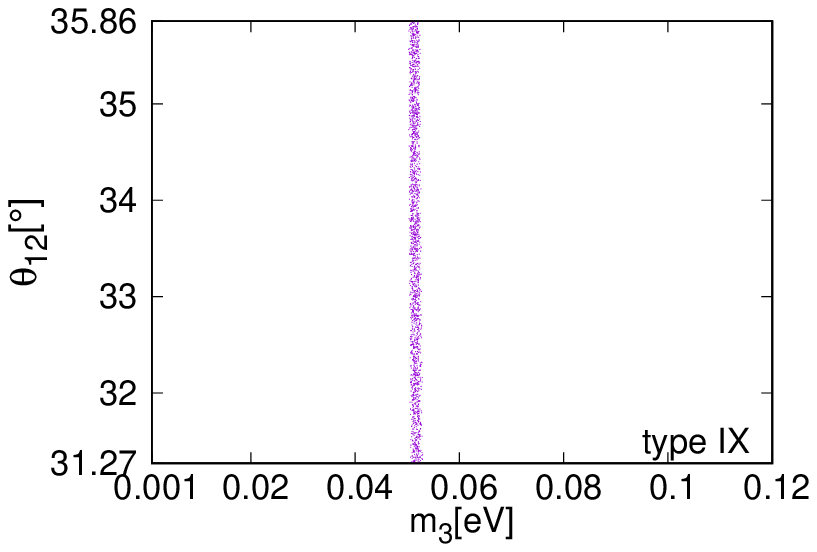}
\includegraphics[scale=0.45]{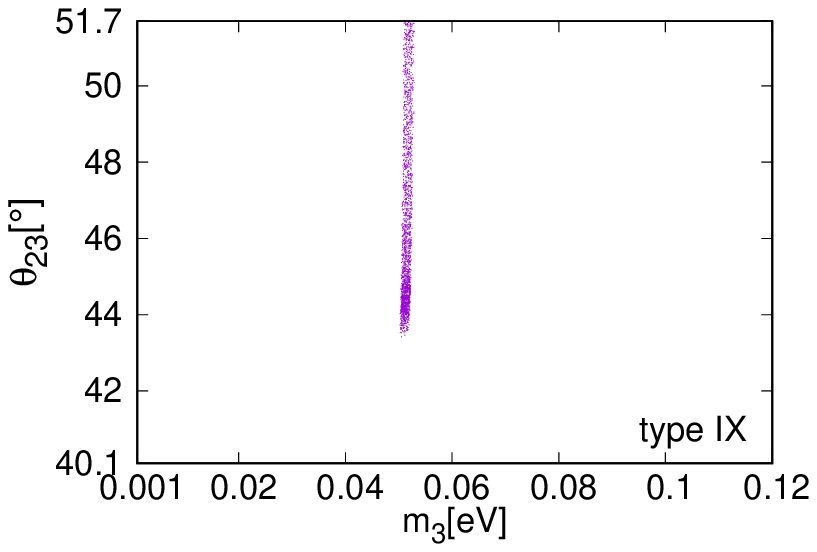}
\includegraphics[scale=0.45]{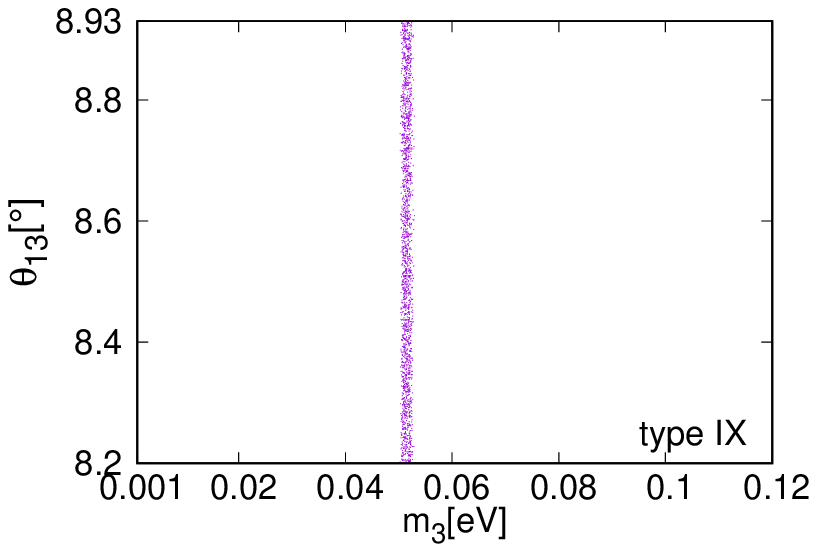}
\includegraphics[scale=0.45]{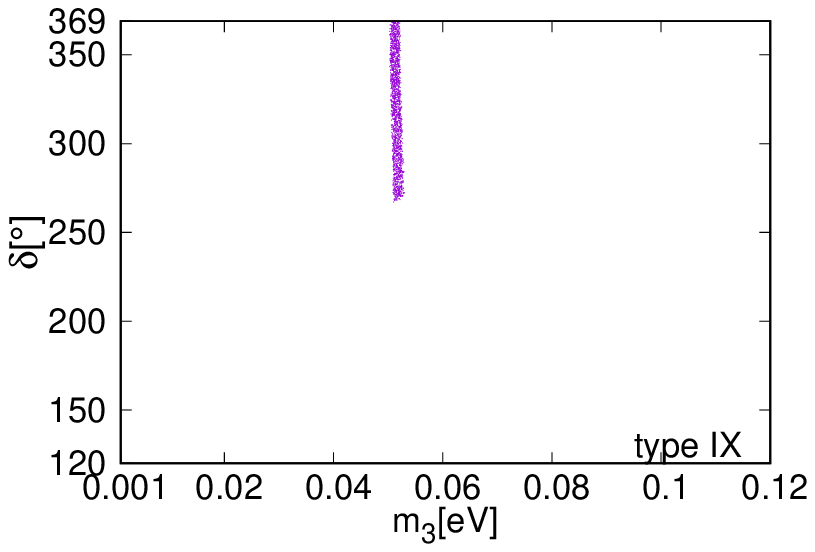}
\includegraphics[scale=0.45]{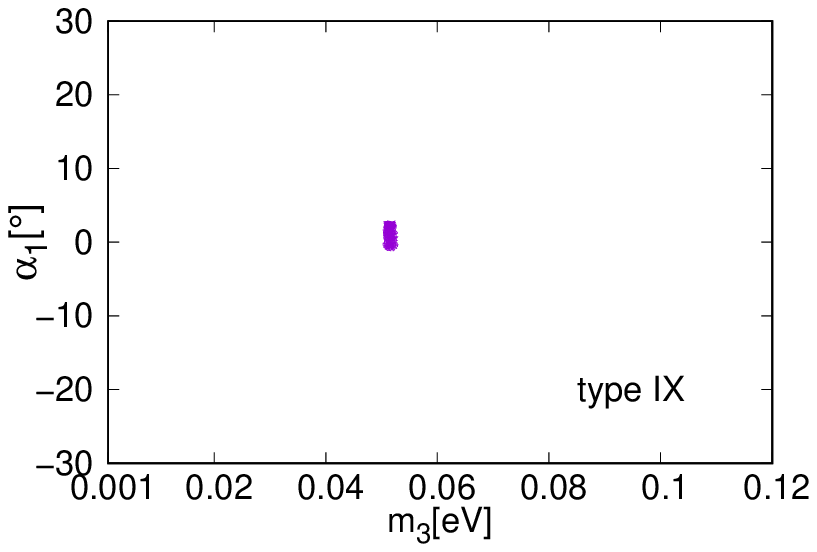}
\includegraphics[scale=0.45]{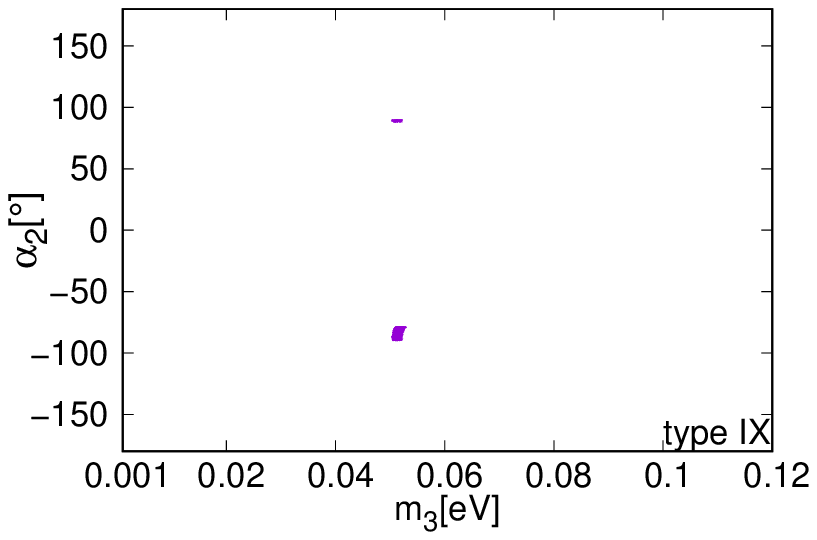}
\caption{Variation in the neutrino oscillation parameters with the masses for type IX magic texture.}
\label{fig:IX_param_mass}
\end{center}
\end{figure}

\begin{figure}[t]
\begin{center}
\includegraphics[scale=0.45]{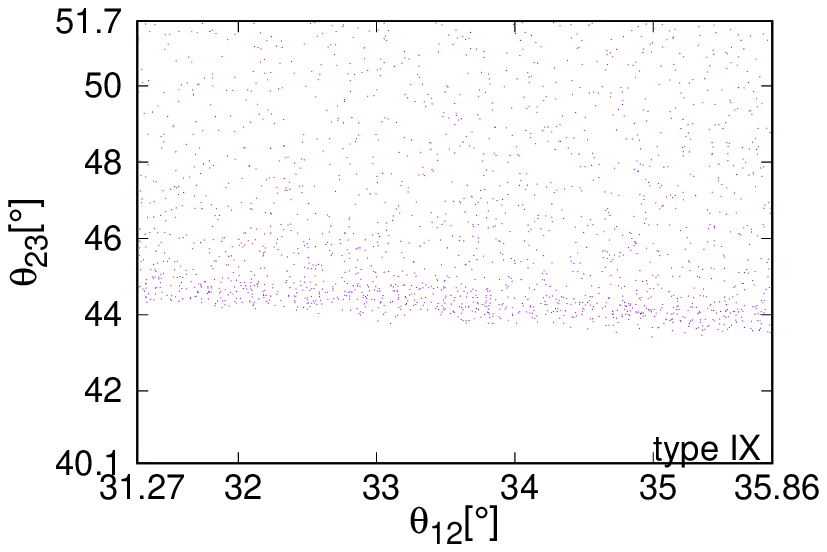}
\includegraphics[scale=0.45]{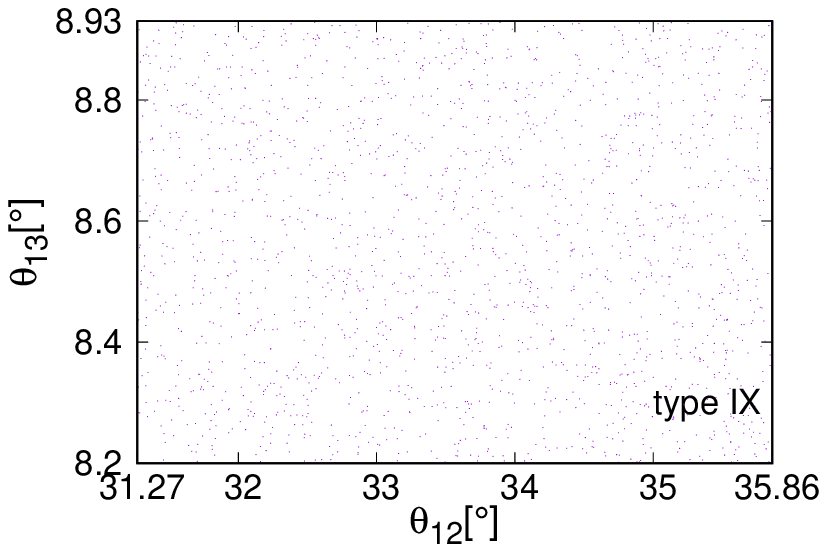}
\includegraphics[scale=0.45]{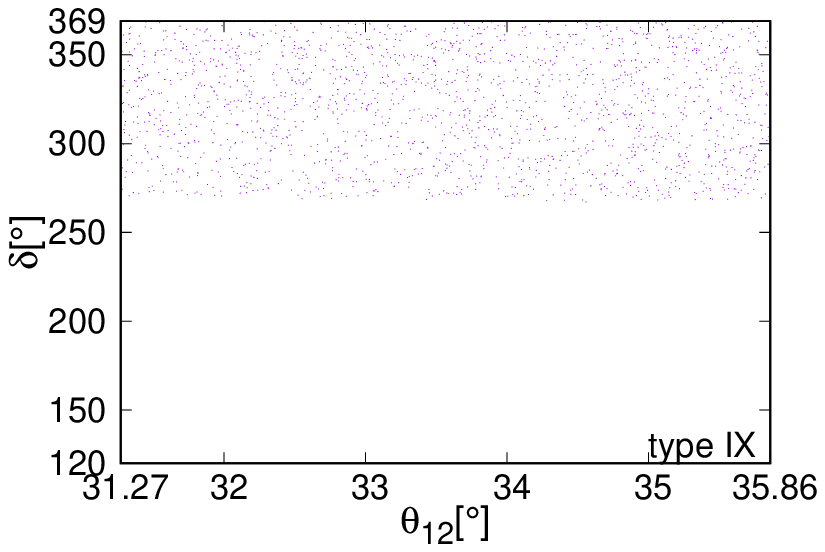}
\includegraphics[scale=0.45]{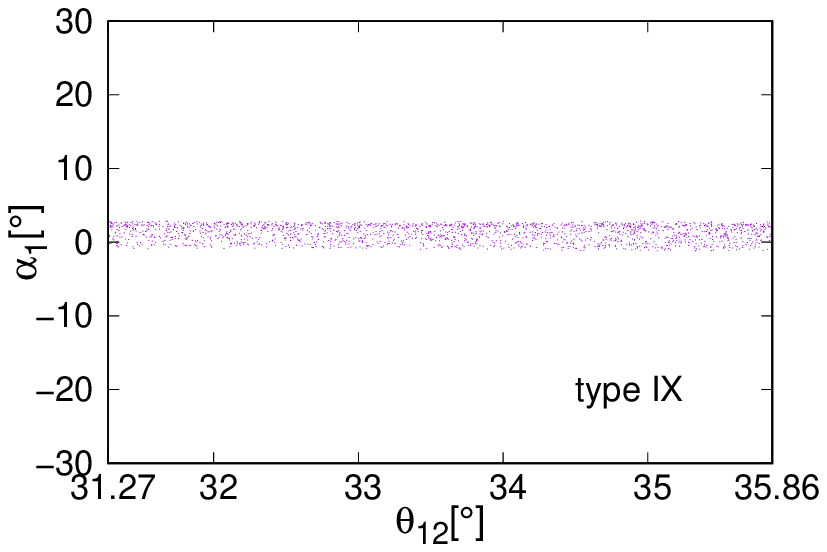}
\includegraphics[scale=0.45]{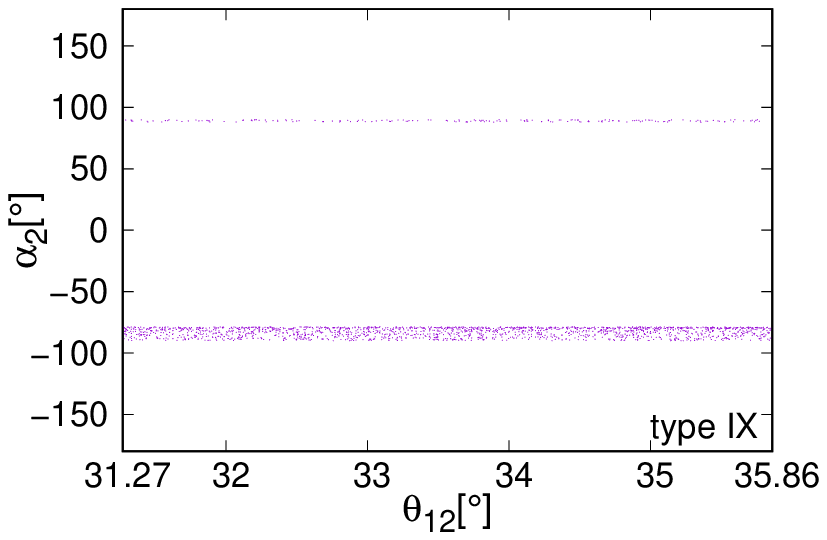}
\includegraphics[scale=0.45]{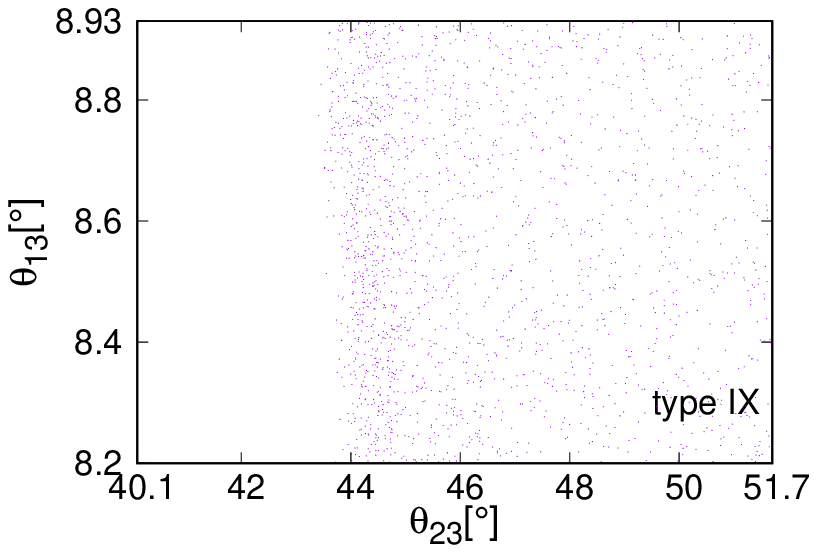}
\includegraphics[scale=0.45]{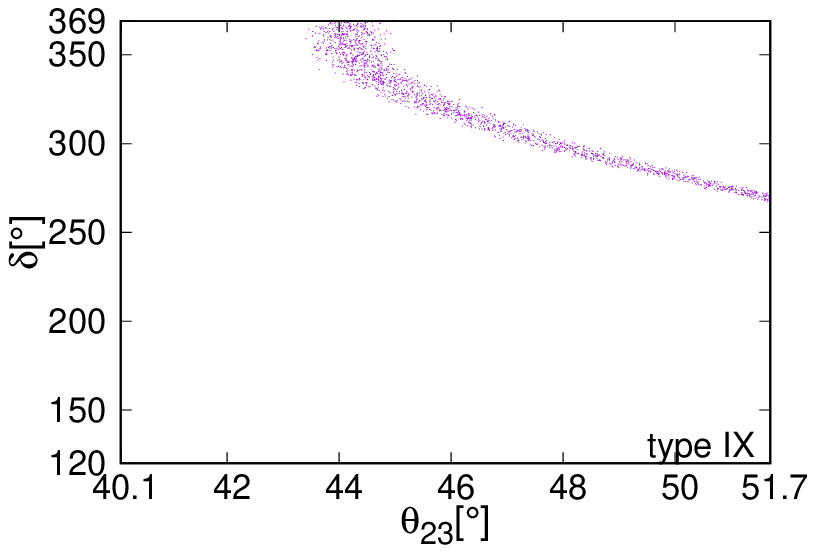}
\includegraphics[scale=0.45]{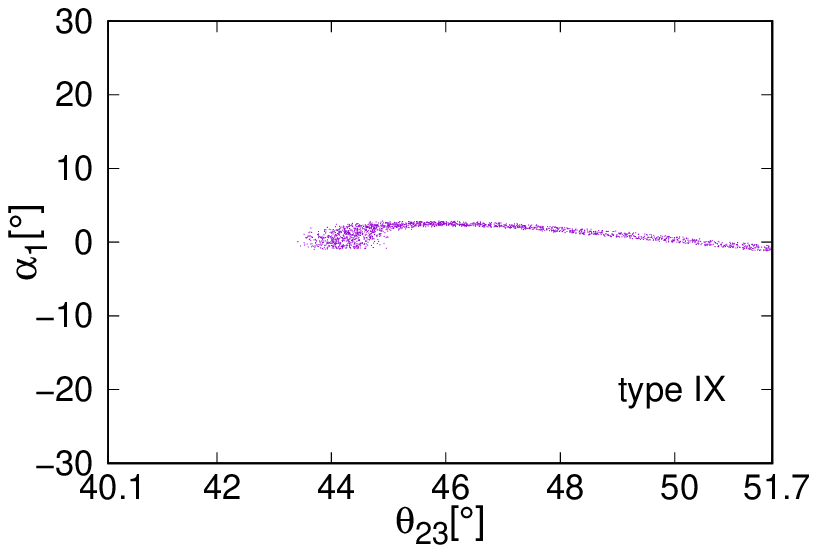}
\includegraphics[scale=0.45]{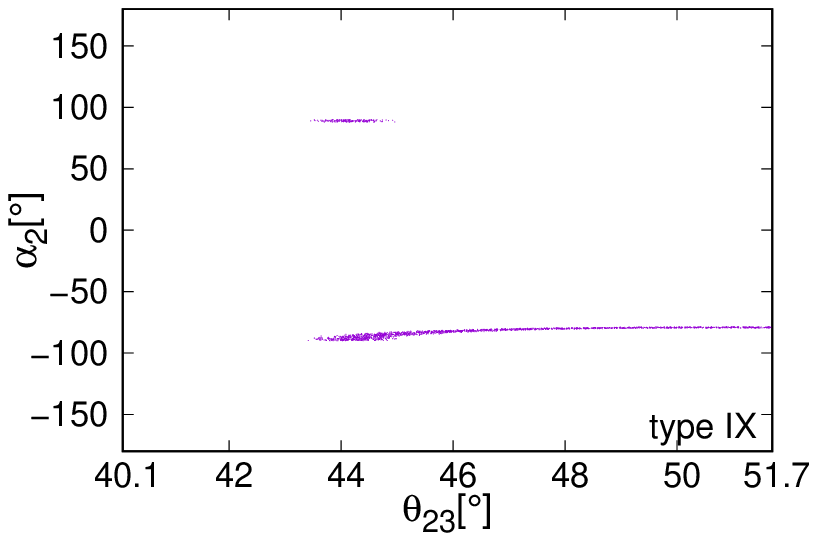}
\includegraphics[scale=0.45]{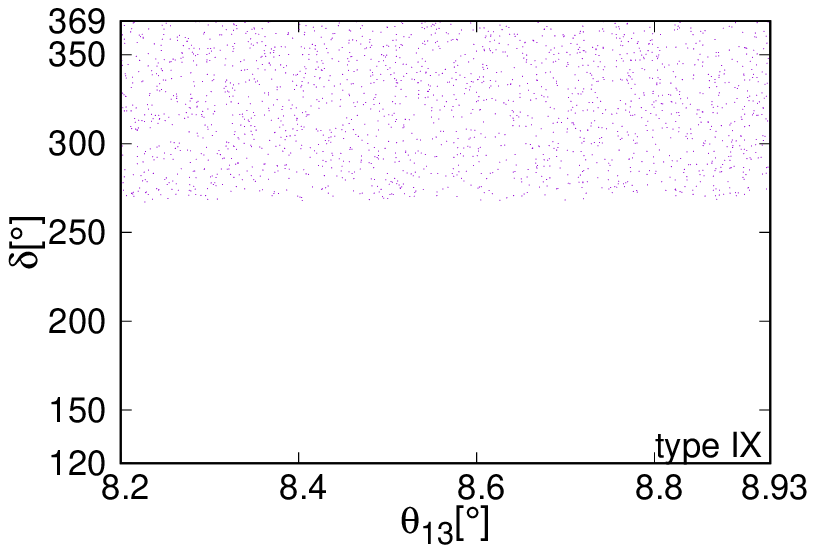}
\includegraphics[scale=0.45]{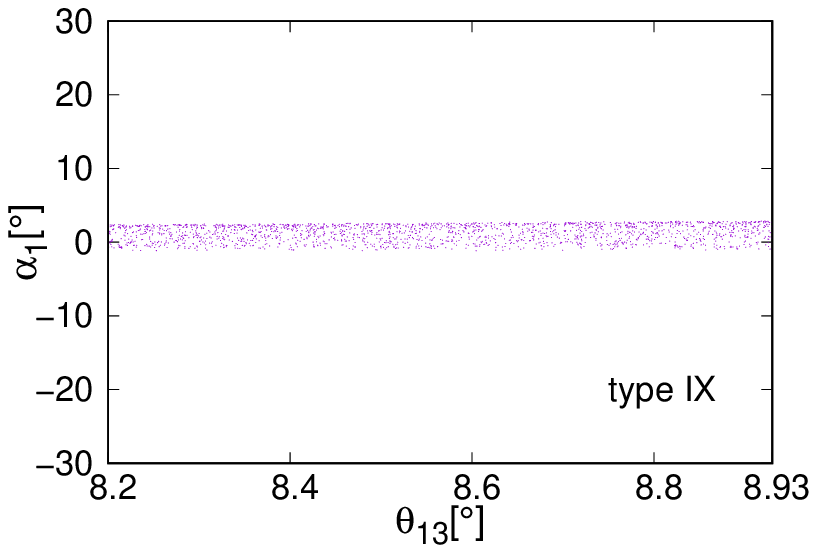}
\includegraphics[scale=0.45]{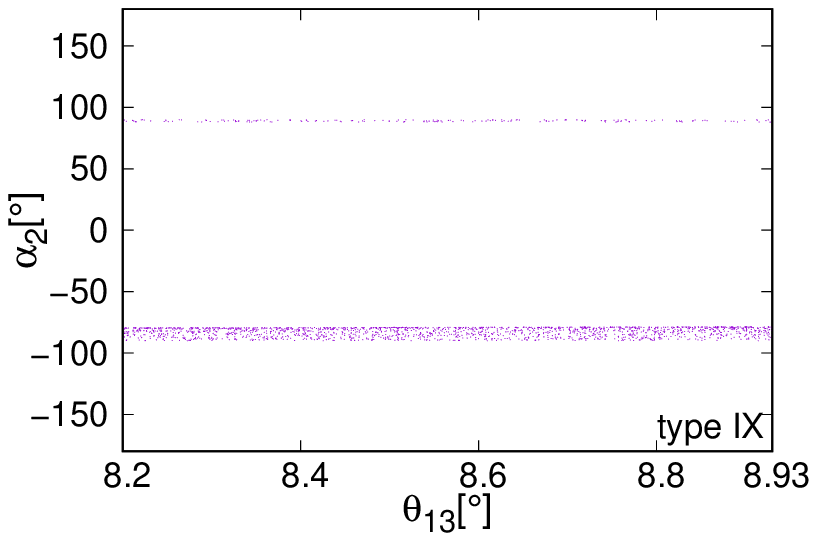}
\includegraphics[scale=0.45]{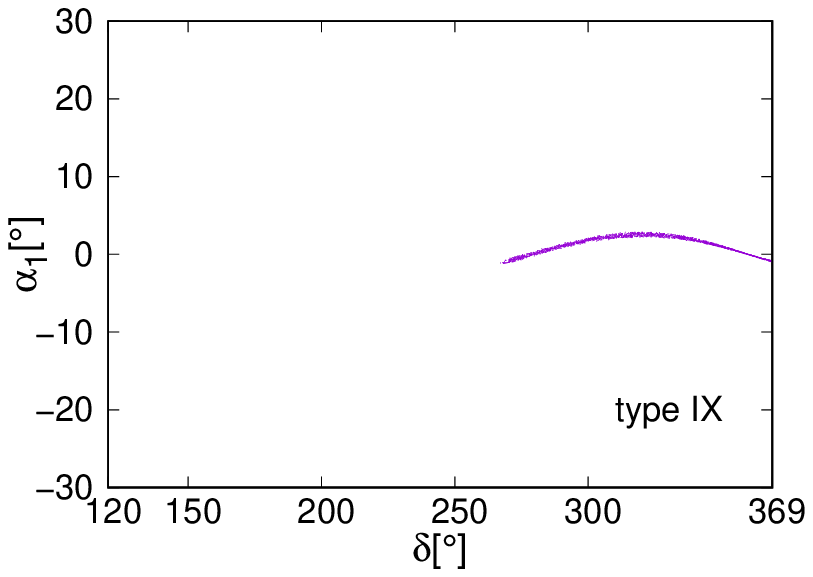}
\includegraphics[scale=0.45]{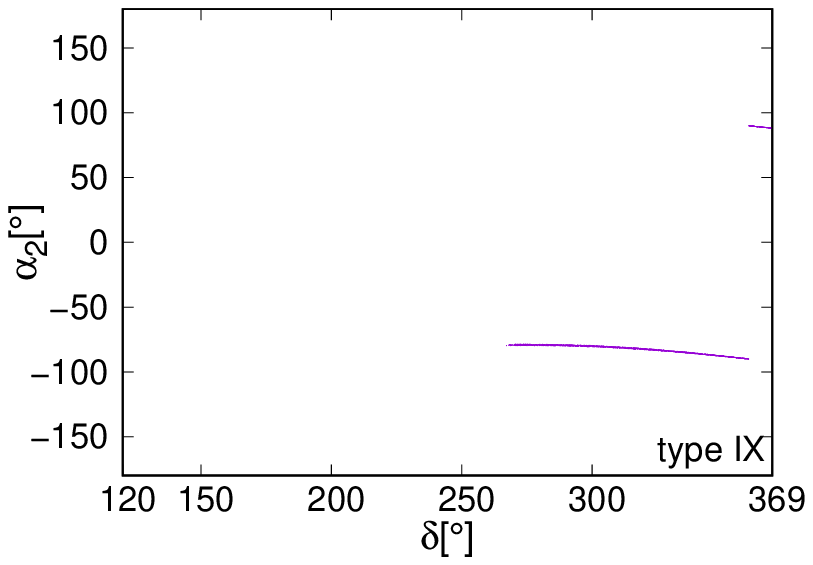}
\includegraphics[scale=0.45]{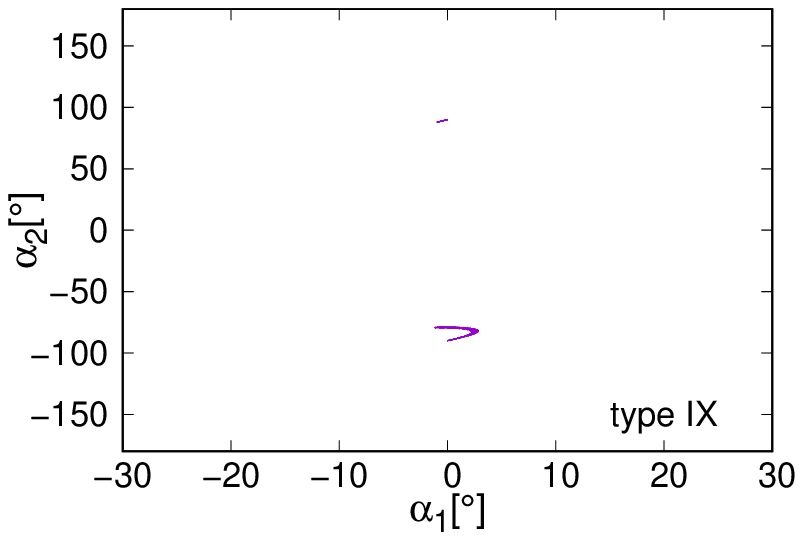}
\caption{Variation in the neutrino oscillation parameters with the mixing angles and CP phases for type IX magic texture.}
\label{fig:IX_param_phases}
\end{center}
\end{figure}

\begin{figure}[t]
\begin{center}
\includegraphics[scale=0.6]{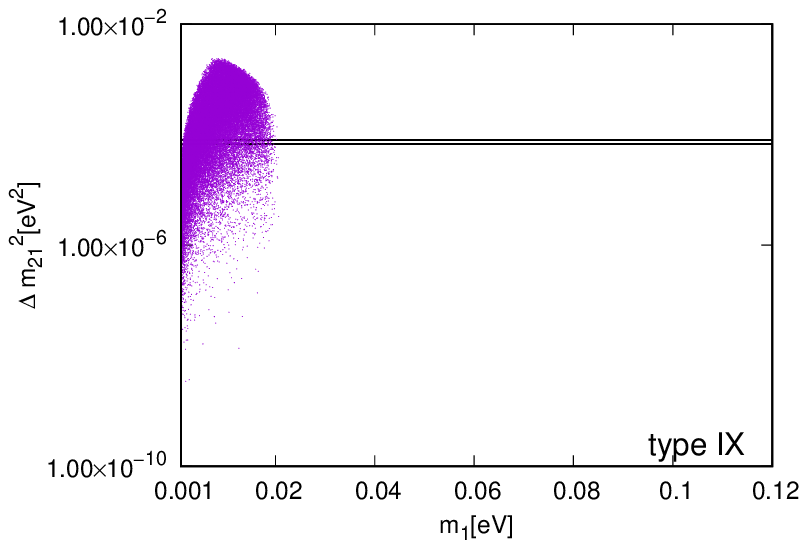}
\includegraphics[scale=0.6]{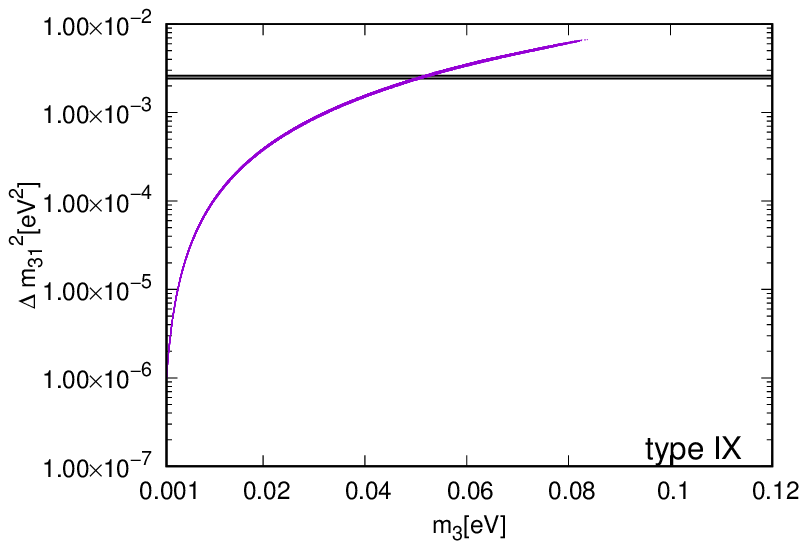}
\caption{Predicted square mass deferences $\Delta m_{ij}^2$ in type IX magic texture. The upper and lower horizontal lines show the observed $\Delta m_{ij}^2$ in 3 $\sigma$ region.}
\label{fig:IX_Dm2}
\end{center}
\end{figure}

\begin{figure}[t]
\begin{center}
\includegraphics[scale=0.6]{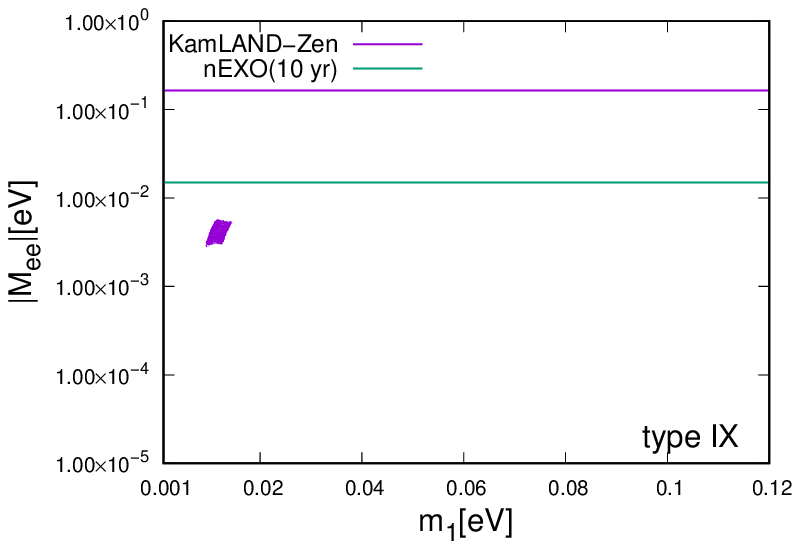}
\includegraphics[scale=0.6]{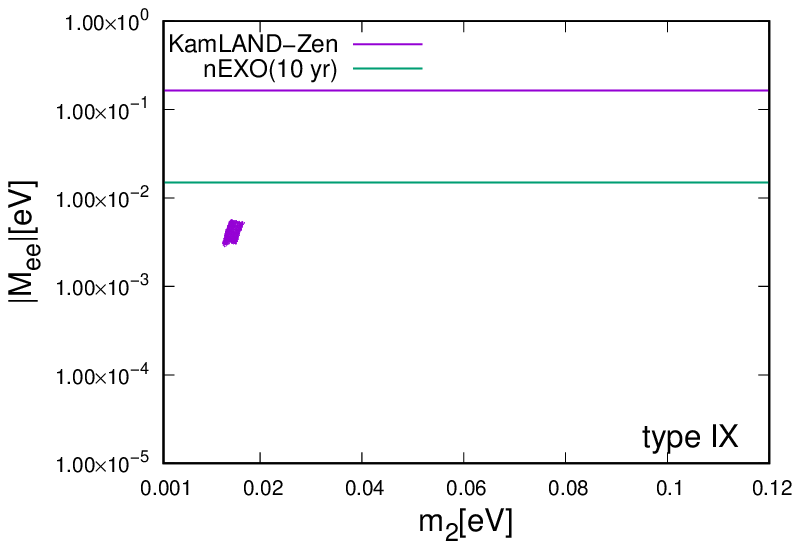}
\includegraphics[scale=0.6]{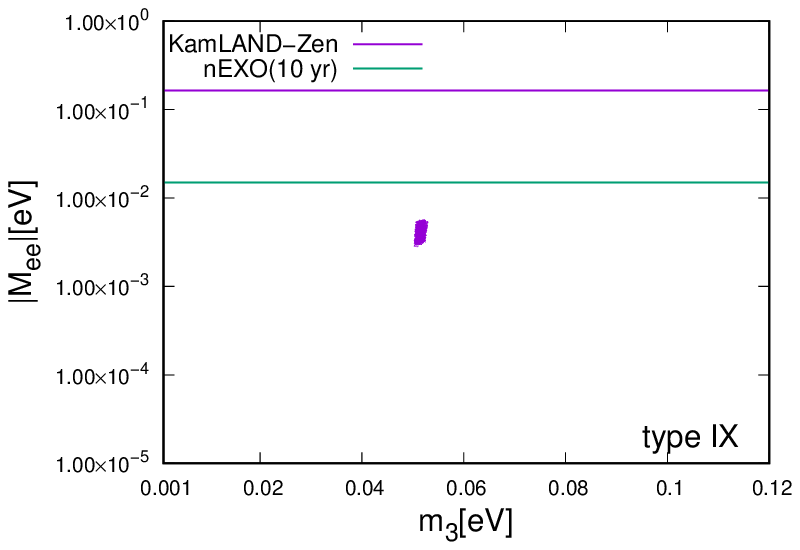}
\caption{Predicted effective neutrino mass for neutrinoless double $\beta$ decay $|M_{ee}|$ in type IX magic texture. The upper  (lower) horizontal line shows the upper bound from KamLAND-Zen (nEXO) experiment.}
\label{fig:IX_Mee}
\end{center}
\end{figure}

\subsection{Type IX \label{subsection:type-IX}}
Type IX magic texture is also consistent with experiments. For example, if we set 
\begin{eqnarray}
(\theta_{12},\theta_{23},\theta_{13},\delta) = (32.56^\circ, 46.65^\circ, 8.89^\circ, 312^\circ), \quad
m_3=0.0515~{\rm eV},
\label{Eq:theta_delta_m3_type9ex}
\end{eqnarray}
which are consistent with observations, then we obtain
\begin{eqnarray}
(m_1,m_2)=(0.0122,0.0148)~{\rm eV},  \quad
(\alpha_1, \alpha_2) =(2.609^\circ, -80.97^\circ),
\label{Eq:m1_m2_alpha_type9ex}
\end{eqnarray}
and 
\begin{eqnarray}
(M_{\rm IX})_{ee} &=& 0.004348-0.001751 i,  \nonumber \\
(M_{\rm IX})_{e\mu} &=& -0.004602-0.006304 i,   \nonumber \\
(M_{\rm IX})_{e\tau} &=& 0.011825-0.002438 i,   \nonumber \\
(M_{\rm IX})_{\mu\mu} &=& 0.024461+0.000298 i,   \nonumber \\ 
(M_{\rm IX})_{\mu\tau} &=& 0.028252+0.001427 i,   \nonumber \\
(M_{\rm IX})_{\tau\tau} &=& 0.019302-0.003126 i.
\label{Eq:M_IX_ex}
\end{eqnarray}
The matrix $M_{\rm IX}$ in Eq. (\ref{Eq:M_IX_ex}) is satisfied with the criteria of type IX magic texture ($S_2=S_4=S_5 \neq S_1 \neq S_2$) as follows: 
\begin{eqnarray}
&& S_2=S_4=S_5=0.048111-0.004578 i, \nonumber \\
&& S_1=0.011571-0.010493 i, \quad
S_3=0.059380-0.004137 i.
\end{eqnarray}
We obtain 
\begin{eqnarray}
&&\Delta m^2_{21} = 6.98 \times 10^{-5}~ {\rm eV^2}, \quad
\Delta m^2_{31} = 2.50 \times 10^{-3} ~{\rm eV^2}, \nonumber \\
&&\sum m_i = 0.07848~{\rm eV}, \quad
|M_{ee}| = 0.004687~{\rm eV},
\label{Eq:DeltaMetc_type9ex}
\end{eqnarray}
from Eqs. (\ref{Eq:m1_m2_alpha_type9ex}) and (\ref{Eq:M_IX_ex}) and these predicted values are consistent with observations for NO case. 

From more general numerical calculations, the following allowed regions and correlations of the neutrino parameters for the type IX magic texture for NO are obtained (see Figs. \ref{fig:IX_param_mass} and \ref{fig:IX_param_phases}):
\begin{itemize}
\item Neutrino masses should be constrained in the following narrow regions (similar to type IV):
\begin{eqnarray}
&&m_1/{\rm eV} = 0.00939 \sim 0.0144, \quad
m_2/{\rm eV} = 0.0126 \sim 0.0169, \nonumber \\ 
&&m_3/{\rm eV} = 0.0503 \sim 0.0530.
\end{eqnarray}
\item Full $3 \sigma$ regions of mixing angles $\theta_{12}$ and of $\theta_{13}$ are allowed  (similar to type IV):
\begin{eqnarray}
\theta_{12}/^\circ = 31.27 \sim 35.86, \quad
\theta_{13}/^\circ = 8.200 \sim 8.930.
\end{eqnarray}
\item Wide region of the mixing angle $\theta_{23}$ is allowed (different from type IV):
\begin{eqnarray}
\theta_{23}/^\circ = 43.41 \sim 51.70.
\end{eqnarray}
\item The allowed region of the Dirac CP phase is not separated into two regions (different from type IV):
\begin{eqnarray}
\delta/^\circ = 267.1 \sim 369.0. 
\end{eqnarray}
\item Majorana CP phases should be constrained in the following narrow and separated regions (similar to type IV):
\begin{eqnarray}
\alpha_1/^{\circ}=-1.163 \sim-0.00426,  \quad
 0.009331 \sim 2.878,
\end{eqnarray}
and
\begin{eqnarray}
\alpha_2/^{\circ}=-89.98 \sim -78.62, \quad
   88.05 \sim 89.99.
\end{eqnarray}
\item The correlation between $\theta_{23}$ and $\delta$ is remarkable  (similar to type IV). For more detail, see the next subsection.
\end{itemize}

Figure \ref{fig:IX_Dm2} shows the predicted square mass deferences $\Delta m_{ij}^2$ in type IX magic texture. The upper and lower horizontal lines show the observed $\Delta m_{ij}^2$ in 3 $\sigma$ region. Figure \ref{fig:IX_Mee} shows the predicted effective neutrino mass for neutrinoless double $\beta$ decay $|M_{ee}|$ in type IX magic texture. The upper  (lower) horizontal line shows the upper bound from KamLAND-Zen (nEXO) experiment \cite{KamLAND-Zen2016PRL,Licciardi2017JPCS}. From these figures we observed the following results.
\begin{itemize}
\item Full 3 $\sigma$ region of the squared mass differences $\Delta m_{21}^{2}$ ($\Delta m_{31}^{2}$) is allowed for $m_1/{\rm eV} = 0.00939 \sim 0.0144$ ($m_3/{\rm eV} = 0.0503 \sim 0.0530$). 
\item The magnitude of the effective neutrino mass of the neutrino less double beta decay $|M_{ee}|$ is predicted in narrow region as
\begin{eqnarray}
|M_{ee}|/{\rm eV} = 2.823 \times 10^{-3}  \sim 5.659 \times 10^{-3}.
\end{eqnarray}
We can expect that this prediction may be tested in the next-to-next future experiments.
\end{itemize}
%
\begin{figure}[t]
\begin{center}
\includegraphics{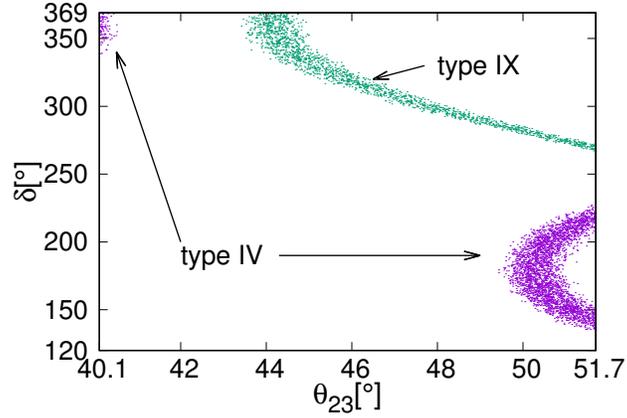}
\caption{Correlation between $\theta_{23}$ and $\delta$ in type IV and type IX magic textures.}
\label{fig:IV_IX}
\end{center}
\end{figure}

\begin{figure}[t]
\begin{center}
\includegraphics{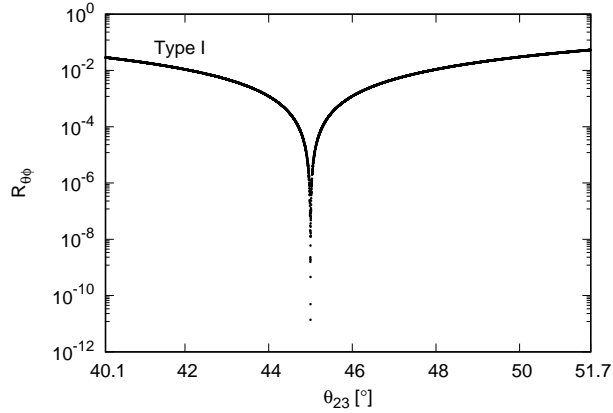}
\caption{Correlation between  $\theta_{23}$ and $R_{\theta\phi}$ for allowed $\{\phi, \theta \}$ for $40.1^\circ \le \theta_{23} \le 51.7^\circ$ in Fig. \ref{fig:I_allowed_phi_delta} in type I magic texture.}
\label{fig:I_R}
\end{center}
\end{figure}

\begin{figure}[t]
\begin{center}
\includegraphics{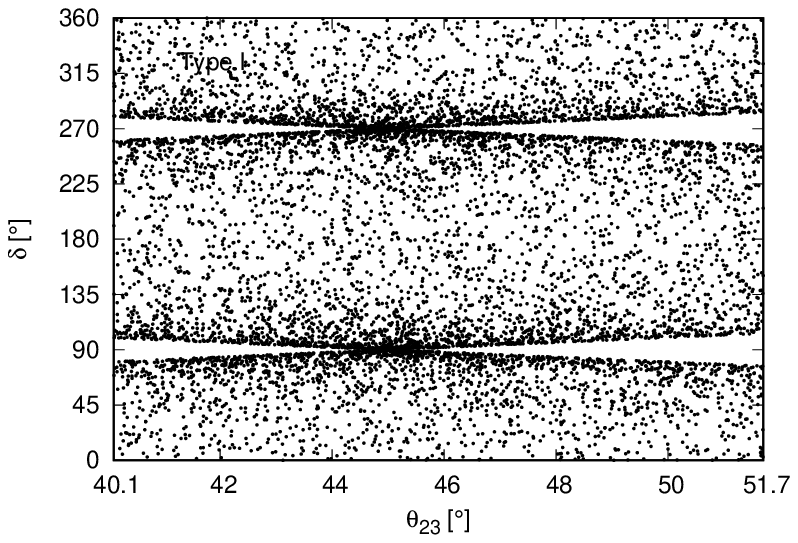}
\caption{Correlation between $\theta_{23}$ and $\delta$ in type I magic texture.}
\label{fig:I_delta_theta23}
\end{center}
\end{figure}
\begin{figure}[t]
\begin{center}
\includegraphics{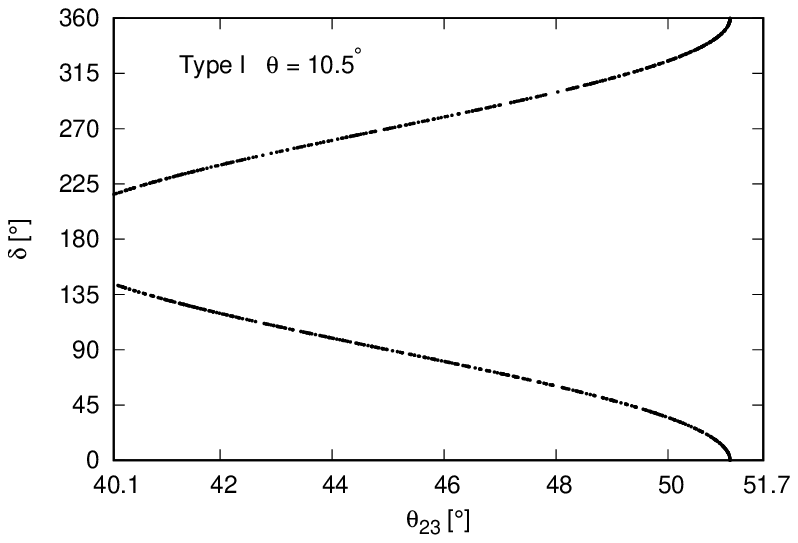}
\caption{Correlation between $\theta_{23}$ and $\delta$ for $\theta=10.5^\circ$ in type I magic texture.}
\label{fig:I_delta_theta23_theta_10_5}
\end{center}
\end{figure}

\subsection{Remarkable difference between type IV and type IX \label{section:typeIV_typeIX}}
Although the overall behavior of the allowed parameter regions is almost the same in type IV and type IX, there is a remarkable difference between these two textures. Figure \ref{fig:IV_IX} shows the correlation between $\theta_{23}$ and $\delta$ in type IV and type IX. The following points are important:
\begin{itemize}
\item Since the best-fit value of the Dirac CP phase $\delta$ is $197^\circ$ for NO, type IV is more favorable than type IX.
\item If future experiments show that $\theta_{23}$ is in lower octant ($\theta_{23} < 45^\circ$), type IV magic square should be severely constrained.
\end{itemize}

These results are very different from the result from type I. In the type I, the CP-violating phase $\delta$ can be calculated from the Jarlskog invariant in Eq. (\ref{Eq:Jarlskog}) with the general relations of the mixing angles in Eq. (\ref{Eq:mixingAngle_from_UPMNS}) and the mixing matrix of trimaximal mixing for $\nu_2$ in Eq. (\ref{Eq:UTM}) as \cite{Gautam2016PRD}
\begin{eqnarray}
\csc^2 \delta = \csc^2 \phi - \frac{3 \sin^2 2\theta \cot^2 \phi}{(3-2\sin^2\theta)^2}.
\label{Eq:delta_typeI}
\end{eqnarray}

Now we define the following ratio
\begin{eqnarray}
R_{\theta\phi} =  \left( \frac{3 \sin^2 2\theta \cot^2 \phi}{(3-2\sin^2\theta)^2}\right)/ \csc^2 \phi .
\label{Eq:delta_typeI}
\end{eqnarray}
Figure \ref{fig:I_R} shows correlation between  $\theta_{23}$ and $R_{\theta\phi}$ for allowed $\{\phi, \theta \}$ for $40.1^\circ \le \theta_{23} \le 51.7^\circ$ which are shown in Fig. \ref{fig:I_allowed_phi_delta} for the type I magic texture. We observe that $R_{\theta\phi} \lesssim 10^{-2}$ for all region of $40.1^\circ \le \theta_{23} \le 51.7^\circ$. In addition, $R_{\theta\phi} \lesssim 10^{-3}$ is satisfied for almost all region of $40.1^\circ \le \theta_{23} \le 51.7^\circ$. Thus we obtain $\csc^2 \delta \sim \csc^2 \phi$ as well as 
\begin{eqnarray}
\delta \sim  \phi.
\end{eqnarray}
Since wide region of $\phi$ is allowed for $40.1^\circ \le \theta_{23} \le 51.7^\circ$ as shown in Fig. \ref{fig:I_allowed_phi_delta}, wide region of $\delta$ is also allowed for $40.1^\circ \le \theta_{23} \le 51.7^\circ$ by the relation of $\delta \sim \phi$. We show the correlation between $\theta_{23}$ and $\delta$ in type I magic texture in Fig. \ref{fig:I_delta_theta23}. As we expected, wide region of $\delta$ is allowed for $40.1^\circ \le \theta_{23} \le 51.7^\circ$. If we fix $\theta$, the result to be more predictable. As an example, we show the correlation between $\theta_{23}$ and $\delta$ for fixed $\theta=10.5^\circ$ in type I magic texture in Fig. \ref{fig:I_delta_theta23_theta_10_5}.

\subsection{Other types}
We have shown that the type I, type IV for NO and type IX for NO magic textures are consistent with observations. In this subsection, we show that the other magic textures should be excluded from observations. 

With the following definitions,
\begin{eqnarray}
R_{\rm NO} = \frac{\Delta m_{21}^2}{\Delta m_{31}^2}, \quad 
R_{\rm IO} = \frac{\Delta m_{21}^2}{-\Delta m_{32}^2},
\label{Eq:R_deff}
\end{eqnarray}
the allowed types of magic texture should be satisfied with
\begin{eqnarray}
0.0280 \le R_{\rm NO} \le 0.0309,
\label{Eq:R_NO}
\end{eqnarray}
for NO
\begin{eqnarray}
0.0264 \le R_{\rm IO} \le 0.0333,
\label{Eq:R_IO}
\end{eqnarray}
for IO from observations. 

Table \ref{tab:R} shows that the predicted ratios of $R_{\rm NO}$ and $R_{\rm IO}$ for the magic textures where we vary the mixing angles and Dirac CP phase within observed $3 \sigma$ range (and we chose one of the neutrino mass within  $0.001 - 0.12$ eV as we mentioned in Sec. \ref{subsection:criteria}). The bold font indicates that the predicted range is consistent with observations. The magic textures other than type I, and type IV for NO, and type IX for NO are excluded from observations.

\begin{table}[t]
\tiny
\caption{Predicted $R_{\rm NO}$ and $R_{\rm IO}$. The bold font indicates that the predicted range is consistent with observations.}
\begin{center}
\begin{tabular}{|c|c|c|}
\hline
type &$R_{\rm NO}$&$R_{\rm IO}$ \\ 
\hline
I & {\bf Any (No prediction)} & {\bf Any (No prediction)} \\
\hline
II & $-24244\sim-3.9362$, $6.5190\sim23542$  & $0.7971\sim 1.1835 $\\
\hline
III & $0.57894\sim0.8858$ & $-7.2618\sim-1.4118$\\
\hline
IV & $-0.0390\sim-4.6698\times 10^{-6}$, ${\bf 2.8406\times10^{-7}\sim0.0389}$ & $-0.0382\sim-5.6224\times 10^{-7}$, $4.1799\times 10^{-7}\sim0.0378$\\
\hline
V & $2.9140\sim3.7114$ & $1.3688\sim1.5228$\\
\hline
VI & $0.9194\sim1.1876$ & $-32478\sim-12.019$, \quad $6.3837\sim163617$\\
\hline
VII & $-8.0021\times 10^{-8}\sim2.9937\times 10^{-5}$ & $-1.1834\times 10^{-12}\sim1.0652\times 10^{-12}$\\
\hline
VIII & $0.9907\sim1.9089$ & $-1638.3\sim-964.01$, $-341.83\sim-149.29$, \\
      && $2.0603\sim745.66$, $934.57\sim2224.3$\\
\hline
IX&$-0.0421\sim-1.0689\times 10^{-5}$,${\bf 2.5382\times 10^{-5}\sim0.7030}$ & $-2.0429\sim-6.0243\times 10^{-6}$,$4.6107\times 10^{-5}\sim0.0401$\\
\hline
X & $-1391.2\sim-508.95$,$-359.35\sim-0.0086$, & $-3414.5\sim-2654.3$, $-656.54\sim-501.68$\\
&$0.9102\sim406.56$, $3325.6\sim10199$ & $-263.42\sim-10.682$, $0.0105\sim402.23$, $617.14\sim785.35$\\
\hline
\end{tabular}
\end{center}
\label{tab:R}
\end{table}

\subsection{Symmetry arguments for the magic textures}
In this subsection, we would like to show some symmetry arguments for magic textures.

A perfect magic texture for Majorana neutrinos, $S_1=S_2=S_3=S_4=S_5$, can be written by using Lucas's formula \cite{Sallows1997MathIntelli},
\begin{eqnarray}
M_0 =\left(
\begin{matrix}
-\alpha + \beta & \alpha + \beta &  \beta \\
\alpha + \beta &  \beta & -\alpha + \beta \\
 \beta & -\alpha + \beta & \alpha + \beta \\
\end{matrix}
\right),
\end{eqnarray}
and this matrix $M_0$ satisfies the following $Z_2$ symmetry \cite{Lam2006PRD,Yang2021arXiv}
\begin{eqnarray}
G M_0 G^T = M_0,
\end{eqnarray}
where 
\begin{eqnarray}
G =\left(
\begin{matrix}
\frac{1}{3} & -\frac{2}{3} &  -\frac{2}{3} \\
-\frac{2}{3} &  \frac{1}{3} & -\frac{2}{3} \\
-\frac{2}{3} & -\frac{2}{3} & \frac{1}{3} \\
\end{matrix}
\right),
\end{eqnarray}
and $G^2 = {\rm diag.}(1,1,1)$. Although the type I magic texture isn't the perfect magic texture, $S_1=S_2=S_3 \neq S_4 \neq S_5$, it is also invariant under this $Z_2$ symmetry;
\begin{eqnarray}
G M_{\rm I} G^T=M_{\rm I}.
\end{eqnarray}
We can understand the origin of this residual symmetry under $Z_2$ transformation by the following discussion. The type I magic texture can be constructed by using the perfect magic texture $M_0$ and a perfect-magic-texture breaking term $ \Delta M_{\rm I}$
\begin{eqnarray}
M_{\rm I} = M_0 + \Delta M_{\rm I},
\end{eqnarray}
where 
\begin{eqnarray}
\Delta M_{\rm I} = \left(
\begin{matrix}
0 & 0 & \delta  \\
0 & 0 & \delta \\
\delta & \delta &-\delta  \\
\end{matrix}
\right) \quad {\rm or} \quad
\left(
\begin{matrix}
0 & 0 & 0  \\
0 & \delta & -\delta \\
0 & -\delta &\delta  \\
\end{matrix}
\right).
\label{Eq:DeltaM_I}
\end{eqnarray}
For example, if we choose the second matrix in Eq.(\ref{Eq:DeltaM_I}) as a perfect-magic-texture breaking term, we obtain 
\begin{eqnarray}
M_{\rm I} =\left(
\begin{matrix}
-\alpha + \beta & \alpha + \beta &  \beta \\
\alpha + \beta &  \beta & -\alpha + \beta \\
 \beta & -\alpha + \beta & \alpha + \beta \\
\end{matrix}
\right)
+
\left(
\begin{matrix}
0 & 0 & 0  \\
0 & \delta & -\delta \\
0 & -\delta &\delta  \\
\end{matrix}
\right),
\end{eqnarray}
and which is satisfied with the type I magic criteria $S_1=S_2=S_3=3\beta, S_4=3\beta+2\delta, S_5=3\beta+\delta$. Since two perfect-magic-texture breaking terms in Eq.(\ref{Eq:DeltaM_I}) are invariant under $Z_2$ symmetry 
\begin{eqnarray}
G\left(
\begin{matrix}
0 & 0 & \delta  \\
0 & 0 & \delta \\
\delta & \delta &-\delta  \\
\end{matrix}
\right) G^T 
=
\left(
\begin{matrix}
0 & 0 & \delta  \\
0 & 0 & \delta \\
\delta & \delta &-\delta  \\
\end{matrix}
\right),
\quad
G\left(
\begin{matrix}
0 & 0 & 0  \\
0 & \delta & -\delta \\
0 & -\delta &\delta  \\
\end{matrix}
\right)G^T
=
\left(
\begin{matrix}
0 & 0 & 0  \\
0 & \delta & -\delta \\
0 & -\delta &\delta  \\
\end{matrix}
\right),
\end{eqnarray}
the type I magic texture is also invariant under $Z_2$ symmetry.

On the contrary, the type II, type III, $\cdots$, type X magic textures are not invariant under $Z_2$ symmetry. The perfect-magic-texture breaking terms for the type II, type III, $\cdots$, type X magic textures are
\begin{eqnarray}
\Delta M_{\rm II} = \left(
\begin{matrix}
0 & 0 & \delta  \\
0 & 0 & \delta \\
\delta & \delta & \delta  \\
\end{matrix}
\right) \quad {\rm or} \quad
\left(
\begin{matrix}
0 & 0 & 0  \\
0 & \delta & -\delta \\
0 & -\delta & -\delta  \\
\end{matrix}
\right),
\end{eqnarray}
\begin{eqnarray}
\Delta M_{\rm III} = \left(
\begin{matrix}
0 & 0 & \delta  \\
0 & 0 & -\delta \\
\delta & -\delta & \delta  \\
\end{matrix}
\right) \quad {\rm or} \quad
\left(
\begin{matrix}
0 & 0 & 0  \\
0 & \delta & \delta \\
0 & \delta & -\delta  \\
\end{matrix}
\right),
\end{eqnarray}
\begin{eqnarray}
\Delta M_{\rm IV} = \left(
\begin{matrix}
0 & 0 & \delta  \\
0 & 0 & -\delta \\
\delta & -\delta & -\delta  \\
\end{matrix}
\right) \quad {\rm or} \quad
\left(
\begin{matrix}
0 & 0 & 0  \\
0 & \delta & \delta \\
0 & \delta & \delta  \\
\end{matrix}
\right),
\end{eqnarray}
\begin{eqnarray}
\Delta M_{\rm V} = \left(
\begin{matrix}
\delta & 0 & 0  \\
0 & \delta & 0 \\
0 & 0 & 0  \\
\end{matrix}
\right) \quad {\rm or} \quad
\left(
\begin{matrix}
0 & 0 & \delta  \\
0 & -\delta & 2\delta \\
0 & 2\delta & 0  \\
\end{matrix}
\right),
\end{eqnarray}
\begin{eqnarray}
\Delta M_{\rm VI} = \left(
\begin{matrix}
0 & 0 & 0  \\
0 & 0 & \delta \\
0 & \delta & -\delta  \\
\end{matrix}
\right),
\end{eqnarray}
\begin{eqnarray}
\Delta M_{\rm VII} = \left(
\begin{matrix}
0 & 0 & \delta  \\
0 & 0 & 2\delta \\
\delta & 2\delta & -\delta  \\
\end{matrix}
\right) \quad {\rm or} \quad
\left(
\begin{matrix}
0 & 0 & 0  \\
0 & \delta & 0 \\
0 & 0 & \delta  \\
\end{matrix}
\right),
\end{eqnarray}
\begin{eqnarray}
\Delta M_{\rm VIII} = \left(
\begin{matrix}
0 & \delta & 0  \\
\delta & \delta & 0 \\
0 & 0 & 0  \\
\end{matrix}
\right) \quad {\rm or} \quad
\left(
\begin{matrix}
0 & 0 & \delta  \\
0 & -\delta & 0 \\
\delta & 0 & 2\delta  \\
\end{matrix}
\right),
\end{eqnarray}
\begin{eqnarray}
\Delta M_{\rm IX} = \left(
\begin{matrix}
\delta & 0 & 0  \\
0 & 0 & 0 \\
0 & 0 & -\delta  \\
\end{matrix}
\right) \quad {\rm or} \quad
\left(
\begin{matrix}
0 & \delta & 0  \\
\delta & 0 & -\delta \\
0 & -\delta & 0  \\
\end{matrix}
\right)\quad {\rm or} \quad
\left(
\begin{matrix}
0 & 0 & \delta  \\
0 & 0 & 2\delta \\
\delta & 2\delta & 2\delta  \\
\end{matrix}
\right),
\end{eqnarray}
and
\begin{eqnarray}
\Delta M_{\rm X} = \left(
\begin{matrix}
0 & 0 & \delta  \\
0 & 0 & -\delta \\
\delta & -\delta & 2\delta  \\
\end{matrix}
\right) \quad {\rm or} \quad
\left(
\begin{matrix}
0 & 0 & 0  \\
0 & \delta & \delta \\
0 & \delta & 0  \\
\end{matrix}
\right).
\end{eqnarray}
These breaking terms are not invariant under $Z_2$ symmetry.

The invariance of the type I magic texture under $Z_2$ symmetry as a general property of a mass independent texture \cite{Lam2006PRD}. In deed, the type I magic texture is independent from the neutrino mass eigenvalues $(m_1,m_2,m_3)$ and is related with only mixing matrix. Thus, we can understand the structure of the type I magic texture by the special structure of the mixing matrix. On the other hand, the type II, type III, $\cdots$, type X magic textures are not invariant under $Z_2$ symmetry. It seems that we can not understand the structure of these new 9 magic textures by some special structure of the mixing matrix. Thus, we study the structure of magic textures by seeing some flavor structure in the mass matrix. 

We recall that the type I, type IV and type IX magic textures are allowed from observations. The perfect-magic-texture breaking terms for the type I, type IV and type IX magic textures, $\Delta M_i (i={\rm I, IV, IX})$, satisfy one of the following forms:
\begin{eqnarray}
\left(
\begin{matrix}
X & 0 & 0  \\
* & 0 & 0 \\
* & * & -X  \\
\end{matrix}
\right),  \quad
 \left(
\begin{matrix}
0 & X & 0  \\
* & 0 & -X \\
* & * & 0  \\
\end{matrix}
\right),
\label{Eq:DeltaM_XX}
\end{eqnarray}
and
\begin{eqnarray}
\left(
\begin{matrix}
0 & 0 & X  \\
* & 0 & X \\
* & * & Y  \\
\end{matrix}
\right), \quad 
\left(
\begin{matrix}
0 & 0 & Y  \\
* & 0 & X \\
* & * & X  \\
\end{matrix}
\right), \quad
\left(
\begin{matrix}
0 & 0 & 0  \\
* & X & X \\
* & * & X  \\
\end{matrix}
\right), \quad
\left(
\begin{matrix}
0 & 0 & 0  \\
* & X & Y \\
* & * & X  \\
\end{matrix}
\right), 
\label{Eq:DeltaM_XXY}
\end{eqnarray}
where ``$*$" denotes symmetric partner in the Majorana (symmetric) mass matrix, ``$X$" and ``$Y$" denote nonzero elements. From Eqs.(\ref{Eq:DeltaM_XX}) and (\ref{Eq:DeltaM_XXY}), we can expect that the allowed magic textures have some flavor symmetries which are related with $ee \leftrightarrow \tau\tau$, $e\mu \leftrightarrow \mu\tau$, $e\tau \leftrightarrow \mu\tau$, $\mu\tau \leftrightarrow \tau\tau$, $\mu\mu \leftrightarrow \mu\tau$  or $\mu\mu \leftrightarrow \tau\tau$ permutations. For the type I magic texture, a kind of these flavor symmetry is studied in ref \cite{Yang2021arXiv}. 
Since our main aim of this study is discovering new types of magic textures which are consistent with observation, we would like to perform  more advanced discussions for the flavor structure of magic textures in our future study. 

\subsection{More six types of magic textures}
Up to now, we have required that the number of equal $S_i$ is three. This requirement is based on the success of the type I magic texture which has three equal $S_i$. However, there no longer is theoretical meaning that the number of equal $S_i$ is just three. We should also investigate the following additional six types:

\begin{description}
\item[Type A-I:  ] $S_1=S_2=S_3=S_4=S_5$
\item[Type A-II:  ] $S_2=S_3=S_4=S_5 \neq S_1$
\item[Type A-III:  ] $S_1=S_3=S_4=S_5 \neq S_2$
\item[Type A-IV:  ] $S_1=S_2=S_4=S_5 \neq S_3$
\item[Type A-V:  ] $S_1=S_2=S_3=S_5 \neq S_4$
\item[Type A-VI:  ] $S_1=S_3=S_3=S_4 \neq S_5$
\end{description}

Unfortunately, our method of making magic texture, equations (\ref{Eq:lambda_i/lambda_j}), (\ref{Eq:ratio_of_masses}), and (\ref{Eq:alpha1_alpha2}), can be only used for a magic texture with three equal $S_i$. We can not use these equations to investigate the type A-I, A-II, $\cdots$, A-VI magic textures. Thus, we use the following relations
\begin{eqnarray}
|S_1 - S_2| \le \epsilon, \  |S_2 - S_3| \le \epsilon, \ |S_3 - S_4| \le \epsilon \  {\rm and} \ |S_4 - S_5|  \le \epsilon,
\label{Eq:A_I}
\end{eqnarray}
for type A-I magic texture where $\epsilon$ denotes the tolerance \cite{Verma2020JPGNP}. If Eq. (\ref{Eq:A_I}) is satisfied with observed neutrino parameters, we recognize the type A-I magic texture is consistent with observations. Otherwise, we understand the type A-I magic texture should be excluded from observations. Similarly, we use 
\begin{eqnarray}
|S_2 - S_3| \le \epsilon, \  |S_3 - S_4| \le \epsilon,  \  {\rm and} \ |S_4 - S_5|  \le \epsilon,
\label{Eq:A_II}
\end{eqnarray}
for type A-II and so on. The tolerance is the key value in this method. In this paper, according to Verma and Kashav \cite{Verma2020JPGNP}, if a type of magic texture is satisfied with the criteria such as Eq. (\ref{Eq:A_I}) with $\epsilon \le 10^{-3}$, we recognize this type of magic texture is consistent with observations. 

We perform our numerical calculations to check the compatibility of the type A-I, A-II, $\cdots$, A-VI magic textures with observations. Table \ref{tab:typeA} shows the compatibility. The abbreviation ``Y" and symbol ``-" indicate ``consistent with observation" and ``excluded from observation", respectively. We observed that these additional six magic textures are excluded from observations.

\begin{table}[t]
\caption{Compatibility of the type A-I, A-II, $\cdots$, A-VI magic texture with observation. The abbreviation``Y" and symbol ``-" indicate ``consistent with observation" and ``excluded from observation", respectively.}
\begin{center}
\begin{tabular}{|c|c|c|c|}
\hline
type &$\epsilon = 10^{-3}$ & $\epsilon = 10^{-2}$ & $\epsilon = 10^{-1}$  \\ 
\hline
A-I & - & Y & Y \\
\hline
A-II & - & Y & Y \\
\hline
A-III & - & Y & Y \\
\hline
A-IV & - & Y & Y \\
\hline
A-V & - & Y & Y \\
\hline
A-IV & - & Y & Y \\
\hline
\end{tabular}
\end{center}
\label{tab:typeA}
\end{table}

\section{Leptogenesis\label{section:leptogeneisis}}

\subsection{Baryon asymmetry of the Universe}
The main aim of this paper is to find new types of magic textures. Although we have already achieved our goal in the previous section, some additional numerical studies may be required to improve our discussions. In this section, we use magic textures to estimate the predicted baryon asymmetry of the Universe via the  leptogenesis scenario \cite{Fukugita1986PLB,Luty1992PRD,Covi1996PLB,Buchmuller1998PLB,Akhmedov2003JHEP,Guo1004PLB,Davidson2008PhysRep}. Since the leptogenesis scenario for the type I magic texture has already been studied in Refs. \cite{Verma2020JPGNP}, we estimate the baryon asymmetry of the Universe via leptogenesis scenario for the type IV and type IX magic textures\footnote{As shown in Eq. (\ref{Eq:UTM}), the type I magic texture could not predict the Majorana CP-violating phases without some additional requirements. For example, $A_4$ symmetry and a broken $\mu-\tau$ symmetry have been employed in Ref. \cite{Verma2020JPGNP} to obtain the prediction of the baryon asymmetry of the Universe with the type I magic texture.}.

Usually, the baryon asymmetry of the Universe is represented in two ways \cite{Davidson2008PhysRep}:
\begin{eqnarray}
\eta_{\rm B} = \left. \frac{n_{\rm B} - n_{\rm \bar{B}}}{n_\gamma} \right|_0, \quad Y_{\rm B} = \left. \frac{n_{\rm B} - n_{\rm \bar{B}}}{s} \right|_0, 
\end{eqnarray}
where $n_{\rm B}$, $n_{\rm \bar{B}}$ and $n_\gamma$ are the number densities of baryons, antibaryons and photons, respectively, and $s$ is entropy density. A subscript 0 indicates ``at present time". The baryon-photon ratio $\eta_{\rm B}$ and co-moving baryon number $Y_{\rm B}$ are related by 
\begin{eqnarray}
\eta_{\rm B} = \left. \frac{s}{n_\gamma}\right|_0 Y_{\rm B} \simeq 7.04 Y_{\rm B}.
\end{eqnarray}
The baryon-photon ratio is also related to the density parameter of baryons $\Omega_{\rm B} = \rho_{\rm B}/\rho_{\rm c}$ as $\eta_B = 2.74 \times 10^{-8} \Omega_{\rm B}h^2$ \cite{Davidson2008PhysRep} where $\rho_{\rm B}$, $\rho_{\rm c}$ and $h$ are the energy density of the baryons, the critical energy density and the dimensionless Hubble parameter, respectively. The observed baryon asymmetry of the Universe is $\Omega_{\rm B} h^2 = 0.0224 \pm 0.0001$ in terms of the density parameter of baryons \cite{PLANCK2018AA}, or equivalently, 
\begin{eqnarray}
\eta_{\rm B} = (6.14 \pm 0.027) \times 10^{-10},
\end{eqnarray}
in terms of the baryon-photon ratio.

\subsection{Type I seesaw mechanism and leptogenesis \label{section:typeI_leptogeneisis}}
The type I seesaw mechanism \cite{Minkowski1977PLB,Yanagida1979KEK,Gell-Mann1979,Glashow1979,Mohapatra1980PRL} provides a very natural explanation of the baryon asymmetry in the Universe through the baryogenesis via leptogenesis scenarios \cite{Fukugita1986PLB,Luty1992PRD,Covi1996PLB,Buchmuller1998PLB,Akhmedov2003JHEP,Guo1004PLB}.  We show a brief review of the type I seesaw mechanism and leptogenesis.

In the type I seesaw mechanism, the right handed heavy Majorana neutrinos $N_i$ $(i=1,2,3)$ with mass $M_i$ are introduced into the particle contents of the standard model. In the type I seesaw mechanism, the flavor neutrino mass matrix $M$ is obtained by
\begin{eqnarray}
M =  m_D^T M_R^{-1} m_D,
\label{Eq:M=mDMRmD}
\end{eqnarray}
where 
\begin{eqnarray}
M_R = {\rm diag.}(M_1,M_2,M_3),
\end{eqnarray}
and $m_D =v \lambda $ denotes the Dirac mass matrix and $v=174$ GeV is the vacuum expectation value of the neutral component of the Higgs doublet. We employ the sigh conventions of Ref. \cite{Pascoli2007PRD} in Eq. (\ref{Eq:M=mDMRmD}).

The Dirac mass matrix can be written by the so-called Casas-Ibarra parametrization \cite{Casas2001NPB}
\begin{eqnarray}
m_D = \sqrt{M_R} R \sqrt{M_{\rm diag}} V^\dag,
\label{Eq:Casas-Ibarra}
\end{eqnarray}
where 
\begin{eqnarray}
\sqrt{M_R} &=& {\rm diag.}(\sqrt{M_1},\sqrt{M_2},\sqrt{M_3}), \nonumber \\
\sqrt{M_{\rm diag}} &=& {\rm diag.}(\sqrt{m_1},\sqrt{m_2},\sqrt{m_3}), 
\end{eqnarray}
here, $V$ is $U {\rm diag.} (e^{i\alpha_1},e^{i\alpha_2},1)$ and $R$ is a complex orthogonal matrix, $RR^T = R^T R=1$. 

In the leptogenesis scenario, the baryon asymmetry is produced from lepton asymmetry via the sphaleron process \cite{Fukugita1986PLB,Luty1992PRD,Covi1996PLB,Buchmuller1998PLB,Akhmedov2003JHEP,Guo1004PLB}. In the framework of the type I seesaw mechanism, the lepton asymmetry from the CP-violating out of equilibrium decay of heavy right-handed neutrinos into Higgs and leptons is given by
\begin{eqnarray}
\epsilon_i = \sum_\ell \frac{\Gamma(N_i \rightarrow H \ell) - \Gamma(N_i \rightarrow \bar{H} \bar{\ell})}{\Gamma(N_i \rightarrow H \ell) + \Gamma(N_i \rightarrow \bar{H} \bar{\ell})}.
\end{eqnarray}
We assume that the heavy right-handed neutrinos have a hierarchical mass spectrum, $M_1 < M_2 < M_3$. In this case, the decay of the lightest right-handed neutrino $N_1$ may be the dominant source of a cosmic lepton asymmetry. The lepton asymmetry generating from the decay of $N_1$ is obtained as \cite{Nardi2006JHEP,Borah2015PRD}
\begin{eqnarray}
\epsilon_1^\ell &=& \frac{1}{8\pi v^2}\frac{1}{(m_D m_D^\dag)_{11}} 
 \left\{ \sum_{j=2}^3 {\rm Im}\left[ (m_D)_{j\ell}(m_D^*)_{1\ell}(m_D m_D^\dag)_{j1}\right]  g(x_j) \right.\nonumber \\
&& \left.+ \sum_{j=2}^3 {\rm Im}\left[ (m_D)_{j\ell}(m_D^*)_{1\ell}(m_D m_D^\dag)_{1j}\right] \frac{1}{1-x_j} \right\},
\end{eqnarray}
where
\begin{eqnarray}
g(x)=\sqrt{x}\left(1+\frac{1}{1-x} - (1+x)\ln\frac{1+x}{x}  \right),
\end{eqnarray}
and $x_j=M_j^2/M_1^2$. 

If we take the sum over flavors, so-called the one-flavor approximation, we obtain the following total asymmetry
\begin{eqnarray}
\epsilon_1 &=& \frac{1}{8\pi v^2}\frac{\sum_{j=2}^3 {\rm Im}\left[ (m_D m_D^\dag)_{j1}^2\right] g(x_j)}{(m_D m_D^\dag)_{11}}. 
\end{eqnarray}
Since the denominator of $\epsilon_1$ should be 
\begin{eqnarray}
\sum_{j=2}^3 {\rm Im}\left[ (m_D m_D^\dag)_{j1}^2\right] =  \sum_{j=2}^3 {\rm Im}\left[ (R R^\dag)_{j1}^2\right],
\end{eqnarray}
by Eq.(\ref{Eq:Casas-Ibarra}), the low energy CP-violating phases, $\delta, \alpha_1, \alpha_2$ in the neutrino sector does not contribute to the origin of the lepton asymmetry \cite{Pascoli2007NPB}.

The one-flavor approximation is rigorously correct only when the interactions mediated by charged lepton Yukawa couplings are out of equilibrium. In this case, indistinguishable leptons propagate between decays and inverse decays and the processes which wash out lepton number are flavor independent. In contrast, if the lepton flavor is distinguishable, the inverse decays from electrons can destroy the lepton asymmetry carried only by the electrons. There are similar phenomena for mu and tau leptons. The asymmetries in each flavor are therefore washed out differently \cite{Barbieri2000NPB,Nardi2006JHEP, Abada2006JCAP, Abada2006JHEP,Dev2014NPB,Borah2015PRD}. When the interactions mediated by the charged tau or mu Yukawa coupling reach equilibrium, the flavors become physical. At temperatures $T < 10^{12}$ GeV, interactions involving tau Yukawa couplings enter equilibrium. Moreover, at $T < 10^9$ GeV, interactions involving muon Yukawa couplings enter equilibrium. Thus, at temperatures $T < 10^{12}$ GeV ($T < 10^9$ GeV), flavor effects become important in the calculation of lepton asymmetry. 

If we include the flavor effects into the leptogenesis scenario \cite{Barbieri2000NPB,Nardi2006JHEP, Abada2006JCAP, Abada2006JHEP,Dev2014NPB,Borah2015PRD}, we can relate the low energy CP-violating phases to the lepton asymmetry \cite{Pascoli2007NPB,Pascoli2007PRD,Moffat2019JHEP}. A nonzero lepton asymmetry can be obtained only when the right-handed neutrino decay is out of equilibrium. Otherwise both the forward and the backward processes will occure at the same rate, resulting in a vanishing asymmetry. If some flavors strongly interact with the right-handed neutrinos, the right-handed neutrinos are brought to thermal equilibrium by inverse decays. Thus, weak or mild wash-out parts are required \cite{Abada2006JHEP}. According to references \cite{Pascoli2007NPB,Borah2015PRD}, we employ the following approximate formulae for baryon asymmetry:
\begin{eqnarray}
Y_B^{(2)} = \frac{-12}{37g_*}\left[\epsilon_2 \eta \left(\frac{417}{589}\tilde{m}_2 \right)  + \epsilon_1^\tau \eta \left(\frac{390}{589}\tilde{m}_\tau \right) \right],
\end{eqnarray}
for the temperature regimes $T < 10^{12}$ GeV (two flavor regime) and 
\begin{eqnarray}
Y_B^{(3)} = \frac{-12}{37g_*}\left[\epsilon_1^e \eta \left(\frac{151}{179}\tilde{m}_e \right)  + \epsilon_1^\mu \eta \left(\frac{344}{537}\tilde{m}_\mu \right)  + \epsilon_1^\tau \eta \left(\frac{344}{537}\tilde{m}_\tau \right) \right],
\end{eqnarray}
for the temperature regimes $T < 10^9$ GeV (three flavor regime) where 
\begin{eqnarray}
\epsilon_2 &=& \epsilon_1^e+\epsilon_1^\mu, \quad \tilde{m}_2 = \tilde{m}_e +\tilde{m}_\mu, \nonumber \\
\tilde{m}_\ell &=& \frac{(m_D^*)_{1\ell}(m_D)_{1\ell}}{M_1},\nonumber \\
\eta(\tilde{m}_\ell)&=&\left[ \left( \frac{\tilde{m}_\ell }{8.25\times 10^{-3} ~{\rm eV}} \right)^{-1}+  \left( \frac{0.2\times 10^{-3} ~{\rm eV}} {\tilde{m}_\ell }\right)^{-1.16} \right]^{-1},
\end{eqnarray}
and $g_*$ denotes the relativistic effective degrees of freedom.

We note about the orthogonal matrix $R$. Since $g(x) \simeq -3/(2\sqrt{x})$ for $x \gg 1$, we obtain \cite{Pascoli2007NPB}
\begin{eqnarray}
\epsilon_1^\ell = -\frac{3M_1}{16\pi v^2} \frac{{\rm Im}\left[ \sum_{i,j} \sqrt{m_i} R_{1i} V^*_{\ell i} \sqrt{m_j}\sqrt{m_j^3} R_{1j} V_{\ell j} \right]}{\sum_i m_i |R_{1i}|^2}.
\nonumber \\
\end{eqnarray}
Thus, if the orthogonal matrix $R$ is not a diagonal matrix, even if $R$ is a real matrix, we can have $\epsilon_1^\ell \neq 0$. Otherwise, e.g., if $R=1$, all three lepton number asymmetries vanish: $\epsilon_1^\ell=0$ \cite{Pascoli2007NPB}.

For sake of simplicity, we take the real and non-diagonal orthogonal matrix 
\begin{eqnarray}
R=
\left(
\begin{array}{ccc}
\cos z  & \sin z& 0 \\
-\sin z & \cos z & 0 \\
0 & 0 & 1\\
\end{array}
\right), 
\end{eqnarray}
as a reference matrix \cite{Mahanta2019JCAP} where $z$ denotes a real parameter.

\begin{figure}[t]
\begin{center}
\includegraphics[scale=0.8]{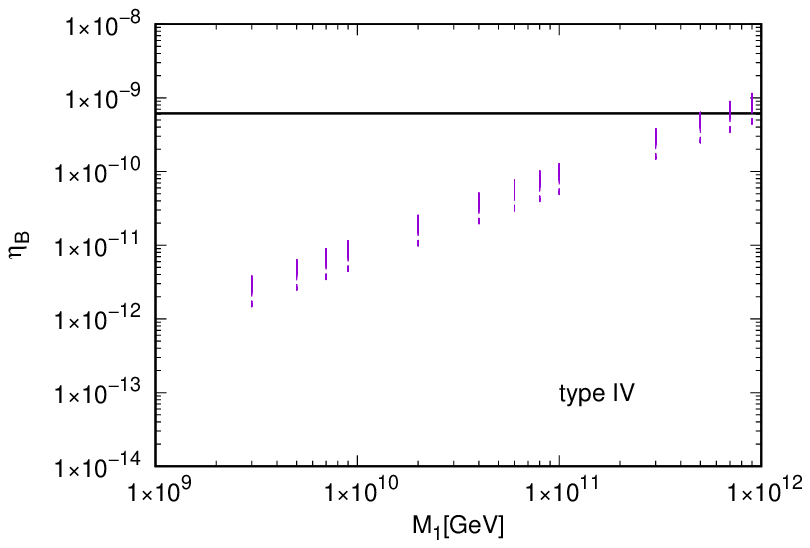}
\includegraphics[scale=0.8]{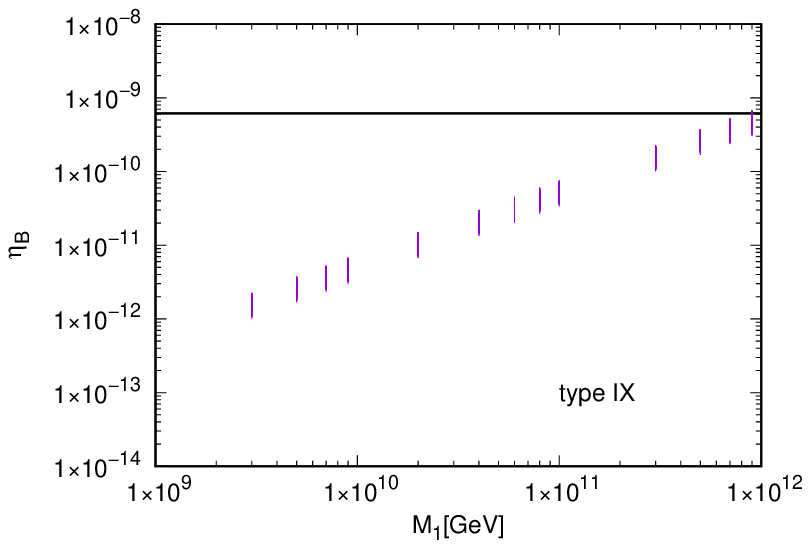}
\caption{Dependence of the baryon-photon ratio $\eta_B$ on the mass of the lightest heavy Majorana neutrino $M_1$ with the type IV (left panel) and type IX (right panel) magic textures. The horizontal line shows the observed magnitude of the baryon-photon ratio. We take $z=\pi/4$ for the orthogonal matrix $R$.}
\label{fig:etaB_M1}
\end{center}
\end{figure}

\begin{figure}[t]
\begin{center}
\includegraphics[scale=0.8]{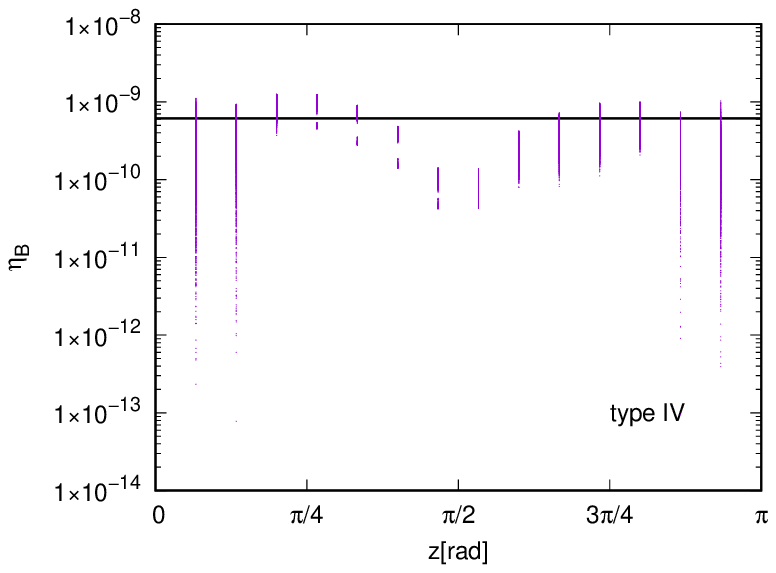}
\includegraphics[scale=0.8]{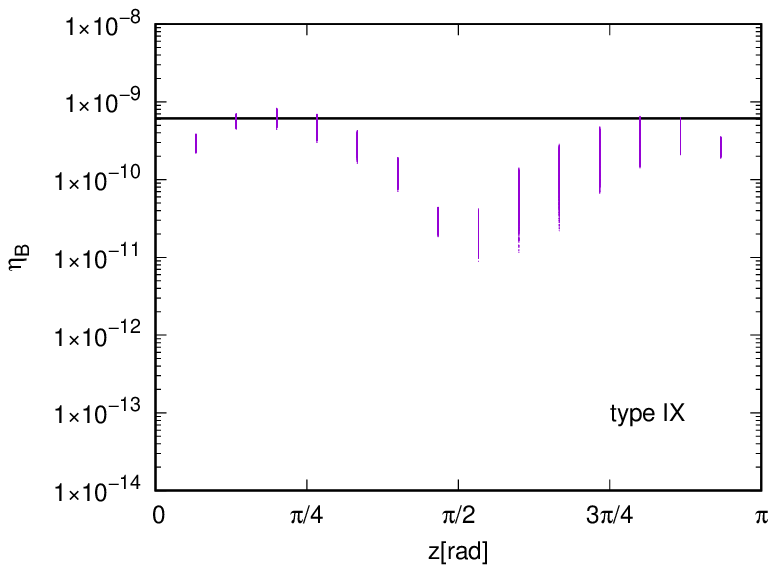}
\caption{Dependence of the baryon-photon ratio $\eta_B$ on the parameter $z$ in the orthogonal matrix $R$  for $M_1 = 10^{12}$ GeV with the type IV (left panel) and type IX (right panel) magic textures. The horizontal line shows the observed magnitude of the baryon-photon ratio.}
\label{fig:etaB_z}
\end{center}
\end{figure}

\subsection{Leptogenesis with type IV and IX}
To study the relationship between the type IV and type IX magic textures and the flavor leptogenesis scenario, we assume that the tiny masses of active neutrinos are generated by the type I seesaw mechanism and the heavy Majorana neutrinos in the type I seesaw model possess a hierarchical mass spectrum $M_1 < M_2 < M_3$, where we take $10^9 ~ {\rm GeV} \le M_1 \le 10^{12} ~{\rm GeV}$, $M_2=10 M_1$ and $M_3=5M_2$.

Figure \ref{fig:etaB_M1} shows the dependence of the baryon-photon ratio $\eta_B$ on the mass of the lightest heavy Majorana neutrino $M_1$ with the type IV (left panel) and type IX (right panel) magic textures. We take $z=\pi/4$ for the orthogonal matrix $R$. The horizontal line shows the observed magnitude of the baryon-photon ratio. To obtain Fig. \ref{fig:etaB_M1}, we use the allowed magnitudes of the neutrino parameters which are shown in Sec. \ref{section:Allowed Magic Textures}. From Fig. \ref{fig:etaB_M1}, we see that the allowed region of $M_1$ is highly constrained with a fixed $z$ in both cases of type IV and type IX magic textures.

One may wonder about the effect of the selection of the orthogonal matrix $R$ on the predicted baryon-photon ratio. Indeed, we can use any orthogonal matrix for the Casas-Ibarra parametrization in Eq.(\ref{Eq:Casas-Ibarra}). The predicted baryon-photon ratio depends on $R$. Figure \ref{fig:etaB_z} shows that the dependence of the baryon-photon ratio $\eta_B$ on the parameter $z$ in the orthogonal matrix $R$  for $M_1 = 10^{12}$ GeV with the type IV (left panel) and type IX (right panel) magic textures. The horizontal line shows the observed magnitude of the baryon-photon ratio. From Fig. \ref{fig:etaB_z}, we see that the allowed region of $z$ for the type IX magic square is more constrained than the type IV magic square with a fixed $M_1$.

Figures \ref{fig:etaB_M1} and \ref{fig:etaB_z} show that the type IV magic texture is slightly more favorable than the type IX magic texture if we include the leptogenesis arguments in our discussion.

\section{Summary\label{section:summary}}
The magic texture, the type I magic texture, has been known as one of the successful textures of the flavor neutrino mass matrix for Majorana neutrinos. The magic texture was obtained as one of the consequences of the so-called trimaximal mixing for $\nu_2$. In this paper, the magic texture is required as the first principle. The magic texture is defined as a Majorana matrix in which three independent sums are the same. Under this definition, not only type I magic texture but also new nine matrices are classified as magic textures. We have shown that two of these new magic textures for Majorana neutrinos, type IV and type IX, are also consistent with the neutrino oscillation experiments, observation of the cosmic microwave background radiation, and neutrinoless double beta decay experiments. Moreover, we observed that if the tiny masses of active neutrinos are generated by the type I seesaw mechanism and the origin of the baryon asymmetry of the Universe is the asymmetric decays of the lightest heavy Majorana neutrino in the type I seesaw model, the type IV magic texture is slightly more favorable than the type IX magic texture.  

Finally, we would like to address the relation between the magic textures and the sterile neutrinos. The effects of the existence of the sterile neutrinos on the trimaximal mixing for $\nu_2$, the type I magic texture, have been studied in Refs. \cite{Dev2016NPB,Dev2019NPB}. The roles of the sterile neutrinos for the type IV and type IX magic textures may be interesting. A detailed analysis of these two topics is required in the future. 




%

\vspace{0.2cm}
\noindent




\end{document}